\documentclass[prd,aps,twocolumn,superscriptaddress,floatfix,preprintnumbers,nofootinbib]{revtex4-2}
\usepackage{amsmath}
\usepackage{graphicx}
\usepackage{xcolor}
\usepackage{amsfonts}
\usepackage{natbib}
\usepackage{tikz}
\usepackage{siunitx}
\usepackage{tikz}
\usepackage{ragged2e}
\usetikzlibrary{math}
\usepackage{amssymb,amsmath,bm}
\usepackage[citecolor=blue,pdfa=true,linktocpage=true,urlcolor=blue,colorlinks=true]{hyperref}
\usepackage[export]{adjustbox}
\usepackage{multirow}
\usepackage{physics}
\usepackage[capitalise]{cleveref}
\usepackage{soul}
\creflabelformat{equation}{#2\textup{#1}#3}

\newcommand{\aap}{{Astron.~Astrophys.}}
\newcommand{\apjl}{{Astrophys.~J.~Lett.}}
\newcommand{\apjs}{{Astrophys.~J.~Supp.}}
\newcommand{\aj}{{Astron.~J.}}
\newcommand{\mnras}{{Mon.~Not.~R.~Astron.~Soc.}}
\newcommand{\jcap}{JCAP}

\newcommand{\be}{\begin{equation}}
\newcommand{\ee}{\end{equation}}

\newcommand{\bea}{\begin{eqnarray}}
\newcommand{\eea}{\end{eqnarray}}
\newcommand{\bdm}{\begin{displaymath}}
\newcommand{\edm}{\end{displaymath}}

\def\fNL{f_{\mathrm{NL}}}

\newcommand{\bk}{\mathbf{k}}
\newcommand{\bn}{\mathbf{n}}

\def\bx{\bold{x}}
\def\bv{\bold{v}}
\def\bq{\bold{q}}

\def\bn{\bold{n}}
\def\mH{\mathcal{H}}

\def\br{\bold{r}}

\def\bDx{\bold{\Delta}}

\begin{document}

\title{Impact and measurability of linear relativistic effects in galaxy surveys}

\author{Robin Y. Wen}
\email{ywen@caltech.edu}

\affiliation{California Institute of Technology, 1200 East California Boulevard, Pasadena, California 91125, USA}

\author{Henry S. {Grasshorn Gebhardt}}
\affiliation{California Institute of Technology, 1200 East California Boulevard, Pasadena, California 91125, USA}
\affiliation{Jet Propulsion Laboratory, California Institute of Technology, Pasadena, California 91109, USA}

\author{Olivier Dor\'e}
\affiliation{California Institute of Technology, 1200 East California Boulevard, Pasadena, California 91125, USA}
\affiliation{Jet Propulsion Laboratory, California Institute of Technology, Pasadena, California 91109, USA}

\begin{abstract}
The three-dimensional galaxy power spectrum is a powerful probe of primordial non-Gaussianity (PNG) and additional general relativistic (GR) effects on large scales, which can be constrained by current and upcoming large-scale structure surveys. In this work, we forecast the measurability of local PNG and linear-order relativistic effects in the spherical Fourier-Bessel (SFB) power spectrum for DESI, Euclid, and SPHEREx surveys. A Chebyshev-decomposition scheme is employed to accelerate the multi-tracer SFB power-spectrum calculations. Fisher forecasts establish baseline constraints and test the sensitivity of \(\fNL\) constraints to SFB mode cuts and to the marginalization over primordial cosmological parameters, while simulated Bayesian inference is used to quantify the impact and measurability of relativistic effects. We find that neglecting GR effects can bias \(\fNL\) constraints at the \(1\)--\(3\sigma\) level for Euclid and SPHEREx. The degeneracy between GR and PNG terms is strongly tracer dependent, with the degradation of \(\sigma(\fNL)\) ranging from a few percent to nearly a factor of two when the GR amplitudes are varied. Lensing can be detected at high significance for several tracers, while multi-tracer analyses substantially improve the measurability of the Doppler term. Assuming GR, we show that relativistic clustering partially breaks the exact \(b_\phi \fNL\) product degeneracy present in Newtonian linear power-spectrum analyses, although the resulting \(b_\phi\) constraints remain weak for the survey configurations considered. The joint PNG--GR inference consistently propagates uncertainty in \(b_\phi\) into the marginalized \(\fNL\) constraint. This firmly establishes the path toward extracting cosmological information from ultra-large-scale galaxy clustering.
\end{abstract}

\maketitle
\tableofcontents

\clearpage
\section{Introduction}

Current Stage-IV galaxy surveys such as Dark Energy Spectroscopic Instrument (DESI) \cite{24DESI_II}, Euclid \cite{24Euclid_overview}, and SPHEREx \cite{26SPHEREx} are tracing the large-scale structure (LSS) of the Universe over increasingly large volumes by measuring the positions and spectra of hundreds of millions of galaxies. The unprecedented constraining power of these surveys offers exciting prospects for probing fundamental physics at scales comparable to the Hubble horizon.

The primary target on these ultra-large scales is local primordial non-Gaussianity (PNG), parametrized by the $f_{\rm NL}$ parameter \cite{01Komatsu_fNL}. Constraining $f_{\rm NL}$ enables discrimination between different inflationary scenarios: multi-field models typically predict $|f_{\rm NL}|\sim 1$ \cite{10Chen_PNG_review,12Senatore_EFT,14Alvarez}, whereas models with a single field generally predict much smaller values \cite{03Maldacena_single_field,04Creminelli_single_field}. Local PNG imprints a characteristic scale-dependent bias in the large-scale galaxy power spectrum \cite{07Dalal_PNG,08Slosar_fnl,08Matarrese_PNG}, a key feature that Stage-IV surveys target.

Additional relativistic effects at the late time also manifest themselves at the ultra-large scales due to the redshift space distortion (RSD)—the mapping between the observed coordinates of an object on the sky and its true position on the rest-frame light cone. Traditionally, the Newtonian modeling of the RSD only considers the impacts of the galaxies' peculiar velocities on the observed redshift \cite{87Kaiser,98Hamilton_RSD}. As we probe galaxy clustering at increasingly larger scales,  the effects of general relativity (GR) at the linear order becomes important \cite{09Yoo_GR,11ChallinorLPS,11Bonvin_GR,12JeongLPS}. These relativistic effects are non-negligible theoretical contaminants for surveys aiming to achieve the precision of $\sigma(f_{\rm NL})\sim1$ \cite{23WAGR,23Foglieni,15Alonso_GR_single,15Alonso_GR_multi}. 

Besides as a theoretical contaminant that can potentially bias the local PNG measurement, these linear relativistic effects also contain additional information on cosmology and galaxy formation. As GR effects contain terms including density, velocity, and potential perturbations, among which the relationships are specified by Einstein’s equations, these relativistic effects potentially allow us to strengthen cosmological parameters and test modified gravity and dark energy models at the perturbation level beyond the standard $\Lambda$CDM paradigm \cite{15Baker_MG,20Duniya_HG,21Blanco_anisotropic_stress,23Blanco_time,24Castello,25Castello}.

These relativistic terms are also proportional to the magnification bias $s$ and evolution bias $b_{\rm e}$, making these effects useful probes of galaxy (tracer) properties \cite{24Blanco_galaxy_property}. The magnification bias quantifies how flux magnification alters the galaxy selection, thereby encoding information on galaxy luminosity functions. Meanwhile, the evoluton bias—defined as the logarithmic derivative of the tracer's comoving number density with respect to the scale factor—captures the physical evolution of the galaxy population.

Motivated by recent work \cite{25Sullivan_bphi_be,25Dalal_bphi}, measuring the evolution bias is especially important for constraining $f_{\rm NL}$ through the scale-dependent bias effect. This is because the evolution bias $b_{\rm e}$ can be considered equivalent to the local PNG bias $b_{\phi}$, which controls the amplitude of the scale-dependent PNG contribution through $b_{\phi} f_{\rm NL}$. Accurate knowledge of $b_\phi$ is essential both for unbiased $f_{\rm NL}$ constraints \cite{20Barreira_bphi_impact_constraint,22Barreira_bphi_BOSS} and for enhancing them through multi-tracer analyses \cite{23Sullivan_bphi,23Barreira_multi}. Forecasts and analyses typically adopt the universality relation for \(b_{\phi}\), since constraining \(b_{\phi}\) has generally been thought to require simulations of realistic tracers, which remain difficult to model robustly due to uncertainties in galaxy formation. The equivalence between $b_{\rm e}$ and $b_{\phi}$ enables a more data-driven approach bypassing galaxy simulations: one can either directly measure the tracer number-density evolution, provided the selection effects are well understood \cite{25Sullivan_bphi_be}, or measure the evolution bias $b_{\rm e}$ through the  relativistic effects in galaxy clustering, which we explore in this work.

To accurately quantify the impact and measurability of linear relativistic effects for Stage-IV surveys, we use the spherical-Fourier Bessel (SFB) power spectrum \cite{93lahav_spherical,95Fisher_SFB,95Heavens_SFB}. By preserving the geometry of the curved sky, the SFB basis naturally includes wide-angle and redshift-evolution effects in 3D clustering, enabling an accurate modeling of relativistic contributions on the light cone \cite{13Yoo_GR_SFB,24Semenzato_SFB_GR,24GR_SFB}. This stands in contrast to Cartesian statistics, where the exact inclusion of GR effects is considerably more challenging \cite{18Tansella-GR,22CatorinaGR-P,23Foglieni}. The SFB power spectrum also accounts for the differing lines-of-sight (LOS) of the galaxies in a galaxy pair and captures the full information from linear-order, anisotropic clustering in 3D. Its recent application to eBOSS galaxy samples further demonstrates that the framework is ready for current and future large-scale-structure analyses \cite{26Bruton}. It is therefore the ideal framework for investigating GR effects. 

The full set of linear-order relativistic contributions to the SFB power spectrum has been calculated in Refs.~\cite{13Yoo_GR_SFB,24Semenzato_SFB_GR,24GR_SFB}. In particular, we build upon Ref.~\cite{24GR_SFB}, which offers a validated calculation\footnote{Their validation is performed by mapping the SFB power spectrum to the angular power spectrum and comparing against results from \texttt{CLASS} \cite{11class,13DiDio_classgal}.}. Their SFB basis is also properly discretized for spherical-shell geometry \cite{19Samushia_SFB} and is directly applicable to data \cite{21Gebhardt_SuperFab}. To enable Markov chain Monte Carlo (MCMC) analyses with the full GR effects, we introduce a new Chebyshev-decomposition method for the SFB power spectrum, reducing the evaluation time of the full GR SFB power spectrum to $O(1{\rm s})$\footnote{Ref.~\cite{24GR_SFB} adopts the Iso-qr integration method (1D spherical Bessel integration along constant $q$-$r$ lines) of Ref.~\cite{23Gebhard_SFB_eBOSS}, requiring tens of seconds per evaluation for the full GR SFB power spectrum.}. To our knowledge, this work provides the first demonstration of Bayesian inference in 3D clustering incorporating the full linear relativistic effects.

This paper is outlined as follows. We begin by summarizing the linear relativistic effects and local PNG in Secs.~\ref{sec:GR} and \ref{sec:local-PNG}, followed by a discussion of the connection between evolution and local PNG biases in Sec.~\ref{sec:bphi-be-equivalence}. We then review the SFB power spectrum formalism in Sec.~\ref{sec:SFB-decomp} and summarize previous GR and SFB forecasting literature in Sec.~\ref{sec:review}. In Sec.~\ref{sec:forecast-formalism}, we describe the simulated inference and Fisher forecast frameworks, along with the DESI, Euclid, and SPHEREx survey specifications adopted in this work. We present our results on the baseline Newtonian forecasts, the impacts of GR effects on $f_{\rm NL}$ constraints, and their measurability in Sec.~\ref{sec:Fisher-baseline}--\ref{sec:measurability-GR}. We discuss the constraints on $b_{\phi}$ through relativistic clustering in Sec.~\ref{sec:bphi-implication} and conclude in Sec.~\ref{sec:conclusion}. 

For SPHEREx galaxy samples where redshift uncertainties can be significant, we provide an exact treatment of redshift error/Fingers-of-God effects in the SFB basis in Appendix~\ref{sec:redshift-error}. The new Chebyshev-decomposition method for computing the multi-tracer SFB power spectra is described in detail in Appendix~\ref{sec:Chebyshev-method}.

In our forecasts, we ignore the impacts of realistic window and integral constraints (IC) \cite{19Mattia_IC} on the theoretical modeling. The effects of partial sky coverage is accounted in covariance through a simple $f_{\rm sky}$ rescaling.  We adopt the  
\textit{Planck} 2018 best-fit $\Lambda$CDM cosmology \cite{18Planck_Parameter} and $\fNL=0$ as the fiducial model, and fix the late-time cosmological parameters ${\Omega_{\rm b}, \Omega_{\rm c}, H_0}$. This choice fixes the redshift-distance relation and avoids the need to model the Alcock-Paczynski effect 
\cite{79AP_effect}. These parameters generally have minimal impacts on $f_{\rm NL}$ constraints. 

We quantify the measurability of individual relativistic contributions by assigning each an independent amplitude parameter (see Sec.~\ref{sec:measurability-GR}). A full assessment of how specific dark energy or modified-gravity models could be constrained would require a self-consistent treatment of both background evolution and perturbations within the SFB framework—an extension we leave to future work.

\section{Background}\label{sec:background}

\subsection{Relativistic effects in galaxy clustering}\label{sec:GR}

We briefly review the linear-order relativistic contributions to galaxy clustering, which are first derived in Refs.~\cite{09Yoo_GR,11ChallinorLPS,11Bonvin_GR,12JeongLPS}. For a galaxy sample selected by a given survey, the galaxy density fluctuation $\delta_{\rm g}(\hat{\bn}, z)$ is gauge-invariant and mostly commonly expressed in the conformal Newtonian gauge:
\begin{align}
dx^2=a(\tau)^2[-(1+2\Psi)d\tau^2+(1-2\Phi)\delta_{ij}dx^idx^j],
\label{eq:CNG}
\end{align}
where $\tau$ denotes the conformal time, and $\Psi$ and $\Phi$ are the temporal and spatial Bardeen potentials.

Ignoring the observer's terms \cite{20GrimmGR,18Scaccabarozzi_GR_TP,22CatorinaGR-P}\footnote{The observer's potential and velocity only impacts the angular monopole and dipole under a full-sky survey \cite{21Ginat_monopole,22Dalang_dipole,24GR_SFB}. We can therefore ignore these terms by imposing $\ell_{\rm min}\geq 2$ for clustering analyses.}, the observed galaxy density contrast is \cite{22CatorinaGR-P,23WAGR,24GR_SFB}
\begin{align}
\delta_{\rm g}(\hat{\bn},z)&=b_1D_{\rm m}-\frac{1} {\mH}\frac{\partial\bv}{\partial x}\cdot\hat{\bn}\nonumber\\
&-(2-5s)\kappa\nonumber\\
&-\mathcal{A}_1\bv\cdot\hat{\bn}\nonumber\\
&+\mathcal{A}_1\Psi-(2-5s)\Phi+\Psi+\frac{1}{\mH}\dot{\Phi}+(b_{\rm e}-3)\mathcal{H}V\nonumber\\
&-\frac{2-5s}{x}\int_{\tau_0}^{\tau(z)}(\Psi(\tau')+\Phi(\tau'))\,d\tau'\nonumber\\
&-\mathcal{A}_1\int_{\tau_0}^{\tau(z)}(\dot{\Psi}(\tau')+\dot{\Phi}(\tau'))\,d\tau'.
\label{eq:GR}
\end{align}
We have grouped the full expression line-by-line as follows: DRSD (the standard density and Newtonian RSD term), lensing, Doppler, NIP (non-integrated potential), Shapiro (time-delay), and ISW (integrated Sachs–Wolfe) effects, with the last three collectively referred to as the GP (gravitational potential) term. 

In the above expression, $D_{\rm m}$ is the matter density fluctuation in the comoving gauge, $\bv$ is the peculiar velocity of the galaxies in the conformal Newtonian gauge (the velocity potential $V$ is define via $\bv=-\nabla V$), and $\kappa$ is the lensing convergence. The dot above the variable refers to a partial derivative with respect to the conformal time, $\mathcal{H}$ is the conformal Hubble parameter, and $b_1$ is the linear galaxy bias. The redshift-dependent coefficient $\mathcal{A}_1$, which directly sets the amplitude of the Doppler and ISW terms, is defined as
\begin{align}
\mathcal{A}_1(z)\equiv\frac{\dot{\mH}}{\mH^2}+\frac{2-5s}{\mH x}+5s-b_{\rm e}.
\label{eq:A1}
\end{align}

For realistic surveys with complex selection functions, the magnification bias $s$ quantifies how the number density of the sample responds to changes in observed flux. Under the simplifying assumption of a luminosity threshold $L_{\rm c}$ (i.e., an absolute magnitude-limited sample), magnification bias reduces to
\begin{align}
s&\equiv-\frac{2}{5}\frac{\partial \ln \bar{n}_g}{\partial \ln L}\bigg|_{\ln L_{\rm c}}\,.
\label{eq:s}
\end{align}
Note that magnification bias is already important for Stage-III and IV photometric surveys, since the lensing magnification term must be included in the angular galaxy clustering analyses in order to achieve unbiased parameter constraints \cite{21von_magnification,24Krolewski_CMB_quadar_fnl,22Euclid_mag_photoz,22Elvin-Poole_DES_MB}. The magnification bias can be directly measured from data or simulation by perturbing galaxy magnitudes and re-running the selection pipeline to determine the resulting changes on the mean galaxy number densities \cite{22Elvin-Poole_DES_MB,23Zhou_DESI_LRG_cross,24Wenzl_mag_bias_SDSS}.

The evolution bias $b_{\rm e}$ measures the evolution of the mean galaxy density in the comoving space
\begin{align}
b_{\rm e}&\equiv \frac{\partial\ln \bar{n}_g}{\mH \partial \tau} = \frac{\partial\ln \bar{n}_g}{\partial\ln a}\,.
\label{eq:be}
\end{align}
The above equation employs a partial derivative with respect to time, such that it only captures the intrinsic evolution of the tracer population, driven by galaxy formation and evolution processes. The evolution bias therefore measures the number density evolution of tracers with respect to their rest-frame absolute luminosity and color, as well as intrinsic properties such as stellar mass and star-formation history.

Consequently, one cannot simply evaluate the total time derivative of the observed radial number density of a galaxy sample \cite{21Maartens_be,25Sullivan_bphi_be}, since survey samples are typically defined based on observed properties such as apparent magnitude or observed-frame color that themselves depend on redshift. This makes a direct determination of the evolution bias from the observed number-density evolution challenging, as it requires accurate modeling and control of the (often complex) selection function.

So far, only a limited number of studies have attempted to measure evolution biases of tracers directly from observational data. Ref.~\cite{20Wang_GR_quasar} infers the evolution bias of eBOSS quasars (QSO) using luminosity functions measured in Ref.~\cite{16Palanque_eBOSS_QSO}, assuming an absolute magnitude threshold for the sample. Ref.~\cite{25Sullivan_bphi_be} corrects for magnification-bias effects in the observed radial number densities of the BOSS luminous red galaxy (LRG) sample, but finds significant residual selection effects. Ref.~\cite{25Shaw_DopplerBias} studies the impact of redshift-induced distortions in observed galaxy colors on the selection of DESI emission line galaxies (ELG) sample\footnote{Observed galaxy colors are measured using fixed photometric bandpasses. The redshifting of galaxy spectra, including Doppler shifts induced by peculiar velocities, moves spectral features such as emission lines and continuum breaks into and out of these bandpasses. This changes the observed colors and can affect whether galaxies pass color-based selection criteria.}. This effect is termed Doppler bias and can be viewed as a generalization of the K-correction \cite{02Hogg_kcorrection}. This effect is shown to be substantial for line-selected samples, with a magnitude comparable to the evolution bias. The Doppler bias effect must therefore be accounted for when inferring the number-density evolution of line-selected samples. These studies illustrate the challenges of measuring evolution bias from the observed radial densities of galaxies.

\subsection{Local primordial non-Gaussianities}\label{sec:local-PNG}

We have so far reviewed the linear-order GR effects in galaxy clustering assuming Gaussian initial conditions. However, many early-Universe scenarios could lead to primordial non-Gaussianities. In particular, the local-type PNG includes a local, quadratic correction to the primordial gravitational potential $\phi_{\rm p}$ parametrized by $f_{\rm NL}$ \cite{01Komatsu_fNL}:
\begin{align}
\phi_{\rm p}(\bx)=\phi_{\rm G}(\bx)+f_{\rm NL}\left[\phi_{\rm G}(\bx)^2-\langle\phi_{\rm G}(\bx)^2\rangle\right]\label{eq:fNL}.
\end{align}
For consistency with the standard sign convention—where positive $f_{\rm NL}$ enhances large-scale clustering—we define $\phi \equiv -\Phi$, with $\Phi$ as the Bardeen potential in \cref{eq:CNG}.

Under local PNG, the relation between biased tracers, such as galaxies and halos, and the underlying matter density field is modified at the linear order as \cite{07Dalal_PNG,08Slosar_fnl}
\begin{align}
\delta_{\rm g}^{\rm LPNG} = \delta_{\rm g}+b_{\phi}f_{\rm NL}\phi \label{eq:bias-PNG},
\end{align}
where the local PNG bias $b_{\phi}$ describes the response of galaxy formation to the large-scale potential $\phi$ \cite{08McDonald_PNG,15Assassi_PNGbias}. The potential at the time of structure formation (the matter dominated epoch $z_{\rm md}$) is
\begin{align}
\phi(z_{\rm md})=\frac{D_{\rm m}(z)}{\alpha(k,z)}=\frac{3H_0^2\Omega_{\rm m0}}{2k^2\tilde{D}(z)T(k)}D_{\rm m}(z)\label{eq:scale-dependent-fnl},
\end{align}
where $\tilde{D}(z)$ is the growth factor normalized to the scale factor $(1+z)^{-1}$ at the matter-dominated epoch,  $T(k)$ is the matter transfer function, and $\Omega_{{\rm m}0}$ and $H_0$ are the matter density and the Hubble parameter today. The $k^{-2}$ scaling in the potential gives the characteristic scale-dependent bias signature for local PNG. At the linear order in \cref{eq:bias-PNG}, the parameters $b_{\phi}$ and $f_{\rm NL}$ are exactly degenerate with each other. Therefore, one has to assume certain relations or priors on $b_\phi$ in order to obtain any direct constraint on $f_{\rm NL}$. 

Assuming halo abundance models with a universal mass function (UMF), one can obtain the universality relation~\cite{08Slosar_fnl,15Tellarini_bias} 
\begin{align}
    b_{\phi}(b_1)=2\delta_{\rm c}(b_1-p),
    \label{eq:bphi_universal}
\end{align}
where $\delta_{\rm c}=1.686$ is the threshold for spherical collapse and $p=1$ under the UMF assumption. The above relation has been adopted in most forecast studies and data analyses of local PNG, as it removes the uncertainty in $b_{\phi}$ by tying it directly to the linear bias $b_1$, which can be tightly constrained from 2-point clustering measurements.

In practice, however, real galaxy samples selected from surveys do not necessarily obey the UMF, causing \cref{eq:bphi_universal} to break down. A commonly used phenomenological remedy is to retain the linear form of \cref{eq:bphi_universal} while allowing the parameter $p$ to deviate from unity. For example, Refs.~\cite{22Cabass_BOSS,25Chudaykin_DESI_PNG} adopt $p=0.55$ for LRG samples\footnote{However, there is currently no consensus on the appropriate value of $p$ for LRG samples, as other studies such as Refs.~\cite{23_DESI_LRG,25Bermejo_LRGCMB_PNG,24_DESI_Y1_PNG} continue to adopt the universal relation ($p=1$) as the default choice for LRG analyses.}, motivated by simulation results for galaxies selected by stellar mass \cite{20Barreira_PNG_bias}. For quasars, it has become customary to use $p=1.6$, following the proposal of Ref.~\cite{08Slosar_fnl} for halos experiencing recent mergers.

While this generalized parametrization partially alleviates the restriction forced by the UMF assumption, varying $p$ alone does not capture the full complexity of the relationship between $b_1$ and $b_{\phi}$ for realistic tracers. In general, this relation depends on redshift, galaxy formation physics, and the properties of the observed galaxies and their host halos.

If realistic galaxy samples can be reliably modeled, for example with hydrodynamical simulations, halo occupation distribution (HOD) prescriptions, or semi-analytic models, \(b_{\phi}\) can in principle be measured from the scale-dependent bias in nonzero-local-PNG simulations \cite{17Biagetti,26Merino_nozeroPNGbias}
or from separate-universe responses \cite{20Barreira_PNG_bias,23Lazeyras_bphi,24Hadzhiyska_abacusPNG}. In the separate-universe framework, the effect of a long-wavelength primordial potential perturbation can be absorbed into a change in the local background density fluctuation \cite{16Baldauf_SU}. Under the peak-background split, this absorption leads to the relation
\begin{equation}
b_{\phi}=4\pdv{\bar{n}_{\rm g}}{\log A_{\rm s}}
=2\pdv{\bar{n}_{\rm g}}{\log \sigma_8}\label{eq:bphi-SU},
\end{equation}
where $\sigma_8 \propto \sqrt{A_{\rm s}}$. In practice, $b_{\phi}$ is estimated by measuring the numerical response of the mean galaxy number densities to small changes in $\sigma_8$ or $A_{\rm s}$ values with different simulations. 

The separate-universe formalism has become the primary method for estimating $b_{\phi}$ for galaxy samples \cite{26Perez_bphi,26Moore_bphi}. However, its applicability hinges on the ability to simulate galaxy populations that are sufficiently realistic and representative of observed tracers. Given the substantial uncertainties associated with galaxy formation physics, this remains a major challenge. As a result, while significant progress has been made, placing robust and reliable priors on $b_{\phi}$ for real survey tracers using simulations has yet to be demonstrated. The uncertainty in $b_{\phi}$ represents one of the major theoretical systematics in directly constraining $f_{\rm NL}$ through LSS.

\subsection{$b_{\phi}$ and $b_{\rm e}$ connection}\label{sec:bphi-be-equivalence}

\begin{figure}[tbp]
\centering
\includegraphics[width=0.4\textwidth]{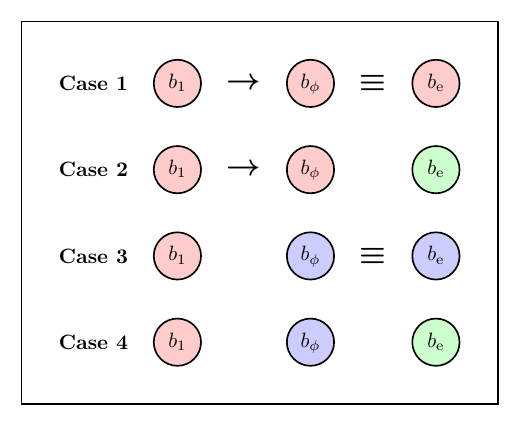}
\caption{Summary of four possible parameterizations of the linear bias $b_1$, local PNG bias $b_\phi$, and evolution bias $b_{\rm e}$. We use \(A\rightarrow B\) to indicate that \(A\) predicts \(B\), and \(A\equiv B\) to indicate that the two parameters are physically equivalent. Case 1: $b_1$ fully determines $b_\phi$ via the universality relation (or its linear generalization with parameter $p$ in \cref{eq:bphi_universal}), and $b_{\rm e}$ is fixed by the $b_\phi$–$b_{\rm e}$ equivalence in \cref{eq:bphi-be-equiv}; only $b_1$ is a free parameter. Case 2: $b_\phi$ is fixed by the universality relation, while $b_{\rm e}$ is treated as an independent parameter unrelated to $b_\phi$. Case 3: No universality relation is assumed; $b_\phi$ and $b_{\rm e}$ share a single degree of freedom through their equivalence relation. Case 4: All three parameters $b_1$, $b_\phi$, and $b_{\rm e}$ are treated as independent.}
\label{fig:biasparams}
\end{figure}

To circumvent the dependence on our incomplete understanding of galaxy formation physics, it is desirable to measure the local PNG bias of tracers directly from data. Applying the separate-universe approach in \cref{eq:bphi-SU} to observational data requires identifying regions that effectively probe different large-scale density environments. A natural strategy is to exploit redshift evolution, since the amplitude of matter fluctuations $\sigma_8(z)$ varies with redshift. Different tracer number densities across redshift slices therefore correspond to different effective $\sigma_8$ environments. Heuristically, the redshift evolution of tracers enables an empirical realization of the separate-universe picture in observational data.

Ref.~\cite{25Sullivan_bphi_be} formalizes this intuition by explicitly identifying an equivalence between the local PNG bias $b_{\phi}$ and the evolution bias $b_{e}$: within the peak-background split framework, the two parameters share a single degree of freedom. An early indication of this equivalence can be traced back to Ref.~\cite{12JeongLPS}, which shows that, under the UMF assumption for halo abundance, the evolution bias is related to the linear bias via
\begin{align}
b_e(z) = \delta_{\rm c} f(z)[b_1(z)-p], \label{eq:be-universal}
\end{align}
where $f(z)$ is the linear growth rate and $p=1$ for the UMF. Noting the similarity between this relation and the UMF prediction for the $b_1$–$b_\phi$ relation in \cref{eq:bphi_universal}, one can obtain
\begin{equation}
b_{e}(z)=\frac{f(z)}{2}b_{\phi}(z)\,,
\label{eq:bphi-be-equiv}
\end{equation}
under the UMF assumption.

Despite this similarity, the connection between the two biases has been largely overlooked. Most previous studies of local PNG and relativistic effects assume the universality relation for $b_\phi$, while treating the evolution bias as an independent quantity inferred from the time derivatives of the expected tracer luminosity functions, which generally yields results that differ from UMF-based predictions. Ref.~\cite{25Sullivan_bphi_be} showed that the connection in \cref{eq:bphi-be-equiv} holds generally within the peak-background split framework, not limited to the UMF assumption, and validated this equivalence on both N-body and hydrodynamical simulations.

The physical origin of this equivalence can be understood from the perspective of local structure formation. Long-wavelength gravitational potentials modify the local expansion history by rescaling the local scale factor in the conformal Fermi coordinates (CFC) \cite{15Dai_CFC,15Dai_SU}, an effect equivalent to a perturbation of the local proper time \cite{18Desjacques_Galaxy_Bias}. Both long-wavelength potentials and proper-time perturbations therefore change the effective age of the local Hubble patch, to which galaxy formation is sensitive\footnote{As argued in Ref.~\cite{12JeongLPS}, at linear order galaxy formation is sensitive only to the local mean density fluctuation and to the evolutionary stage of the Universe, quantified by the linear bias $b_1$ and the evolution bias $b_e$, respectively.}.

Although these effects originate from distinct physical sources—primordial non-Gaussianity in the initial conditions versus relativistic effects at late times—both act as a local rescaling of the scale factor. Consequently, both $b_\phi$ and $b_{\rm e}$ quantify the response of the tracer number density to a rescaling of the local time evolution. This equivalence is ultimately rooted in the equivalence principle of general relativity: long-wavelength gravitational potentials and proper time perturbations are indistinguishable for local physical processes.

Motivated by this equivalence, we summarize in \cref{fig:biasparams} four possible relations among $b_1$, $b_\phi$, and $b_{\rm e}$. Under the standard assumption of the universality relation for the local PNG bias, the evolution bias should also follow its universal form given in \cref{eq:be-universal}, corresponding to Case~1 in \cref{fig:biasparams}. Throughout this work, we adopt Case~1 as the baseline assumption in our forecasts.

This choice contrasts with Case~2, which is implicitly adopted in much of the previous literature, where $b_\phi$ is fixed by the universality relation while $b_{\rm e}$ is treated as an independent parameter and allowed to take different values. In light of the $b_\phi$–$b_{\rm e}$ equivalence, we therefore take Case~1 to be the preferred default over Case~2\footnote{There may nevertheless be motivations for Case~2 and 4. For example, Doppler contributions arising from velocity effects on observed colors can introduce an additional bias beyond the standard evolution bias for tracers selected by observed colors (see footnote~4) \cite{25Shaw_DopplerBias}. If such Doppler bias is non-negligible, the effective total evolution bias (evolution bias plus Doppler bias) will deviate from the values implied by the equivalence relation of \cref{eq:bphi-be-equiv}, leading to Cases 2 and 4 where the two biases are not equivalent.}.

The equivalence between $b_\phi$ and $b_{\rm e}$ also opens up the possibility of constraining $b_\phi$ directly from data for real tracers, where the universality relation may not hold. This scenario corresponds to Case~3 in \cref{fig:biasparams}. As discussed at the end of Sec.~\ref{sec:GR}, the evolution bias has already been measured in observational data—most notably for eBOSS quasars \cite{20Wang_GR_quasar}—through the redshift evolution of luminosity functions. Such measurements therefore already provide data-driven estimates of $b_\phi$\footnote{Interpreting the evolution-bias measurements of Ref.~\cite{20Wang_GR_quasar} with the $b_\phi$–$b_{\rm e}$ equivalence implies that the inferred local PNG bias for eBOSS quasars at $z<2.2$ deviates significantly from the standard parametrization in \cref{eq:bphi_universal} for both $p=1$ and $p=1.6$.}.

Alternatively, since the evolution bias enters directly into the relativistic effects, measuring these effects in galaxy clustering provides an alternative data-driven avenue for constraining $b_\phi$. This constitutes one of the central goals of this work, and we present our findings in Sec.~\ref{sec:bphi-implication}. Accordingly, we focus on Cases~1 and 3 in this paper.

\subsection{SFB power spectrum}\label{sec:SFB-decomp}
The spherical Fourier-Bessel basis is composed of
eigenfunctions of the Laplacian in spherical coordinates, namely the product of spherical harmonics and spherical Bessel functions. The SFB basis was first introduced to cosmology and in the early 90s \cite{91Binney_GaussianRF,93lahav_spherical} and used in earlier galaxy surveys \cite{95Heavens_SFB,95Fisher_SFB,99Tadros_PSCz,04Percival_2DF_SFB}. Here we adopt the discrete SFB basis proposed in Ref.~\cite{19Samushia_SFB}, which solves the Laplacian eigenequation over a spherical shell spanning some finite comoving distance range $[x_{\rm min},x_{\rm max}]$. 

The orthogonality relation over the spherical shell leads to discrete radial modes $k_{nl}$, and the radial basis functions becomes \cite{19Samushia_SFB,21Gebhardt_SuperFab}
\begin{align}
g_{n\ell}(x) = c_{n\ell} \, j_\ell(k_{n\ell}x) + d_{n\ell} \, y_\ell(k_{n\ell}x),
\label{eq:gnl_basis}
\end{align}
which are linear combinations of spherical Bessel functions of the first and second kinds, satisfying the orthonormal relations under the velocity (Newmann) boundary condition \cite{23Gebhard_SFB_eBOSS}. Here $\ell$ denotes the angular multipole of spherical harmonics, while the index $n$ characterizes the radial modes by counting the number of half-cycles in the oscillations of the radial basis functions \cite{25ppn0}. In practice, we can compute the radial functions $g_{n\ell}(x)$ with the code \href{https://github.com/hsgg/SphericalFourierBesselDecompositions.jl}{\texttt{SuperFaB}}. 

The SFB transform of the density fluctuation field $\delta(\bx)$ is then
\begin{align}
\delta_{n \ell m}
&= \int_{\bx}\,g_{n\ell}(x)\,Y^*_{\ell m}(\hat{\bn})\delta(\bx)\,,
\label{eq:sfb_discrete_fourier_pair_b}
\end{align} 
where the integral of \cref{eq:sfb_discrete_fourier_pair_b} goes over the finite volume within $x_{\rm min}\leq x \leq x_{\rm max}$. We can form the SFB power spectrum (PS) for the density fluctuations of two tracers $A$ and $B$
\begin{align}
   \langle\delta_{n_1 \ell_1 m_1}^{A}\delta_{n_2 \ell_2 m_2}^{B*}\rangle = C_{\ell_1 n_1n_2}^{AB}\delta_{\ell_1\ell_2}^{\rm K}\delta_{m_1m_2}^{\rm K}
   \label{eq:SFB-discrete-PS}\,.
\end{align}
The Kronecker deltas in angular modes follow from statistical isotropy on the full sky; in the absence of masking, the spherical-shell geometry retains the azimuthal symmetry. 

The SFB PS nevertheless remains a matrix in the radial mode indices $n_1$ and $n_2$, and we define the radial off-diagonal separation as $\Delta n\equiv |n_1-n_2|$. In realistic surveys, radial selection effects, redshift evolution, and redshift-space distortions induce radial-mode mixing and generate off-diagonal components with $n_1\neq n_2$. However, these effects are modest, and the SFB PS remains close to diagonal in radial mode space \cite{22Khek_SFB_fast,24GR_SFB}. This motivates the off-diagonal $\Delta n_{\rm max}$ cuts considered in Sec.~\ref{sec:forecast-mode-cut}, which retain only PS components satisfying $|n_1-n_2|\leq\Delta n_{\rm max}$.

According to Refs.~\cite{23Gebhard_SFB_eBOSS,24GR_SFB}, the SFB power spectrum can be computed as the following
\begin{align}
    C_{\ell n_1n_2}^{AB}&=\int_{0}^{\infty}dq\, \mathcal{W}_{n_1\ell}^{A}(q) \mathcal{W}_{n_2\ell}^{B}(q) P_{\rm m,0}(q)\label{eq:dSFB_compute}\,,
\end{align}
where $P_{\rm m,0}(q)$ is the present-day matter power spectrum. The SFB kernel $\mathcal{W}^{A}_{n\ell}(q)$ for a tracer $A$ is given by
\begin{align}
\mathcal{W}_{n\ell}^{A}(q)\equiv\sqrt{\frac{2}{\pi}}q\int_{x_{\rm min}}^{x_{\rm max}}dx\,x^2g_{n\ell}(x)R^{A}(x)\Delta^{A}_{\ell}(x,q)\label{eq:Wnlq_kernel}\,,
\end{align}
where $R(x)$ is the radial selection function, and $\Delta_{\ell}(x,q)$ is the angular kernel that can contain various physical effects including the Newtonian RSD, GR, PNG, and Fingers-of-God effects. We summarize the angular kernels for these different effects in Appendix~\ref{sec:angular-kernel}. 

The numerical strategy for evaluating SFB PS is to first integrate over $x$ to compute the SFB kernel in \cref{eq:Wnlq_kernel}, and then integrate over $q$ to obtain the power spectrum following \cref{eq:dSFB_compute}. The principal computational challenge lies in evaluating the SFB kernel in Eq. \ref{eq:dSFB_compute}, which involves two spherical Bessel functions and two-dimensional integrations. In this work, we employ a Chebyshev decomposition to accelerate the evaluation of the SFB kernel in \cref{eq:Wnlq_kernel}, achieving sub-second computation of the SFB power spectrum under the full linear relativistic effects. Detailed descriptions of this Chebyshev method is provided in Appendix~\ref{sec:Chebyshev-method}.

\begin{table*}[tbp]
\begin{tabular}{p{0.18\textwidth}<{\centering}p{0.12\textwidth}<{\centering}p{0.08\textwidth}<{\centering}p{0.11\textwidth}<{\centering}p{0.08\textwidth}<{\centering}p{0.18\textwidth}<{\centering}p{0.15\textwidth}<{\centering}}
\hline
\hline
\noalign{\vskip 2pt}
Reference & Cosmology & RSD & GR & PNG & Surveys & Other Statistics \\
\hline
\noalign{\vskip 1pt}
Nicola et al. \cite{14Nicola_SFB} & $w$CDM & yes & no & no & Custom & TSH\\
Lanusse et al. \cite{15Lanusse_SFB} & $w_{0}w_{a}$CDM & no & no & no & E & TSH\\
Passaglia et al. \cite{17Passaglia_2D_3D} & $w$CDM & no & no & no & D, L & TSH, SFB$\times$TSH\\
Zhang et al. \cite{21Zhang_SFB_TSH} & $\Lambda$CDM+$\gamma$  & yes & only lensing & yes & D, E, L, SP+CLens & TSH, SFB$\times$TSH \\
Khek et al. \cite{22Khek_SFB_fast} & $\Lambda$CDM  & yes & no & yes & SP & none\\
Semenzato et al. \cite{24Semenzato_SFB_GR} & $\Lambda$CDM & yes & yes & yes & SP & none\\
this work &  $\Lambda$CDM & yes & yes & yes & D, E, SP & none\\
\hline
\end{tabular}
\caption{Summary of literature on Fisher forecasts using SFB PS. These works consider different cosmological models\footnote{The $w_{0}w_{a}$CDM model extends the standard $\Lambda$CDM cosmology parameters with a dark-energy equation of state following $w(a)=w_0+(1-a)w_a$ \cite{01CPL}, and the $w$CDM model corresponds to $w_a=0$. The $\Lambda$CDM+$\gamma$ model parametrizes modified gravity via a modified linear growth rate $f(z)=\Omega_{\rm m}(z)^{\gamma}$ \cite{05Linder_gamma}, with $\gamma\simeq 0.55$ under GR.} and include different physical effects. Here, RSD denotes the linear Newtonian RSD effect, while GR denotes the full linear relativistic corrections beyond the standard density and Newtonian RSD terms. The labels D, E, L, and SP refer to the DESI, Euclid, LSST, and SPHEREx galaxy surveys respectively, and CLens shorthands for CMB lensing. Survey labels should be interpreted as survey-like, since different studies may adopt somewhat different analysis configurations for the same survey, and the resulting forecasts are not always directly comparable. We also indicate other summary statistics considered in these works, including tomographic spherical harmonics (TSH) and cross-correlations between SFB and TSH modes.}
\label{tab:SFB-Fisher-literature}
\end{table*}

\begin{table*}[tbp]
\begin{tabular}{p{0.2\textwidth}<{\centering}p{0.15\textwidth}<{\centering}p{0.15\textwidth}
<{\centering}p{0.15\textwidth}<{\centering}}
\hline
\hline
\noalign{\vskip 2pt}
Reference & Interest & Surveys & Statistics \\
\hline
\noalign{\vskip 1pt}
Di Dio et al. \cite{14Dio_Cl_forecast} & $\Lambda$CDM & E & TSH \\
Camera et al. \cite{15Camera_GR_SKA1,15Camera_GR_forecast} & $f_{\rm NL}$ & SK & TSH\\
Raccanelli et al. \cite{16Raccanelli} & $f_{\rm NL},\alpha_s, w_0, w_a$ & E, SK & TSH \\
Alonso et al. \cite{15Alonso_GR_single,15Alonso_GR_multi} & $f_{\rm NL},b_{\rm e},\epsilon_{\rm GR}$ &  E, L, SK & TSH\\
Lorenz et al. \cite{18Lorenz_GR} & $f_{\rm NL},w_0,w_a,\sum m_{\nu},c_{\rm H}$ & L & TSH \\
Contreras et al. \cite{19Contreras_kSZ} & $f_{\rm NL},b_{\rm e},\epsilon_{\rm GR}$ & L+kSZ & TSH\\
Viljoen et al. \cite{21Viljoen_Cl_PNG_GR,21Viljoen_GR_bias} & $f_{\rm NL}, \epsilon_{\rm GR}, w$ & D, E, SK & TSH\\
Addis et al. \cite{25Addis_GR} & $f_{\rm NL},b_{\rm e}$ & D, E, M, R, SK & PSM\\
Semenzato et al. \cite{24Semenzato_SFB_GR}  & $f_{\rm NL}$ & SP & SFB\\
this work & $f_{\rm NL},b_{\rm e}, \epsilon_{\rm GR}$ & D, E, SP & SFB\\
\hline
\end{tabular}
\caption{Summary of literature on Fisher forecasts including the full linear relativistic effects for galaxy surveys. We use the same labels and conventions as in \cref{tab:SFB-Fisher-literature}. Here $\alpha_s$ is the running of the primordial power spectrum, $\sum m_{\nu}$ is the total sum of neutrino masses, $\epsilon_{\rm GR}$ represents a set of amplitude parameters for individual or combined GR effects (see Sec.~\ref{sec:measurability-GR}), and $c_{H}$  denotes the coefficients of the Horndeski gravity theories \cite{16Bellini_Horndeski}. The additional labels M, R, and SK refer to the MegaMapper \cite{Megamapper}, Roman \cite{20Eifler_Roman}, and SKA surveys \cite{20SKA}, and kSZ stands for the kinetic Sunyaev-Zeldovich effects.}
\label{tab:GR-Fisher-literature}
\end{table*}

\subsection{Literature review}\label{sec:review}

We contextualize our work by reviewing previous literature on forecasts for the SFB power spectrum and relativistic effects. Owing to the computational cost of evaluating the SFB PS, most existing SFB-based forecasts rely on the Fisher formalism. \Cref{tab:SFB-Fisher-literature} summarizes this literature, covering a range of cosmological models, physical effects, survey configurations, and scientific targets. To our knowledge, Ref.~\cite{24Semenzato_SFB_GR} is the only previous study to include all linear GR contributions in SFB-based Fisher forecasts. Compared to their work, we adopt the discrete SFB basis for spherical shells proposed in Ref.~\cite{19Samushia_SFB}, which employs spherical Bessel functions of both kinds rather than only the first kind appropriate for a full spherical volume.

Beyond Fisher forecasts, previous work has performed SFB power-spectrum inference using Bayesian inference. In particular, Ref.~\cite{26Bruton}, building on the validation of the SFB inference framework in Ref.~\cite{23Gebhard_SFB_eBOSS}, analyzed eBOSS LRG and quasar samples including Newtonian RSD and local PNG effects. The inference performed in those studies is more directly applicable to data analyses, whereas our work is theory-focused and uses simulated inference to study the impact of GR effects while neglecting observational systematics.

Besides SFB PS, a large body of literature has forecasted the impact of individual or full relativistic effects using other two-point statistics, including tomographic spherical harmonics (TSH, also known as the angular power spectrum), Fourier-space power spectrum multipoles (PSM), and configuration-space two-point correlation function multipoles (2CFM). Many studies have focused on Doppler and lensing terms individually, as Doppler effect is the most prominent at low redshift ($z\lesssim 0.5$), while lensing dominates the high-redshift regime ($z\gtrsim 1$) \cite{11ChallinorLPS,22CatorinaGR-P,24GR_SFB}. Examples assessing the impact of lensing terms through Fisher forecasts include Refs.~\cite{15Montanari_lensing,16Wilmar_lensing,20Bellomo_Cl_forecast_error,20Bernal_Cl_forecast_bias,21Jelic-Cizmek,22Euclid_mag_photoz,24Euclid_mag_specz}, while studies focusing on Doppler terms include Refs.~\cite{17Abramo_Doppler,18Raccanelli_doppler,24Montano_Doppler,24Blanco_galaxy_property}. Refs.~\cite{12Yoo_GR,20Ginzburg_shotnoise} further uses Fisher to assess the measurability of non-integrated GR terms, including both Doppler and NIP, under the plane-parallel approximation.

To our best knowledge, we summarize existing Fisher-forecast studies including the full set of linear GR effects for galaxy surveys in \cref{tab:GR-Fisher-literature}. Almost all of these works focus on local PNG, while other parameters of interest and survey specifications vary. The vast majority have employed TSH, since the inclusion of GR effects in this statistic is the most straightforward and has been accurately implemented in well-established codes such as \texttt{CAMB} \cite{CAMB,11ChallinorLPS} and \texttt{CLASS} \cite{11class,13DiDio_classgal}. In comparison, evaluation of GR effects in Cartesian-space PSM and 2CFM is more challenging due to the geometric mismatch between Cartesian coordinates of the statistics and the underlying spherical light cone. Robust implementations of these statistics have only become available more recently \cite{18Tansella-GR,18Tansella_coffe,22CatorinaGR-P,23Foglieni}. While including GR effects in the SFB basis is conceptually straightforward, the evaluation of the SFB PS itself had been challenging, and the SFB basis remains significantly less familiar in the literature than TSH or PSM.

Compared to Cartesian-space PSM, the SFB basis provides a natural separation between angular and radial modes. It therefore avoids the need for explicit wide-angle modeling and is also more robust for systematics mitigation \cite{25ppn0}. This systematics-mitigation aspect has been demonstrated in Ref.~\cite{26Bruton}, where SFB angular and radial mode cuts were used to identify and isolate potentially contaminated modes in the eBOSS LRG and quasar samples. Moreover, the PSM can be viewed as a weighted sum—i.e. a compression—of the SFB PS \cite{24PSM_SFB,26Wen_PSM}, implying that the SFB PS retains the full information content of 3D clustering and is naturally compatible with and beneficial for a PSM-based analysis. 

Compared to TSH, the SFB basis explicitly defines radial modes and allows straightforward Fourier-scale and radial-mode cuts to be robust against uncertainties in non-linear modeling or astrophysical systematics. To capture more information content of spectroscopic surveys, TSH requires very fine redshift binning, leading to large data vectors with high correlation and mixing of non-linear radial modes. Therefore, the SFB basis is more naturally suited for spectroscopic surveys, on which we focus in this work. Nevertheless, since most previous work has focused on TSH, it remains of interest to compare the two bases for PNG and GR effects and to understand their respective advantages and limitations in practical analyses, especially for wide-field photometric surveys with reasonably good redshift such as LSST \cite{19LSST} or photometric surveys in Euclid \cite{24Euclid_overview}. We leave such detailed comparisons to future work.

Beyond these Fisher-based studies, Refs.~\cite{22Martinelli_Cl_forecast,25Addis_GR} have investigated the impact of GR effects using MCMC inference. These studies examined how GR effects may bias PNG constraints and therefore repeatedly evaluated approximate theoretical models during inference—either using the Limber approximation with density-only contributions \cite{22Martinelli_Cl_forecast} or neglecting the integrated relativistic effects including lensing, ISW, and Shapiro terms \cite{25Addis_GR}, due to the computational complexities of these integrated GR terms. In contrast, our analysis will evaluate the full GR theory with the Chebyshev-based method at every step of the inference.

Finally, as discussed in Sec.~\ref{sec:bphi-be-equivalence}, previous PNG forecast studies have treated the evolution bias and the local PNG bias as independent parameters in forecasts, implicitly assuming the Case 2 scenario of \cref{fig:biasparams}, and adopting fiducial values that do not satisfy the $b_{\phi}$–$b_{\rm e}$ equivalence. In contrast, we always adopt fiducial bias values that obey this equivalence and consider Cases~1 and~3 for inference.

\begin{figure}[tbp]
\centerline{\includegraphics[width=0.88\hsize]{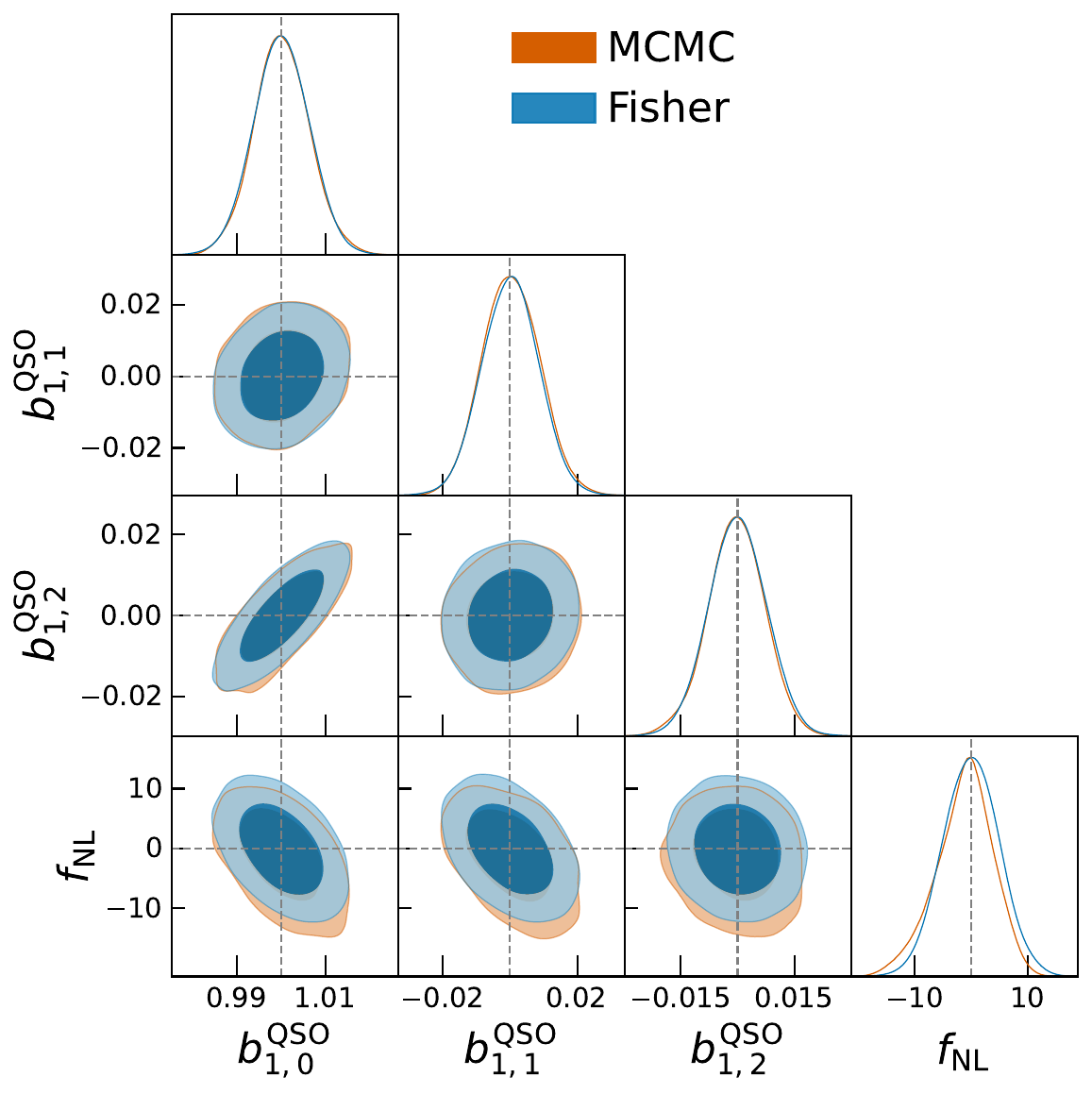}}
\caption{Comparison of constraints from Fisher forecast with MCMC posterior for DESI QSO under Newtonian modeling at $k_{\rm max}=0.06\,h\,{\rm Mpc}^{-1}$. Contours denote the 68\% and 95\% confidence regions, corresponding to the 1 and 2$\sigma$ levels for a Gaussian likelihood. Dashed lines indicate the fiducial parameter values. The Fisher and MCMC constraints are broadly consistent, validating our results. The MCMC posterior is close to Gaussian but slightly skewed toward negative with $f_{\rm NL} = -1.3^{+5.6}_{-4.4}$, compared to the Gaussian constraint with $\sigma(\fNL)=5.0$ from Fisher.}
\label{fig:fisher_mcmc_overlay}
\end{figure}

\section{Forecast Formalism}\label{sec:forecast-formalism}

Our main results are based on simulated inference using MCMC and nested sampling. This approach allows an accurate assessment of biases in $f_{\rm NL}$ and of the measurability of $b_{\phi}$ from GR effects. Neither can be easily assessed using Fisher-based methods, due to the degeneracy between PNG and GR effects and the highly non-Gaussian posterior of $b_\phi$ (see \cref{fig:p-bphi-QSO} for example). However, when the posterior is approximately Gaussian and the inference is unbiased, the Fisher forecast and simulated-inference results agree, as demonstrated in \cref{fig:fisher_mcmc_overlay} using DESI QSO as an example. 

We therefore use the Fisher formalism primarily to establish forecast baselines in Sec.~\ref{sec:Fisher-baseline}. Because Fisher forecast is significantly faster than repeated MCMC analyses, we also use it to explore the impacts of SFB mode cuts and including the primordial PS parameters. These Fisher-based scans complement the main inference results and inform the parameter choices used in inference in Secs.~\ref{sec:ignore-GR} to \ref{sec:bphi-implication}. 

In the modeling, we ignore the impacts of survey mask, so the theoretical SFB PS exactly follows from \cref{eq:dSFB_compute}. We adopt a minimum angular multipole given by
\begin{align}
\ell_{\min} = \max\!\left\{2,\ \frac{1}{f_{\rm sky}}\right\}.
\label{eq:lmin}
\end{align}
The $\ell\ge 2$ cut avoids systematics associated with the angular monopole and dipole, which includes contributions from the observer's gravitational potential and peculiar velocities. In addition, integral constraints are known to suppress large-scale angular modes, with their dominant impacts at $\ell\lesssim 1/f_{\rm sky}$ \cite{23Gebhard_SFB_eBOSS}, since angular modes larger than the survey footprint cannot be measured. Our choice of \(\ell_{\rm min}\) therefore avoids the need for explicit integral-constraint modeling.

Since galaxies are discrete tracers, the total SFB PS entering the likelihood
includes shot noise for auto-correlations
\begin{align}
\widetilde{C}_{\ell n_1n_2}^{AB}=
C_{\ell n_1n_2}^{AB} + \delta^{\rm K}_{AB}N_{\ell n_1n_2}^{A}\,.
\end{align}
The shot-noise contribution is given by \cite{21Gebhardt_SuperFab,22Khek_SFB_fast}
\begin{align}
N^{A}_{\ell n_1 n_2}&=\frac{1}{\bar{n}^{\rm max}}\int dx\,x^2g_{n_1\ell}(x)g_{n_2\ell}(x)R^{A}(x)\,,
\label{eq:sn-nic-sfb}
\end{align}
where the radial selection function $R(x)$ for a tracer is normalized by its maximum comoving number density $\bar{n}^{\rm max}$ such that ${\rm max}_x\,R(x)=1$. We assume the shot noise for cross-correlation between different samples to vanish. Generally, cross-stochasticity between different tracer samples can be nonzero, especially when the samples occupy overlapping halo populations \cite{11Hamaus,13Baldauf_stochasity,20Ginzburg_shotnoise}. We leave such concern to more realistic data analyses.

\subsection{Simulated inference}\label{sec:mcmc-formalism}

For simulated inference, we work at the power-spectrum level and, for simplicity, assume a Gaussian likelihood for the estimated SFB power spectra. This is an approximation for the Wishart distribution of PS at the large angular scales \cite{08Hamimeche_likeCMB}. The likelihood used in Bayesian inference is then
\begin{subequations}
\label{eq:ps_like}
\begin{align}
\ln \mathcal{L}(\theta)
&={\rm const}-\frac{1}{2}\Delta(\theta)^T \boldsymbol{\Sigma}^{-1}\Delta(\theta)\\
\Delta(\theta)
&= \hat{C}_{\ell nn'}^{AB,\mathrm{obs}} - \widetilde{C}_{\ell nn'}^{AB}(\theta)\,.
\end{align}
\end{subequations}
where $\theta$ denotes the set of inferred parameters. In our forecasts, the simulated data vector is taken to be noiseless in the sense that it is set equal to the theory prediction under fiducial parameters
\begin{align}
    \hat{C}_{\ell nn'}^{AB,\mathrm{obs}}=\widetilde{C}_{\ell nn'}^{AB}(\theta_{\rm fid})\,,
\end{align}
so that we expect unbiased parameter recovery provided
the model is correctly specified.

Ignoring masking and integral constraints such that the azimuthal symmetry is preserved, the Gaussian covariance
of the multi-tracer SFB PS on the full sky
is
\begin{align}
    &\mathrm{cov}^{\rm f.s.}\!\left(\hat C_{\ell nn'}^{AB},\hat C_{LNN'}^{CD}\right)=\langle \hat{C}_{\ell n n'}^{AB}\hat{C}_{L N N'}^{CD} \rangle - \widetilde C_{\ell nn'}^{AB}\widetilde C_{LNN'}^{CD}\nonumber\\
    &=\frac{1}{(2\ell+1)(2L+1)}\sum_{m,M}\langle\delta^{A}_{\ell m n}\delta^{B*}_{\ell m n'}\delta^{C}_{L M N}\delta^{D*}_{L M N'}\rangle\nonumber\\
    &\qquad-\widetilde{C}_{\ell nn'}^{AB}\widetilde{C}_{LNN'}^{CD}\nonumber\\
    &=\frac{\delta^{\rm K}_{\ell L}}{2\ell+1}\left[\widetilde{C}_{\ell n N'}^{AD}\widetilde{C}_{\ell n'N}^{BC}+\widetilde{C}_{\ell n N}^{AC}\widetilde{C}_{\ell n'N' }^{BD}\right]\,,
\end{align}
where ``f.s'' denotes full-sky. We then apply the standard $f_{\rm sky}$ approximation for angular mode-counting,
\begin{align} \boldsymbol{\Sigma}\simeq\frac{1}{f_{\rm sky}}\mathrm{cov}^{\rm f.s.}\,,
\end{align}
which is consistent with Ref.~\cite{21Gebhardt_SuperFab} for the single tracer case upto mode binning. The above approximation avoids explicit treatment of
window-induced mode coupling in covariance. Since our analysis focuses on large scales with linear modeling, we consider only the Gaussian (disconnected) covariance. The covariance is evaluated and fixed at the fiducial parameters during sampling.

To make likelihood evaluations tractable, we organize the data vector hierarchically as follows: first by angular multipoles, then by auto and cross-spectra, and finally by radial indices, keeping the block-diagonal structure of the covariance in terms of $\ell$. We sample the posterior 
using an Adaptive Metropolis-Hastings (AMH) sampler \cite{09Roberts_adaptiveMC} following the implementation of Ref.~\cite{23Gebhard_SFB_eBOSS}. We adopt the AMH sampler for results in Secs.~\ref{sec:ignore-GR} and \ref{sec:measurability-GR}, and assume flat priors for all parameters in these MCMC runs. We use \texttt{GetDist} \cite{25GetDist}
for contour plots to visualize sampling results. 

To study the measurability of $b_\phi$ from GR effects in Sec.~\ref{sec:bphi-implication}, we use nested sampling \cite{04Skilling_NS} rather than the AMH sampler, since $b_\phi$ is highly degenerate with $\fNL$ and the resulting posterior can be strongly non-Gaussian. Although nested sampling is typically more computationally expensive than AMH, it provides a more robust characterization of curved degeneracies and possible multi-modal structure among parameters, which can be difficult for MH-type samplers to traverse efficiently. In particular, we use the code \texttt{NAUTILUS} \cite{23Lange_nautilus}, a fast implementation of importance nested sampling \cite{19Feroz_INS} employing neural networks to construct efficient proposal distributions. Nested sampling requires explicit priors because it explores the posterior by compressing a finite prior volume into nested likelihood contours. We use the broad uniform priors summarized in \cref{tab:nautilus-priors}.

\subsection{Fisher forecasts}\label{sec:fisher-formalism}

We complement the simulated inference with Fisher-matrix forecasts. Starting from the Gaussian likelihood for the estimated SFB coefficients
$\hat{\delta}_{n\ell m}^{A}$ of all tracers
\begin{align}
\label{eq:like_field}
\mathcal{L}(\hat{\delta}\,|\,\theta)
\sim
\frac{1}{\sqrt{|\widetilde{\boldsymbol{C}}(\theta)|}}
\exp\!\left[
-\frac{1}{2}\hat{\delta}^{\dagger}\,
\widetilde{\boldsymbol{C}}^{-1}(\theta)\,
\hat{\delta}
\right],
\end{align}
where $\widetilde{\boldsymbol{\mathcal{C}}}(\theta)\equiv\langle\delta^{A}_{n\ell m}\delta^{B*}_{n'\ell' m'}\rangle$ is the full power spectrum of the discrete SFB modes, including shot noise, in matrix form. The Fisher matrix is then \cite{22Khek_SFB_fast}
\begin{align}
\label{eq:fisher_general}
F_{\alpha\beta}
=
\frac{1}{2}\mathrm{Tr}\!\left(
\boldsymbol{\widetilde{C}}^{-1}\boldsymbol{\widetilde{C}}_{,\alpha}
\boldsymbol{\widetilde{C}}^{-1}\boldsymbol{\widetilde{C}}_{,\beta}
\right),
\end{align}
with $C_{,\alpha}\equiv \partial C/\partial\theta_\alpha$. 

Assuming azimuthal symmetry under the full sky, $\widetilde{\boldsymbol{C}}=\delta_{\ell\ell'}^{\rm K}\delta_{mm'}^{\rm K}\widetilde{C}^{AB}_{\ell nn'}$. The $f_{\rm sky}$ approximation simplifies the Fisher matrix to
\begin{align}
\label{eq:fisher_ell}
F_{\alpha\beta}
=f_{\rm sky}
\sum_{\ell=\ell_{\rm min}}^{\ell_{\rm max}}
\frac{2\ell+1}{2}\,
\mathrm{Tr}\!\left[
\widetilde{\boldsymbol{C}}_{\ell}^{-1}
\widetilde{\boldsymbol{C}}_{\ell,\alpha}
\widetilde{\boldsymbol{C}}_{\ell}^{-1}
\widetilde{\boldsymbol{C}}_{\ell,\beta}
\right],
\end{align}
where $\boldsymbol{\widetilde{C}}_{\ell}$ is the $\ell$-block of the multi-tracer SFB PS in a matrix form organized by tracers and radial indices. The parameter covariance is then approximated by
\begin{align}
\mathrm{Cov}(\theta_\alpha,\theta_\beta) \simeq (F^{-1})_{\alpha\beta}.
\end{align}

Derivatives of the SFB PS are computed
numerically using finite differences around the fiducial parameters. We have verified that these derivatives agree
well with results obtained using automatic differentiation enabled by the Chebyshev method implemented in \texttt{Julia}.

Compared to simulated inference, the above Fisher forecast is derived from the field-level likelihood. Under linear theory, the likelihood of the density fluctuation is exactly Gaussian, whereas a Gaussian likelihood for the PS is only an approximation to its true sampling distribution. If the PS covariance is fixed at fiducial parameters and Gaussian likelihood is assumed (\cref{eq:ps_like}), both field-level and PS-level likelihoods yield equivalent Fisher information \cite{20Bellomo_Cl_forecast_error,08Hamimeche_likeCMB}. For the off-diagonal \(\Delta n_{\rm max}\) cut, we use PS-level Fisher forecasts based on the likelihood in \cref{eq:ps_like}. In this case, the mode selection is applied to the SFB PS components rather than to the underlying SFB modes, so \cref{eq:like_field} can not be applied.

\begin{figure*}[tbp]
\centerline{\includegraphics[width=0.92\hsize]{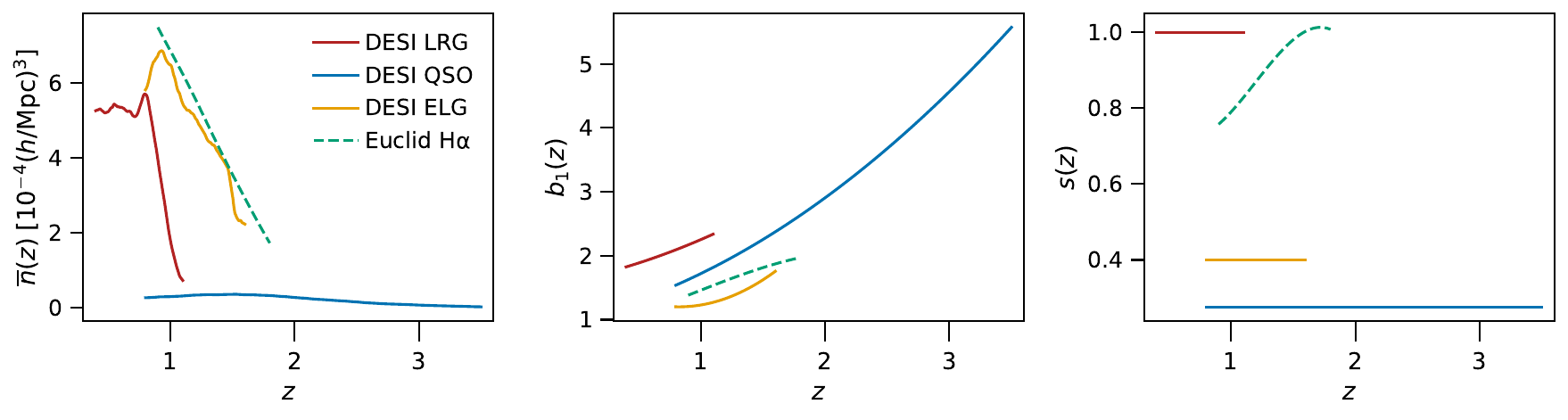}}
\caption{Survey parameters for DESI LRG, ELG, QSO, and Euclid H$\alpha$ samples. Shown are the comoving number density (left), linear galaxy bias (center), and magnification bias (right) as functions of redshift. Curves reflect the latest data releases and modeling assumptions. All functions are plotted over the full redshift range of each tracer.}
\label{fig:DESI_Euclid_Spec}
\end{figure*}

\begin{figure*}[tbp]
\centerline{\includegraphics[width=0.92\hsize]{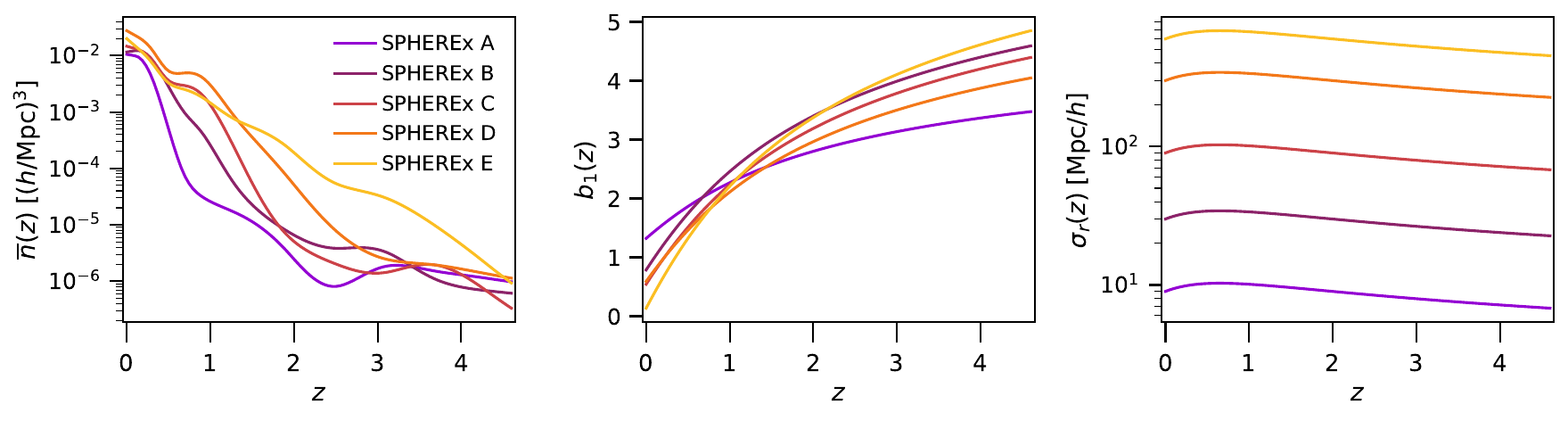}}
\caption{Survey parameters for the SPHEREx galaxy samples used in our forecasts. Shown are the comoving number density (left), linear galaxy bias (middle), and redshift error on comoving distance (right) as functions of redshift for the five SPHEREx samples (A--E). These five samples are characterized by redshift uncertainties $\sigma_z/(1+z) = 0.003,\, 0.01,\, 0.03,\, 0.1,$ and $0.2$, respectively, ranging from spectroscopic-like (Sample~A) to photometric-like (Sample~E) qualities. Samples from A to E have progressively higher number densities but larger redshift uncertainties. These specifications follow Ref.~\cite{22Khek_SFB_fast} and are constructed from product provided in \cite{SPHEREx_data}.}
\label{fig:SPHEREx_Spec}
\end{figure*}

\subsection{Survey specification}\label{sec:survey-spec}

We focus on three current wide-field surveys with spectroscopic or spectroscopic-like redshift, for which the SFB basis is well-suited. DESI has already released its first-year cosmology data \cite{25DESI_DR1}, so its tracer specifications are directly informed by observational measurements. In contrast, the Euclid and SPHEREx specifications remain based on forecasts and simulations, with their first cosmology-grade products expected in the coming years. We describe below our choices for the number densities, linear biases, and magnification biases for the three surveys. We assume the universality relation with $p=1$ for the fiducial $b_{\phi}$ for all tracers except DESI QSO where $p=1.6$. We use the equivalence relation discussed in Sec.~\ref{sec:bphi-be-equivalence} to set the fiducial $b_{\rm e}$ values according to the fiducial $b_\phi$.  

\subsubsection{DESI}

We focus on LRG, ELG, and QSO samples of the DESI survey. We exclude the bright galaxy sample (BGS) due to its low-redshift coverage and limited survey volume, which will have poor constraining power on $f_{\rm NL}$ and GR effects without being further split into sub-samples for multi-tracer analyses. We assume a final footprint of $14{,}000\,\mathrm{deg}^2$ with $f_{\rm sky}\approx 0.34$. Since DESI is a spectroscopic survey with resolution $\Delta \lambda/\lambda>1000$ \cite{22DESI_instrument}, the redshift errors for all DESI samples are negligible for our purpose.

The redshift ranges and survey specifications for the DESI LRG ($0.4<z<1.1$), ELG ($0.8<z<1.6$), and QSO ($0.8<z<3.5$) samples are shown in \cref{fig:DESI_Euclid_Spec}. The redshift-dependent number densities for all three samples are taken from the NGC catalogs of DESI Year~1 release \cite{24DESI_II} after completeness corrections. For the linear bias, we use the quadratic fits inferred from DESI data in Ref.~\cite{24_DESI_Y1_PNG} for LRG and QSO. For ELG, we fit a quadratic function to the linear-bias measurements from Ref.~\cite{25Shaw_DopplerBias}.

For the magnification bias, we take $s=1$ for LRG\footnote{This is further supported by the BOSS LRG sample, which also yields $s\simeq 1$ under magnitude perturbation and reselection \cite{24Wenzl_mag_bias_SDSS}.}, consistent with measurements finding uncertainties $\sigma(s)\sim 0.015$ across the four sub-redshift bins of the LRG redshift range and no strong redshift evolution \cite{23Zhou_DESI_LRG_cross}. We choose $s=0.4$ for ELG according to measurements in Ref.~\cite{25Shaw_DopplerBias}, also finding minimal redshift evolution. For QSO we adopt the fiducial value $s=0.2764$ based on measurements from Ref.~\cite{24Krolewski_CMB_quadar_fnl}\footnote{Ref.~\cite{24Krolewski_CMB_quadar_fnl} provides only a single effective magnification-bias value for the full QSO sample, without resolving its redshift evolution. We therefore expect our fiducial QSO magnification-bias model to be less certain than the corresponding LRG and ELG values.}.

In this work, we only consider cross-correlation between different tracers when they cover the same redshift range, that is we consider ELG+QSO for $0.8<z<1.6$ and LRG+ELG+QSO for $0.8<z<1.1$. We leave the study of 3D cross-correlations between tracers with non-overlapping redshift ranges to future work, which may be interesting for constraining integrated relativistic effects such as lensing and ISW \cite{24Semenzato_SFB_GR,25Sailer_DESIcross}.

\subsubsection{Euclid spectroscopic}

Euclid measures H$\alpha$ emission-line galaxies in its spectroscopic survey spannning $0.9<z<1.8$, with survey area of around $14{,}500\,\mathrm{deg}^2$ corresponding to $f_{\rm sky}\approx 0.35$.
\cite{24Euclid_overview}. The exact specifications for the H$\alpha$ sample vary across different forecast studies. Here we adopt the number densities and the linear bias from Ref.~\cite{20Euclid_forecast}, and the polynomial fits for the magnification bias provided in Ref.~\cite{24Euclid_mag_specz}. These functions are plotted in \cref{fig:DESI_Euclid_Spec}. We note that the Euclid H$\alpha$ sample has number densities and linear bias similar to those of DESI ELGs, but with somewhat higher redshift coverage and significantly larger magnification bias. We assume a constant redshift uncertainty $\sigma_z = 0.002$ over the full redshift range, consistent with the expected slitless spectroscopic redshift performance of Euclid.

\subsubsection{SPHEREx}

 We follow Ref.~\cite{22Khek_SFB_fast} for the SPHEREx forecast specifications. We assume a sky fraction of $f_{\rm sky}=0.75$ for the full-sky galaxy survey. The radial number density is constructed as a cubic spline in log-space using the values in \cite{SPHEREx_data}, and the linear bias function is obtained via linear regression between the comoving distance and the measured linear bias values.

As a spectro-photometric survey, SPHEREx samples spanning $0<z<4.6$ are characterized by redshift uncertainties of the form $\sigma_z(z) = (1+z)\sigma_i$, with $\sigma_i = 0.003, 0.01, 0.03, 0.1,$ and $0.2$, corresponding to five samples ranging from spectroscopic-like to photometric-like redshift qualities. Here we name the five samples as A-E in order of increasing redshift uncertainty. Owing to the large redshift uncertainties of the photometric-like samples, we implement an exact modeling of redshift error for SFB PS in Appendix~\ref{sec:redshift-error}, improving over the approximate treatments in Refs.~\cite{22Khek_SFB_fast,21Gebhardt_SuperFab}.

The survey parameters for all five SPHEREx samples are shown in \cref{fig:SPHEREx_Spec}. For the Fisher forecasts, we include all five samples over the full redshift range, $z\in[0,4.6]$. For the Bayesian inference, we restrict the analysis to $z\in[0,2]$, since extending to the full range requires a much larger set of SFB modes, dramatically increasing the computational cost and memory usage. We denote the combined fives samples including all the auto and cross-correlations as SPHEREx 5T hereafter.

Estimating the magnification bias for SPHEREx would require perturbing the spectro-photometric flux measurements and rerunning the redshift-measurement pipeline to determine the resulting number-density response for each sample. We leave this dedicated effort to future work. Without a magnification bias measurement from simulation or data, we set $s=1$ for all five samples across redshift for simplicity, mimicking the magnification bias for the DESI LRG sample. We show the impacts of shifting the fiducial magnification bias to different values on our results in \cref{fig:ignore-GR-vary-s} and \cref{tab:gr_amp_spherex_c}.

\subsection{Forecast parametrization}\label{sec:bias-param}
All bias functions are redshift-dependent, and it is important to model their redshift evolution, to which the SFB power spectrum is sensitive. To marginalize over uncertainties in the bias functions, we adopt a parametrization in which the fiducial bias functions are modulated by a Chebyshev expansion
\begin{align}
F(z) &= \left(\sum_{n=0}^{N}F_{n}T_n(\tilde{x}(z))\right)F^{\rm fid}(z)\label{eq:bias-general-chebyshev}\,,
\end{align}
where $F^{\rm fid}(z)$ is the fiducial bias function set for each tracer. We use Chebyshev polynomials $T_n$ up to order $N$ to marginalize over uncertainties in the redshift evolution. The expansion is performed over the normalized comoving distance $\tilde{x}\in [-1,1]$
\begin{align}
\tilde{x}(x(z)) = \frac{x(z)-x_{\rm mid}}{\Delta x}\,,
\end{align}
where $x_{\rm mid}=\frac{1}{2}(x_{\rm max}+x_{\rm min})$ and $\Delta x=\frac{1}{2}(x_{\rm max}-x_{\rm min})$. The Chebyshev coefficients $F_n$ are the parameters marginalized in forecasts. We refer readers to \cref{sec:Chebyshev-method} for more discussion on Chebyshev polynomials.

For the linear galaxy bias $b_1(z)$, we include Chebyshev components up to the second order, such that
\begin{align}
b_1(z) &= b_{1,0} + b_{1,1} , \tilde{x}(z)+ b_{1,2}(2\tilde{x}(z)^2-1)\,.
\end{align}
A quadratic function has been shown to fit the redshift evolution of linear bias well for tracers such as LRG and QSO \cite{24_DESI_Y1_PNG,26Bruton}. The baseline parameter set used in the forecast is therefore $\{f_{\rm NL},b_{1,0}, b_{1,1}, b_{1,2}\}$, and the three linear bias parameters for each tracer are always marginalized.

For the magnification bias and LPNG bias, with the evolution bias tied to $b_{\phi}$ as previously described, we use only the zeroth-order Chebyshev components $s_{,0}$ and $b_{\phi,0}$, such that
\begin{align}
b_\phi(z)=b_{\phi,0}\times2\delta_{\rm c}(b_1(z)-p)\,.
    \label{eq:bphi-cheb}
\end{align}
In other words, we marginalize only over the overall amplitude of these bias functions and neglect the uncertainties in their redshift evolution, given the relatively weaker constraining power on these parameters compared to linear bias.

We do not further marginalize over uncertainties in redshift errors or velocity dispersions, since they primarily affect smaller radial scales, as shown in Appendix~\ref{sec:redshift-error}, and have minimal impacts on large scales.

\begin{table}
  \centering
\begin{tabular}{lc}
    \hline\hline
    Tracer            & $\sigma(f_{\rm NL})$ \\
    \hline
    DESI LRG          &  7.76 \\
    DESI ELG          &  9.95 \\
    DESI QSO          &  4.76 \\
    \hline
    Euclid H$\alpha$  &  5.27 \\
    \hline
    SPHEREx A         &  4.61 \\
    SPHEREx B         &  2.28 \\
    SPHEREx C         &  2.20 \\
    SPHEREx D         &  1.50 \\
    SPHEREx E         &  0.91 \\
    SPHEREx 5T        &  0.78 \\
    \hline\hline
\end{tabular}
\caption{Baseline $1\sigma$ constraints on $f_{\rm NL}$ from the SFB Fisher forecast under Newtonian modeling. We use $k_{\rm max}=0.08\,h/\mathrm{Mpc}$ and the full redshift range for each tracer, with $\ell_{\rm min}=2$ for SPHEREx (full-sky) and $\ell_{\rm min}=3$ for DESI and Euclid, reflecting their different sky coverages. }
  \label{tab:fnl_baseline}
\end{table}

\section{Results}\label{sec:results}

\subsection{Baseline Fisher forecasts}\label{sec:Fisher-baseline}

We present our baseline Newtonian constraints on $f_{\rm NL}$ through Fisher forecasts in \cref{tab:fnl_baseline}, marginalizing over the linear bias parameters. We adopt $k_{\rm max}=0.08\,h/{\rm Mpc}$ as a conservative cutoff for our linear-theory modeling. Pushing to smaller scales would require one-loop corrections to the SFB PS, which remain to be developed. We set $\ell_{\rm min}=2$ for SPHEREx and $\ell_{\rm min}=3$ for DESI and Euclid following \cref{eq:lmin}. For each tracer, we use the full available redshift range. We include all SFB PS components satisfying these cuts, with no off-diagonal truncation.

We find that DESI QSO, Euclid H$\alpha$, and the SPHEREx Sample A have comparable constraining power, with $\sigma(f_{\rm NL})\sim 5$. This is comparable to the current best \textit{Planck} CMB constraint, $\sigma(f_{\rm NL})=5.1$, suggesting that individual spectroscopic or spectroscopic-like samples can reach \textit{Planck}-level sensitivity in the near term. DESI LRG is weaker because of its lower redshift coverage, while the photometric-like SPHEREx D and E samples reach $\sigma(f_{\rm NL})\sim1$. This improvement is driven by their high number densities and extended redshift coverage. Combining all five SPHEREx samples gives $\sigma(f_{\rm NL})=0.78$, further improving on the SPHEREx Sample E constraint by about $15\%$.

Despite both being emission-line galaxy samples, DESI ELG and Euclid H$\alpha$ have substantially different constraining power on $f_{\rm NL}$. This difference is mainly driven by Euclid H$\alpha$ enjoying slightly higher redshift coverage, larger bias, and higher number density. Results in \cref{tab:fnl_baseline} illustrate the statistical constraining power of each tracer, and should be interpreted as a theoretical baseline. In practice, real-data constraints will depend critically on observational systematics and uncertainties in $b_{\phi}$, which already limit existing LSS measurements of local PNG~\cite{22Barreira_bphi_BOSS,24Cagliari_eBOSS_quasar_fnl,24_DESI_Y1_PNG,26Bruton}. In contrast, optimized sub-sample splits based on $b_1$ and $b_\phi$ could potentially further improve the constraints over the baseline~\cite{23Barreira_multi,23Sullivan_bphi}.

Because previous forecasts differ in redshift range, bias parametrization, summary statistics, and nuisance parameter treatments, the literature comparisons below should be viewed as approximate consistency checks rather than one-to-one validations. Among SFB PS forecasts, our SPHEREx constraints are generally consistent with Refs.~\cite{22Khek_SFB_fast}\footnote{The constraints in Table I of \cite{22Khek_SFB_fast} are slightly tighter than those in \cref{tab:fnl_baseline} for the two lower-accuracy SPHEREx samples. This difference is likely due to their marginalizing over fewer linear-bias parameters, together with our improved treatment of redshift errors.} and \cite{21Zhang_SFB_TSH}\footnote{Relative to Ref.~\cite{21Zhang_SFB_TSH}, our SPHEREx A and B constraints are generally consistent, while their Sample C constraint is roughly a factor of two weaker because their first three SPHEREx samples are restricted to $z\in[0,1.4]$. For the shared DESI ELG and Euclid H$\alpha$ tracers, their constraints are slightly tighter, mainly due to broader assumed redshift ranges.}. For DESI LRG and QSO samples, our SFB Fisher constraints are about $20\%$ tighter\footnote{This difference likely reflects our modeling choices such as the full QSO redshift range used here, $z\leq3.5$ rather than $z\leq3.1$, the absence of a $k_{\rm min}$ cut, and the use of the full anisotropic SFB information.} than the forecasts of Ref.~\cite{24_DESI_Y1_PNG}, which use Fourier-space PSM as the summary statistic and perform inference on mock catalogs. For Euclid H$\alpha$, our constraint is consistent with previous Cartesian Fourier-space forecasts \cite{12Giannantonio_Euclid_PNG,14SPHEREx,18Karagiannis_PNG}\footnote{The significantly tighter $P(k)$-only constraint reported in Ref.~\cite{25Eulcid_IC} partly reflects their use of $b_\phi=2\delta_c(b_1-p)$ with $p=0.55$, instead of our more conservative $p=1$. This increases the PNG bias response and leads to a more optimistic $\sigma(f_{\rm NL})$.}, which found $\sigma(f_{\rm NL})\simeq4$--$6$ depending on their particular assumptions.

\begin{figure*}[tbp]
\centerline{\includegraphics[width=0.72\hsize]{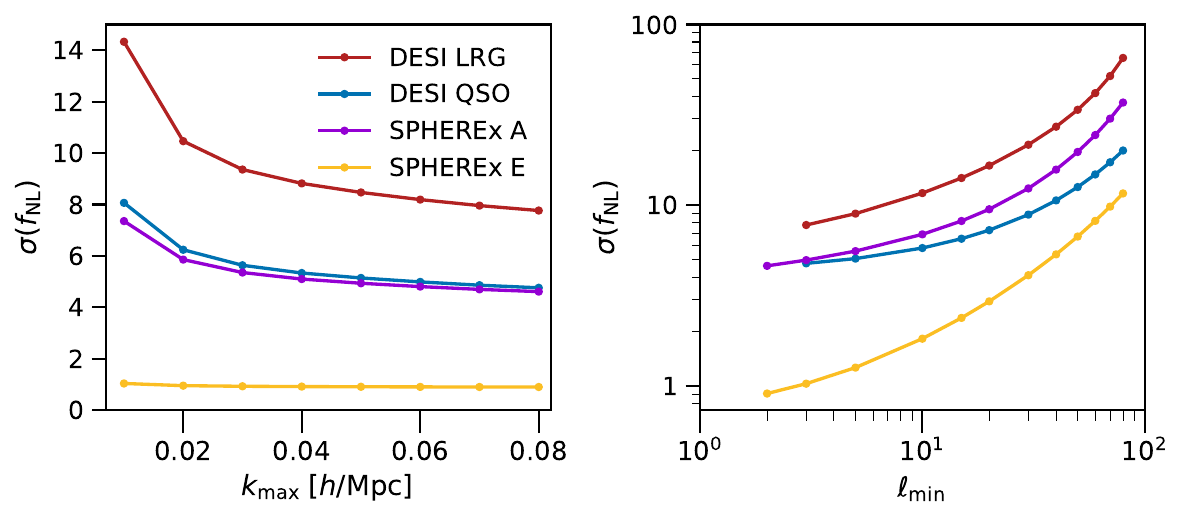}}
\caption{Sensitivity of $\sigma(f_{\rm NL})$ to the minimum angular multipole $\ell_{\rm min}$ and the maximum Fourier scale $k_{\rm max}$ under Newtonian modeling for the DESI LRG and QSO samples, and for the SPHEREx A (spectroscopic-like) and E (photometric-like) samples. The $\ell_{\rm min}$ sweep starts at $\ell_{\rm min}=2$ for SPHEREx and at $\ell_{\rm min}=3$ for DESI. Across all tracers, $\sigma(f_{\rm NL})$ depends only mildly on $k_{\rm max}$ but is highly sensitive to $\ell_{\rm min}$.}
\label{fig:fnl-sensitivity-kl}
\end{figure*}

\begin{figure}[tbp]
\centerline{\includegraphics[width=0.8\hsize]{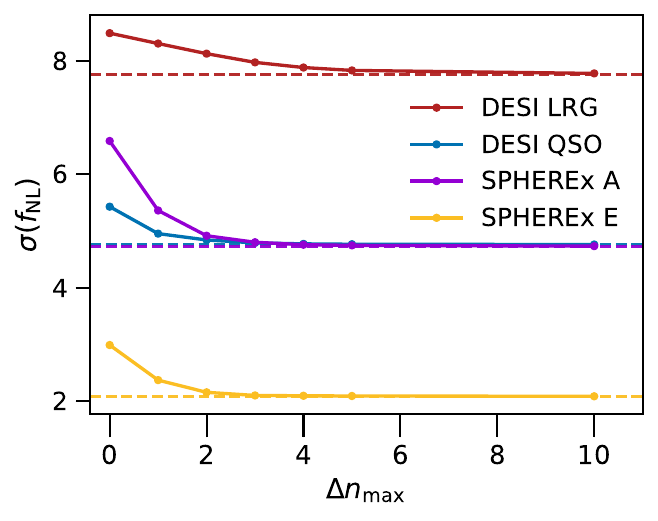}}
\caption{Sensitivity of $\sigma(f_{\rm NL})$ to the off-diagonal cut $\Delta n_{\rm max}$, where only SFB PS components with radial modes satisfying $|n_1-n_2|\leq \Delta n_{\rm max}$ are retained. For computational efficiency, the SPHEREx samples here are restricted to $z\in[0,2]$. Dashed horizontal lines show the un-cut results where all radial components are included. We find that $\sigma(f_{\rm NL})$ is generally insensitive to $\Delta n_{\rm max}$, and the diagonal SFB PS already captures most of the $f_{\rm NL}$ information.}
\label{fig:fnl-sensitivity-dnmax}
\end{figure}

\subsubsection{Sensitivity to SFB mode cuts}\label{sec:forecast-mode-cut}

We investigate how the baseline $\fNL$ constraints depend on mode cuts on the SFB power spectra. We vary the maximum Fourier scale $k_{\rm max}$, the minimum angular multipole $\ell_{\rm min}$, and the maximum radial off-diagonal separation $\Delta n_{\rm max}$ for a few selected tracers, as shown in \cref{fig:fnl-sensitivity-kl,fig:fnl-sensitivity-dnmax}.

The left panel of \cref{fig:fnl-sensitivity-kl} shows that $\sigma(\fNL)$ is mildly sensitive to $k_{\rm max}$. A lot of the PNG information comes from large scales $k\lesssim 0.04\,h/{\rm Mpc}$, as expected from the local PNG's scaling as $k^{-2}$. Among the four tracers considered, DESI LRG shows the strongest dependence on $k_{\rm max}$, while SPHEREx E has the weakest. This weak dependence is due to its large photometric-like redshift error, which damps smaller radial scales, corresponding to higher radial mode numbers $n$, and suppresses the extra information at higher $k$. Extrapolating trends in the plot beyond $k_{\rm max}=0.08\,h/{\rm Mpc}$ suggests that linear modeling captures substantial part of the local PNG information, although extending to smaller scales can still improve the $\fNL$ constraints for spectroscopic tracers such as DESI LRG, motivating the further incorporation of one-loop corrections to local PNG analysis \cite{22Cabass_BOSS,25DAmico_PNG,Moore_inprep}.

The $\fNL$ constraint is much more sensitive to $\ell_{\rm min}$, as shown in the right panel of \cref{fig:fnl-sensitivity-kl}. Compared to other tracers, DESI QSO is relatively more robust to this angular cut, because its high-redshift coverage maps a fixed $\ell$ to larger physical scales. Increasing the cut from $\ell_{\rm min}=3$ to $\ell_{\rm min}=10$ degrades the DESI QSO constraint from $\sigma(\fNL)=4.76$ to $5.79$, only a $22\%$ increase. In contrast, SPHEREx E degrades from $\sigma(\fNL)=1.03$ to $1.82$, a $77\%$ increase, because photometric-like redshift errors erase radial information, making the constraint more dependent on the lowest angular multipoles.

This angular-mode sensitivity is important for mitigating observational systematics, since the largest angular scales are often most affected by stellar contamination, dust extinction, and instrumental systematics. The $\ell_{\rm min}$ sweep therefore quantifies the trade-off between statistical constraining power and robustness to systematics. This trade-off is tracer-dependent: high-$z$ tracers probe larger physical scales at fixed $\ell$, but they can also suffer from stronger systematics, as illustrated by the larger residual systematics present in the QSO sample compared to LRG \cite{24_DESI_Y1_PNG,26Bruton}. The DESI QSO--SPHEREx E comparison in \cref{fig:fnl-sensitivity-kl} further shows that photometric samples can achieve strong statistical constraints while remaining more dependent on the lowest, most systematics-sensitive angular modes.

The $\Delta n_{\rm max}$ sweep in \cref{fig:fnl-sensitivity-dnmax} shows that only a small number of radial off-diagonal components are needed to recover most of the $\fNL$ information. For DESI QSO and SPHEREx samples, $\Delta n_{\rm max}=2$ already saturates the constraints, while DESI LRG requires a higher cut with $\Delta n_{\rm max}\simeq 5$. The diagonal components alone capture most of the PNG information for DESI QSO, although the first few off-diagonal components still provide a noticeable improvement for SPHEREx samples.

These behaviors are likely related to the smoothness of the radial selection functions. As shown in \cref{fig:DESI_Euclid_Spec,fig:SPHEREx_Spec}, the DESI QSO and SPHEREx samples have comparatively smooth radial selection functions, whereas the DESI LRG selection is roughly uniform over $z\in[0.4,0.8]$ but drops steeply over $z\in[0.8,1.1]$. This sharper radial transition can leak power from the diagonal SFB PS into more distant off-diagonal components, potentially explaining why the DESI LRG constraint saturates only at larger $\Delta n_{\rm max}$. This also suggests that additional radial weightings could help concentrate the observed SFB PS closer to the diagonals and reduce the necessary size of the truncated SFB data vectors, although they would not change the constraints in the absence of an off-diagonal cut.

Motivated by the mild sensitivity to $k_{\rm max}$ shown in \cref{fig:fnl-sensitivity-kl}, we adopt $k_{\rm max}=0.06\,h/{\rm Mpc}$ for all tracers in inference runs. This choice preserves most of the local PNG information while reducing the cost of the exact SFB PS calculations, whose computational complexity scales approximately as $k_{\rm max}^3$.

The off-diagonal cut with $\Delta n_{\rm max}$ plays a different role. Since it is applied at the power-spectrum level rather than the mode level, it does not accelerate the theoretical calculation, which is dominated by the SFB kernel evaluation in \cref{eq:Wnlq_kernel}. Instead, $\Delta n_{\rm max}$ reduces the data-vector size of PS and hence the covariance matrix. This reduction is especially important for the SPHEREx 5T inference, which contains 10 auto- and cross-power spectra; without an off-diagonal cut, the covariance matrix becomes impractically large. We therefore impose $\Delta n_{\rm max}=2$ for SPHEREx 5T, since SPHEREx samples retain most of their $\fNL$ information within this cut, illustrated in \cref{fig:fnl-sensitivity-dnmax}.

For the other inference analyses, the covariance memory footprint is manageable under our $f_{\rm sky}$ approximation, where the covariance remains block diagonal in angular modes. We therefore use the full SFB PS data vector. In realistic survey analyses, however, angular masks require explicit window-mixing calculations and break this block-diagonal covariance structure, making both window-mixing and covariance storage substantially more memory intensive. Radial off-diagonal cuts can therefore become necessary even for single-tracer PS-level inference \cite{23Gebhard_SFB_eBOSS,26Bruton}.

\begin{figure}
\centerline{\includegraphics[width=0.88\hsize]{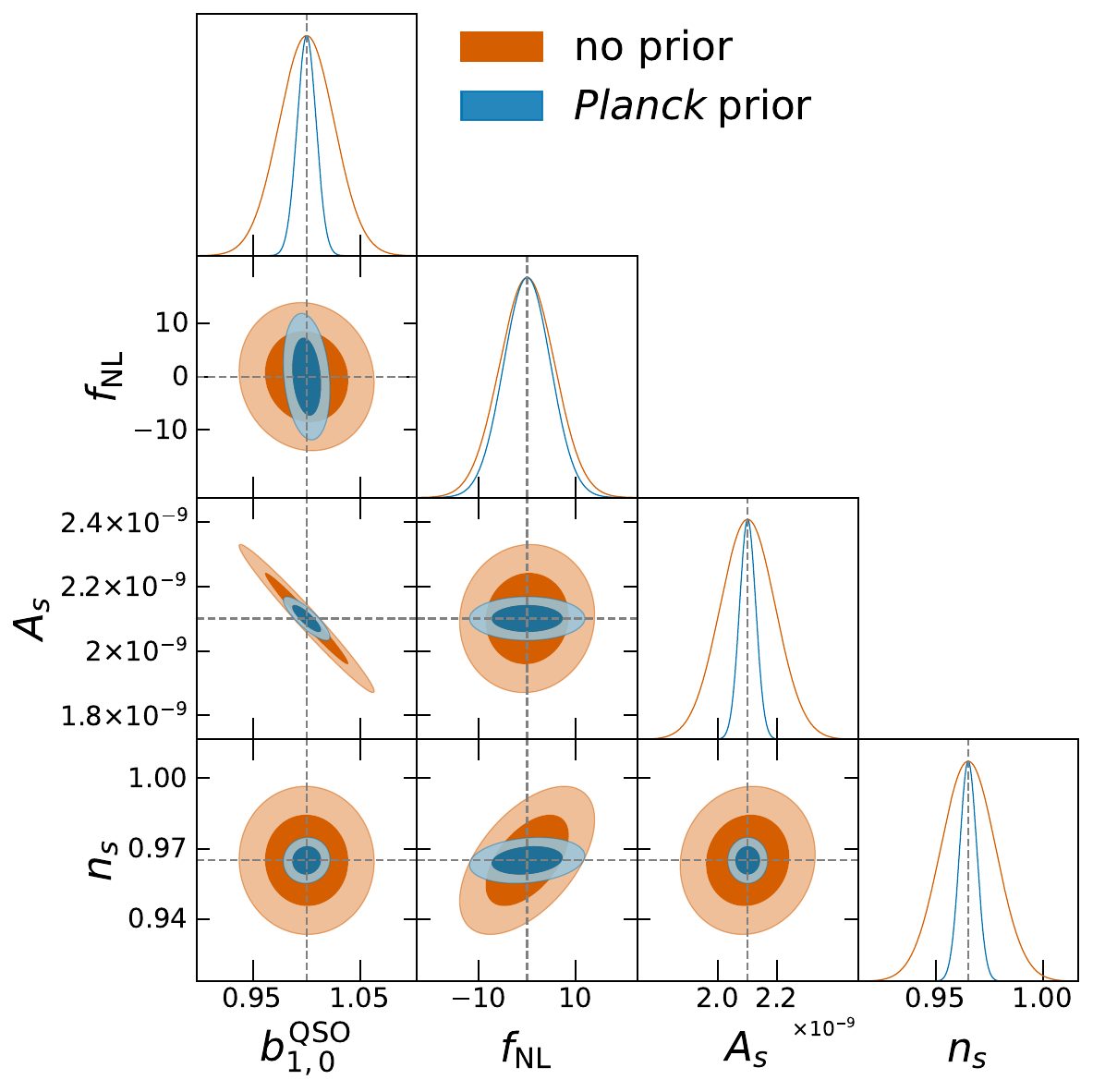}}
\caption{Fisher forecast for the DESI QSO sample under Newtonian modeling, shown in the $\{f_{\rm NL}, b_{1,0}, A_s, n_s\}$ parameter subspace. Orange contours show constraints without external priors on $A_s$ or $n_s$, while blue contours include Gaussian priors from \textit{Planck} 2018 with $\sigma(A_s)=0.029\times10^{-9}$ and $\sigma(n_s)=0.0042$ \cite{18Planck_Parameter}.}
\label{fig:fnl-AsNs-triangle}
\end{figure}

\begin{figure}
\centerline{\includegraphics[width=0.92\hsize]{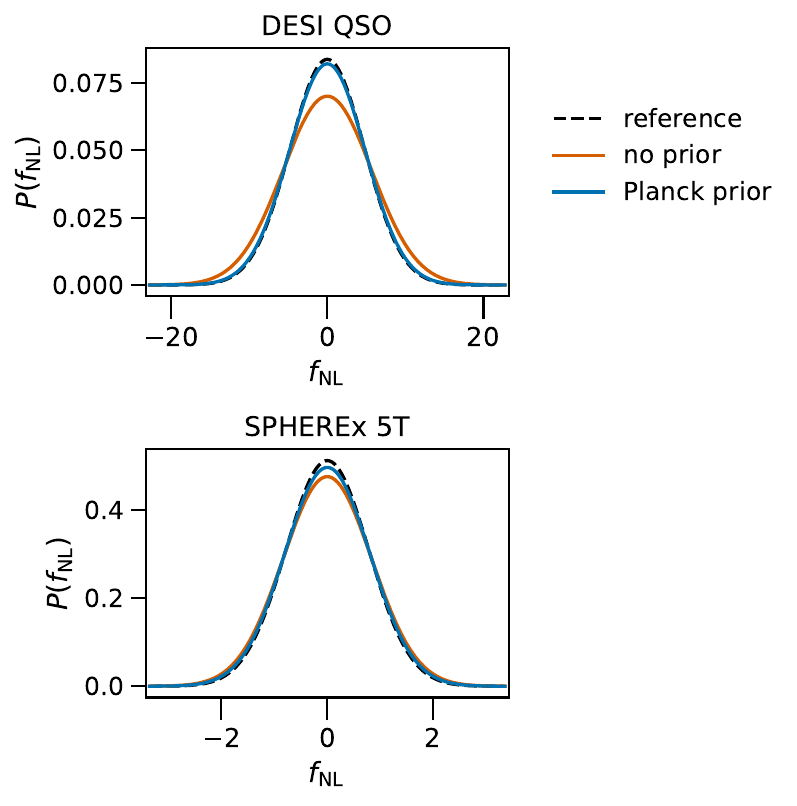}}
\caption{One-dimensional marginalized constraints on $f_{\rm NL}$ for DESI QSO and SPHEREx 5T (full redshift range), shown in the top and bottom panels, respectively, using the same configuration as \cref{fig:fnl-AsNs-triangle}. Black dashed lines show the baseline constraints with $A_s$ and $n_s$ fixed. Orange and blue curves show results after marginalizing over $A_s$ and $n_s$ without external priors and with \textit{Planck} 2018 Gaussian priors, respectively. The \textit{Planck} priors bring the constraints close to the fixed-$A_s,n_s$ baseline.}
\label{fig:fnl-1d-AsNs}
\end{figure}

\begin{figure*}[!t]
\centering
\begin{minipage}[t]{0.28\textwidth}
\includegraphics[width=\linewidth]{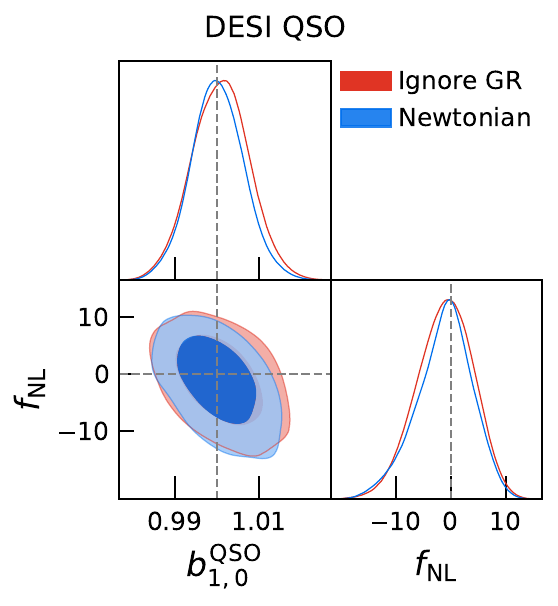}
\end{minipage}\hfill
\begin{minipage}[t]{0.28\textwidth}
\includegraphics[width=\linewidth]{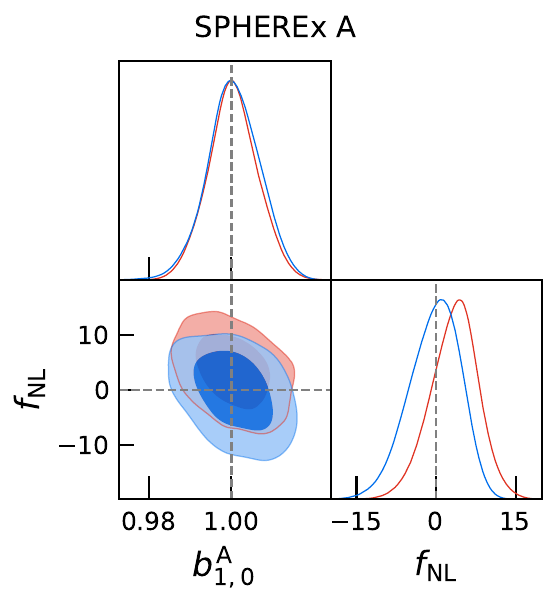}
\end{minipage}\hfill
\begin{minipage}[t]{0.28\textwidth}
\includegraphics[width=\linewidth]{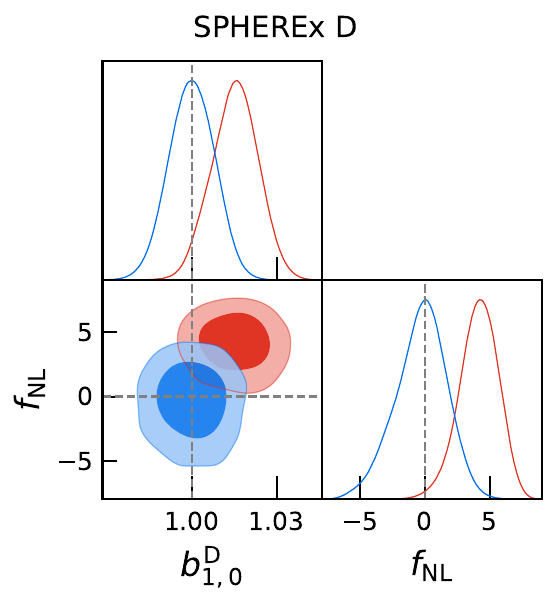}
\end{minipage}

\vspace{1ex}

\makebox[\textwidth][c]{%
\begin{minipage}[t]{0.28\textwidth}
\includegraphics[width=\linewidth]{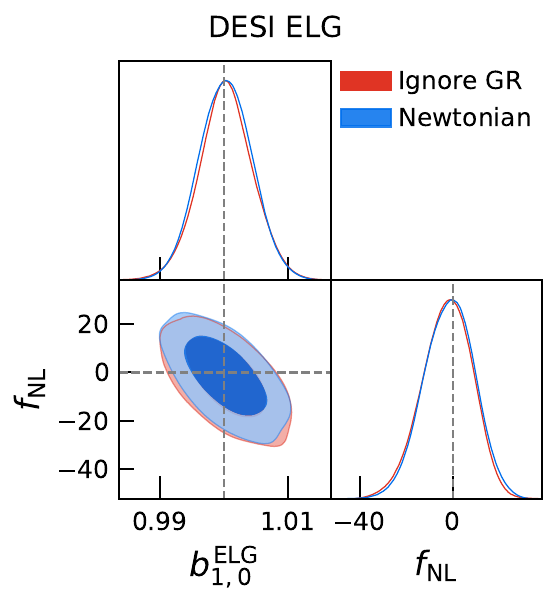}
\end{minipage}\hspace{0.02\textwidth}%
\begin{minipage}[t]{0.28\textwidth}
\includegraphics[width=\linewidth]{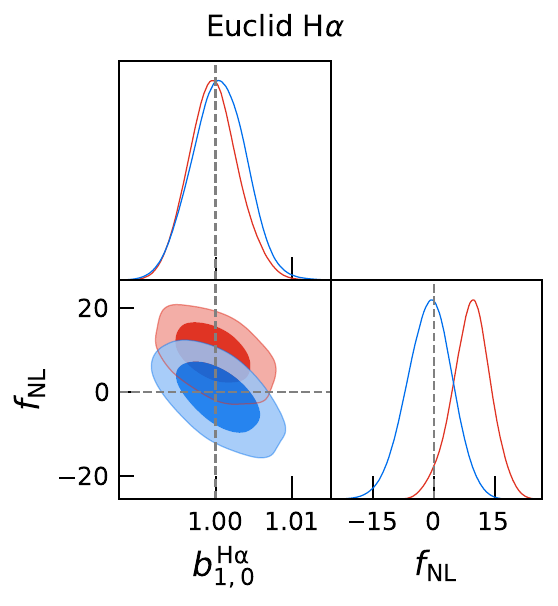}
\end{minipage}%
}
\caption{Impact of ignoring linear relativistic effects in local PNG inference for DESI, Euclid, and SPHEREx single tracers. We show the joint marginalized posterior in $\{b_{1,0}, f_{\rm NL}\}$, with dashed lines indicating the fiducial parameter values. Blue contours show the baseline Newtonian case, where both the simulated SFB PS data vector and the inference model use Newtonian theory. Red contours show the GR-mismatch case, where the simulated data vector is generated with the full linear GR model but the inference is performed with Newtonian modeling. Shifts of the red contours away from the dashed lines indicate parameter biases induced by neglecting GR effects. Depending on the galaxy sample, both linear bias and local PNG parameters can become biased.}
\label{fig:ignore-GR}
\end{figure*}

\begin{figure*}
\centerline{\includegraphics[width=0.7\hsize]{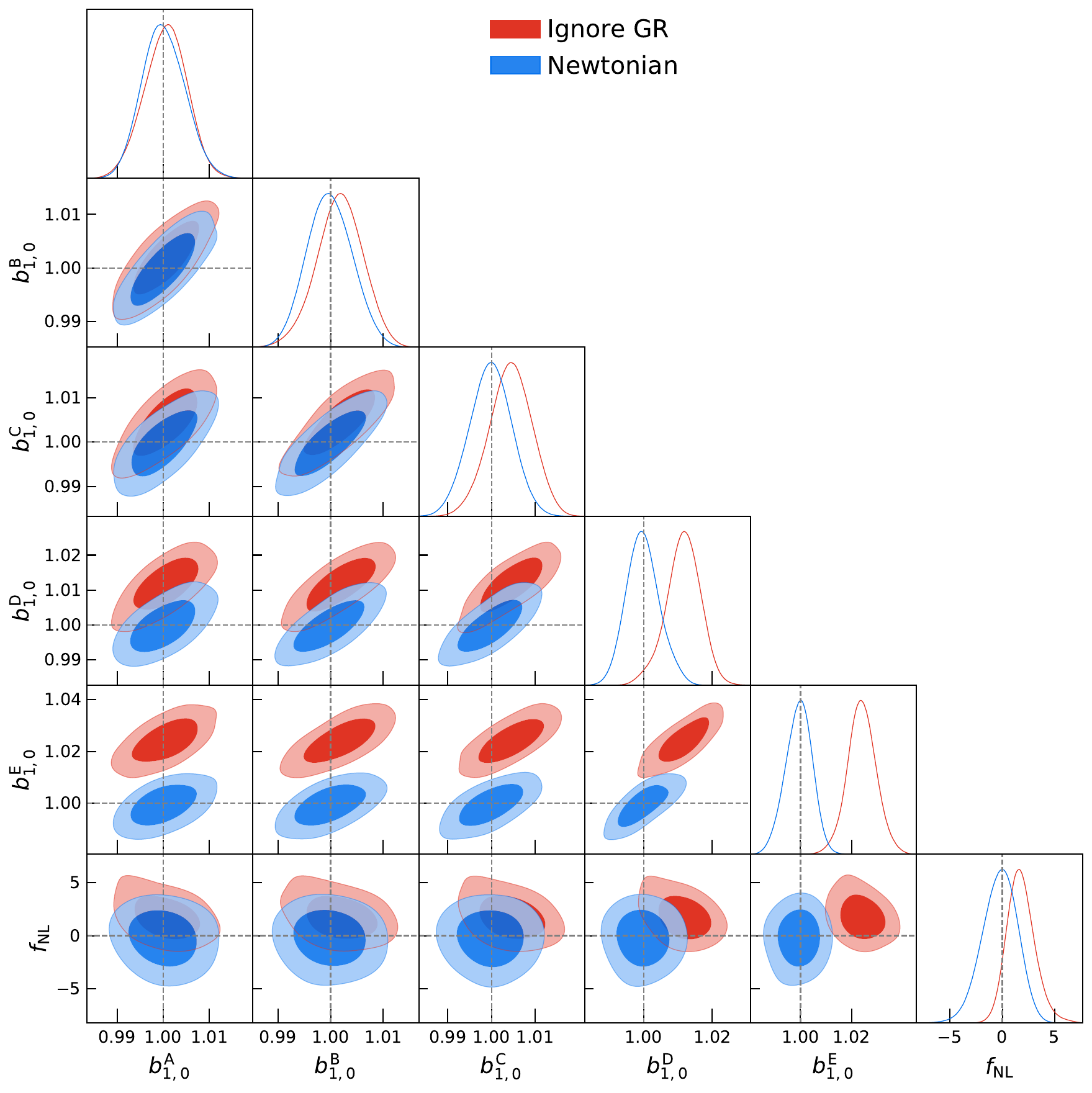}}
\caption{Impact of ignoring relativistic effects in local PNG inference for the combined SPHEREx five samples (SPHEREx 5T, $z\in[0,2]$), plotted with the same setup as \cref{fig:ignore-GR}. Ignoring GR effects biases the $\fNL$ constraint by $1.3\sigma$ ($\Delta f_{\rm NL} = +1.8$). The constraints of the linear bias also become significantly more biased for more photometric-like samples with larger redshift errors, due to the similarity of lensing and density terms in angular modes.}
\label{fig:ignore-GR-5T}
\end{figure*}

\subsubsection{Impact of $A_s$ and $n_s$}

We next examine the impact of marginalizing over the primordial power spectrum parameters, the scalar amplitude $A_s$ and the scalar spectral index $n_s$, on the $f_{\rm NL}$ constraints. These parameters are expected to be more relevant for local PNG inference than other $
\Lambda$CDM parameters, because they directly control the amplitude and shape of the underlying matter power spectrum.

Using DESI QSO, \cref{fig:fnl-AsNs-triangle} provides a representative example of the relevant degeneracies. The most important degeneracy for $f_{\rm NL}$ is with $n_s$: the local PNG bias introduces a strong low-$k$ enhancement to the galaxy power spectrum, which can partially mimic the effect of changing the primordial spectral tilt. The amplitude $A_s$ is instead primarily degenerate with the linear bias amplitude, since the density contribution to the galaxy power spectrum is controlled by the combination $b_1^2 A_s$. For spectroscopic tracers this $A_s$--$b_1$ degeneracy is broken by RSD, because the density and velocity contributions depend differently on $b_1$ and the growth rate $f$. The degeneracy is expected to be stronger for photometric-like tracers, where redshift errors suppress radial modes and weaken the RSD information.

For DESI QSO, the galaxy-only constraint on $n_s$ is about a factor of four times weaker than the \textit{Planck} constraint, leaving a residual $f_{\rm NL}$--$n_s$ degeneracy that noticeably degrades the marginalized $f_{\rm NL}$ constraint, as seen in \cref{fig:fnl-AsNs-triangle}. Pushing to smaller Fourier scales (higher $k_{\rm max}$) would improve the galaxy-only constraints on $A_s$ and $n_s$, but would require nonlinear modeling beyond the linear SFB PS used here.

The direct impact on the marginalized $f_{\rm NL}$ constraint is shown in \cref{fig:fnl-1d-AsNs}. Marginalizing over $A_s$ and $n_s$ without external priors typically degrades $\sigma(f_{\rm NL})$ at the $10$--$30\%$ level for all the tracers considered in this work. Once Gaussian priors from \textit{Planck} 2018 results are imposed, however, the degradation is reduced to the few-percent level ($\lesssim 5\%$), bringing results close to the baseline case with $A_s$ and $n_s$ fixed. This suggests that fixing the primordial power spectrum parameters, as commonly done in local PNG analyses, is a reasonable approximation. It does not artificially tighten the $f_{\rm NL}$ constraint, provided external CMB priors are effectively imposed. We therefore fix $A_s$ and $n_s$ to their fiducial values in the following inference runs.

\begin{figure}
\centerline{\includegraphics[width=0.75\hsize]{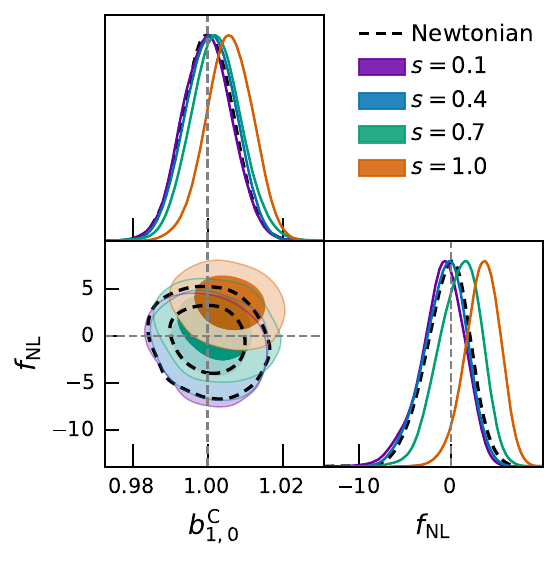}}
\caption{Impact of ignoring GR effects on local PNG inference for the SPHEREx C sample, shown for different fiducial values of the magnification bias. The induced biases in $\fNL$ and $b_{1,0}$ depend strongly on the assumed magnification bias, which controls the amplitude of the lensing magnification contribution.}
\label{fig:ignore-GR-vary-s}
\end{figure}

\subsection{Impact of GR effects on constraining $f_{\rm NL}$}\label{sec:ignore-GR}

With the inference configuration set, we now quantify the bias induced by ignoring linear relativistic effects for inferring $\fNL$. 

In the baseline Newtonian runs, both the simulated data vector and the inference model use Newtonian theory. The recovered posteriors are centered on the fiducial values, as shown by the blue contours in \cref{fig:ignore-GR}, confirming that our simulated-inference pipeline can recover the true values. We then generate the simulated data vector with the full linear relativistic corrections, using the fiducial parameters previously specified, but fit it with Newtonian model. The resulting shifts in the $f_{\rm NL}$ posterior then quantify the bias induced by omitting GR effects from the theory model.

The different panels of \cref{fig:ignore-GR} show that the GR-induced bias is strongly tracer dependent. The largest shifts occur for SPHEREx D and Euclid H$\alpha$, reaching $2.8\sigma$ ($\Delta f_{\rm NL}=+4.2$) and $1.9\sigma$ ($\Delta f_{\rm NL}=+9.2$), respectively. SPHEREx A shows a smaller shift of $0.76\sigma$ ($\Delta f_{\rm NL}=+3.4$), while the shifts for DESI ELG and DESI QSO are both minimal, below $0.25\sigma$. Thus, ignoring GR effects has little impact on the $f_{\rm NL}$ constraints for DESI ELG and QSO, but can lead to significant biases for Euclid H$\alpha$ and the photometric-like SPHEREx samples. The combined SPHEREx 5T experiences a bias of $1.3\sigma$ for $\fNL$ when ignoring GR effects (\cref{fig:ignore-GR-5T}).

Besides the local PNG parameter, neglecting GR effects can also bias the inferred linear galaxy bias parameters, especially for the photometric-like SPHEREx samples, as shown in \cref{fig:ignore-GR,fig:ignore-GR-5T}. The largest shift occurs for the SPHEREx E sample, where the zeroth-order Chebyshev coefficient of the linear galaxy-bias function is biased by $1.4\sigma$ in the single-tracer case and by $4.3\sigma$ in the multi-tracer case. By contrast, the linear bias parameters remain largely unbiased for the spectroscopic-like samples in DESI, Euclid, and SPHEREx. This is because spectroscopic-like samples retain radial modes, allowing the galaxy bias to be constrained by smaller-scale radial information that is mostly unaffected by lensing magnification. Photometric-like samples instead rely primarily on angular modes, where neglecting lensing induces an amplitude mismatch in the clustering signal that can be partially absorbed by the galaxy bias.

The GR-induced shift in $\fNL$ is not set only by the statistical constraining power of the tracer. DESI QSO, Euclid H$\alpha$, and SPHEREx A have comparable baseline sensitivity to $f_{\rm NL}$ (\cref{tab:fnl_baseline}), but show very different shifts when the full-GR data vector is fit with Newtonian model. This tracer dependence arises because the relativistic contributions depend on the fiducial magnification bias, evolution bias, redshift range, and radial selection. For high-redshift tracers considered here, the dominant GR contribution is lensing \cite{11ChallinorLPS,22CatorinaGR-P,24GR_SFB}. The small shift for DESI QSO is largely due to its magnification bias being close to $s=0.4$, where the lensing prefactor ($2-5s$) diminsihes. In contrast, Euclid H$\alpha$ has a larger lensing magnification contribution and therefore a larger GR-induced shift in $f_{\rm NL}$. This explanation is further supported by \cref{fig:ignore-GR-vary-s}, where we vary the fiducial magnification bias for SPHEREx Sample C and find that the GR-induced shift increases when the lensing prefactor \(|2-5s|\) is larger. The smaller shift for SPHEREx A compared to Euclid H$\alpha$ is likely driven by its relatively lower effective redshift, which reduces the lensing contribution.

This interpretation identifies lensing magnification as the dominant source of the GR-induced shifts. This may seem surprising, since lensing is not highly degenerate with local PNG\footnote{Ref.~\cite{24GR_SFB} showed that the local PNG and lensing signals have distinct angular dependence in SFB space, and in Sec.~\ref{sec:measurability-GR} we directly show the weak correlation between the two terms.}. The $f_{\rm NL}$ shifts should therefore not be interpreted as lensing simply mimicking the PNG signal. Instead, when lensing is omitted from the inference model, the best-fit $f_{\rm NL}$ can shift in an attempt to absorb part of the missing lensing contribution, even if the resulting Newtonian model remains a poor fit to the full-GR data vector. Thus, omitted relativistic terms, especially lensing magnification, can significantly bias local PNG inference even when they are not exactly degenerate with the PNG scale-dependent bias effect.

Taken together, these results show that the impact of linear relativistic effects is strongly tracer dependent. For local PNG analyses approaching $\sigma(f_{\rm NL})\sim 5$, especially in samples with large lensing magnification contributions, GR effects must be assessed on a tracer-by-tracer case to ensure unbiased inference at near- or beyond-\textit{Planck} sensitivity.

\begin{figure}
\centerline{\includegraphics[width=\hsize]{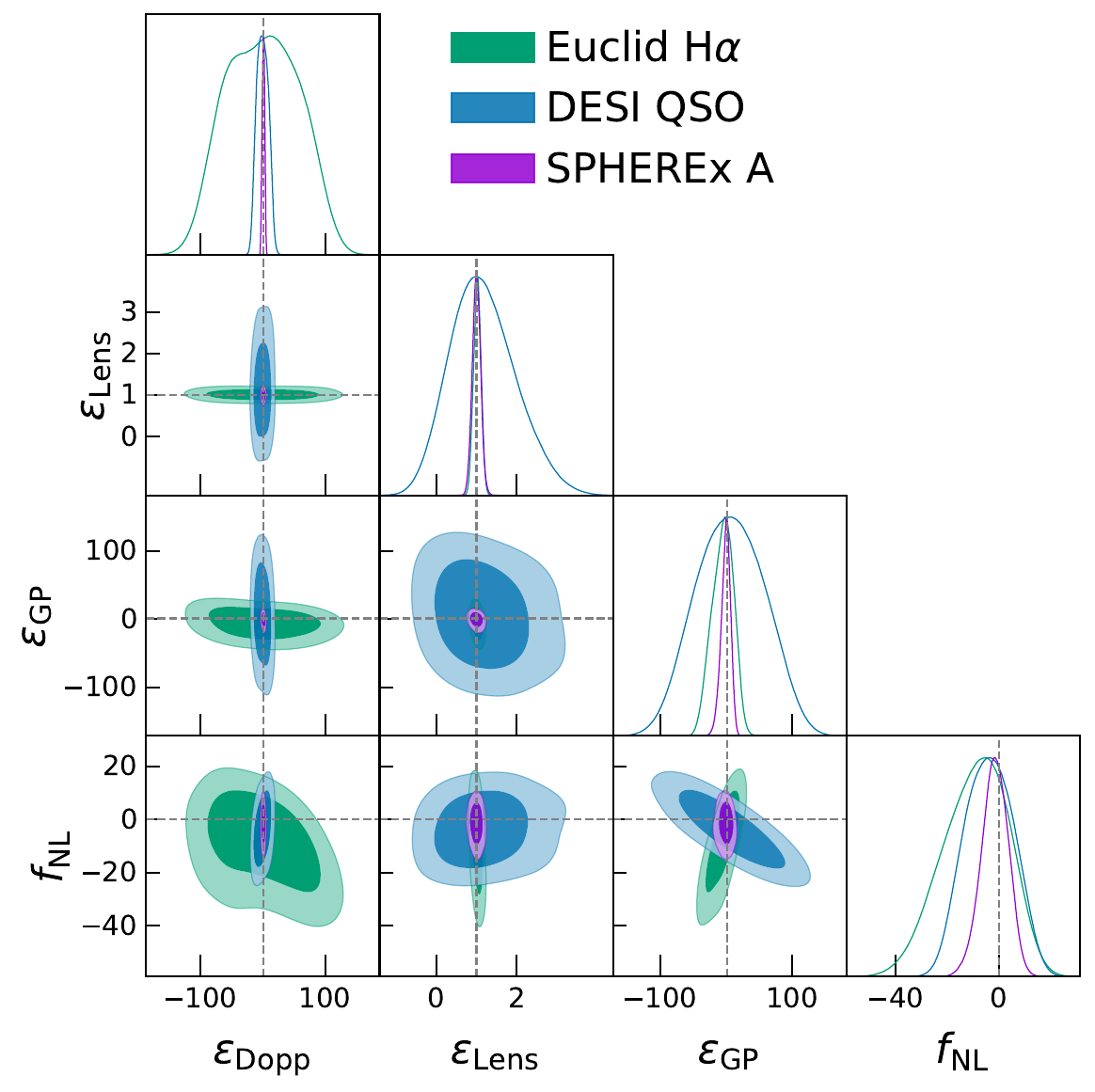}}
\caption{Measurability of individual relativistic terms (Doppler, lensing, and gravitational potential) through $\{\epsilon_{\rm Dopp},\,\epsilon_{\rm Lens},\,\epsilon_{\rm GP}\}$ and their degeneracy with $\fNL$. Three individual tracers are considered: Euclid H$\alpha$ (green), DESI QSO (purple), and SPHEREx A ($z\in [0,2]$, blue). Dashed lines mark the fiducial values with $\epsilon_{\rm Dopp} = \epsilon_{\rm Lens} = \epsilon_{\rm GP} = 1$ and $f_{\rm NL}=0$. For Euclid H$\alpha$ and DESI QSO, the GP and PNG terms are mildly degenerate with each other.}
\label{fig:amp-EQA}
\end{figure}

\subsection{Measurability of GR effects}\label{sec:measurability-GR}

We quantify the measurability of individual relativistic contributions by introducing phenomenological amplitudes $\{\epsilon_{\rm Dopp},\,\epsilon_{\rm Lens},\,\epsilon_{\rm GP}\}$, which scale the Doppler, lensing, and gravitational-potential terms in \cref{eq:GR}, respectively, with unity as their fiducial values. These amplitudes provide a simple diagnostic of whether a given survey can detect each relativistic term, without committing to a specific modified-gravity or dark-energy model. Their interplay with $\fNL$ also quantifies the potential degeneracy between local PNG and individual GR contributions.

When varying these amplitude parameters, we fix the magnification bias $s$ and set the evolution bias $b_{\rm e}$ through the linear bias using the Case 1 scenario of \cref{fig:biasparams}. This choice is necessary because the two parametrizations, $\{\epsilon_{\rm Dopp},\,\epsilon_{\rm Lens},\,\epsilon_{\rm GP}\}$ and $\{s,b_{\rm e}\}$, control closely related amplitude information. The $\epsilon$ parameters rescale combinations of relativistic terms directly, whereas $s$ and $b_{\rm e}$ change the tracer-dependent bias functions multiplying those same terms. Therefore, without external informative priors, varying both sets of parameters at the same time would introduce redundant degrees of freedom.

The two parametrizations nevertheless have different interpretations. The $\epsilon$ parameters are shared amplitudes of the physical relativistic contributions, and are therefore useful for asking how well a survey could detect these terms, or test for deviations from GR predictions. In contrast, $s$ and $b_{\rm e}$ are tracer-specific properties of the observed sample, tied to the selection function and redshift evolution of the tracer. This distinction is especially important in multi-tracer analyses: the amplitude parameters are common to all tracers, while each tracer has its own $s$ and $b_{\rm e}$.

In a standard PNG analysis, one would fix the GR amplitudes to their fiducial values, effectively assuming general relativity, and marginalize over uncertainties in $s$ and $b_{\rm e}$, or over the associated uncertainty in $b_\phi$ under the $b_\phi$-$b_{\rm e}$ equivalence. This case is discussed in Sec.~\ref{sec:bphi-implication}. Conversely, if one wants to use large-scale relativistic effects to constrain deviations from GR, then the magnification and evolution biases must be known sufficiently well from direct measurements of the selection function and time evolution. How well and trustworthy $s$ and $b_{\rm e}$ can be directly measured for realistic tracers remains an open question, especially for $b_{\rm e}$, while existing measurements of $s$ can depend on the exact method chosen \cite{22Elvin-Poole_DES_MB}. 

For these reasons, we first fix the bias parameters when varying the GR amplitudes in this section, and later vary the sample bias parameters themselves with the GR amplitudes fixed.

\begin{figure}
\centerline{\includegraphics[width=\hsize]{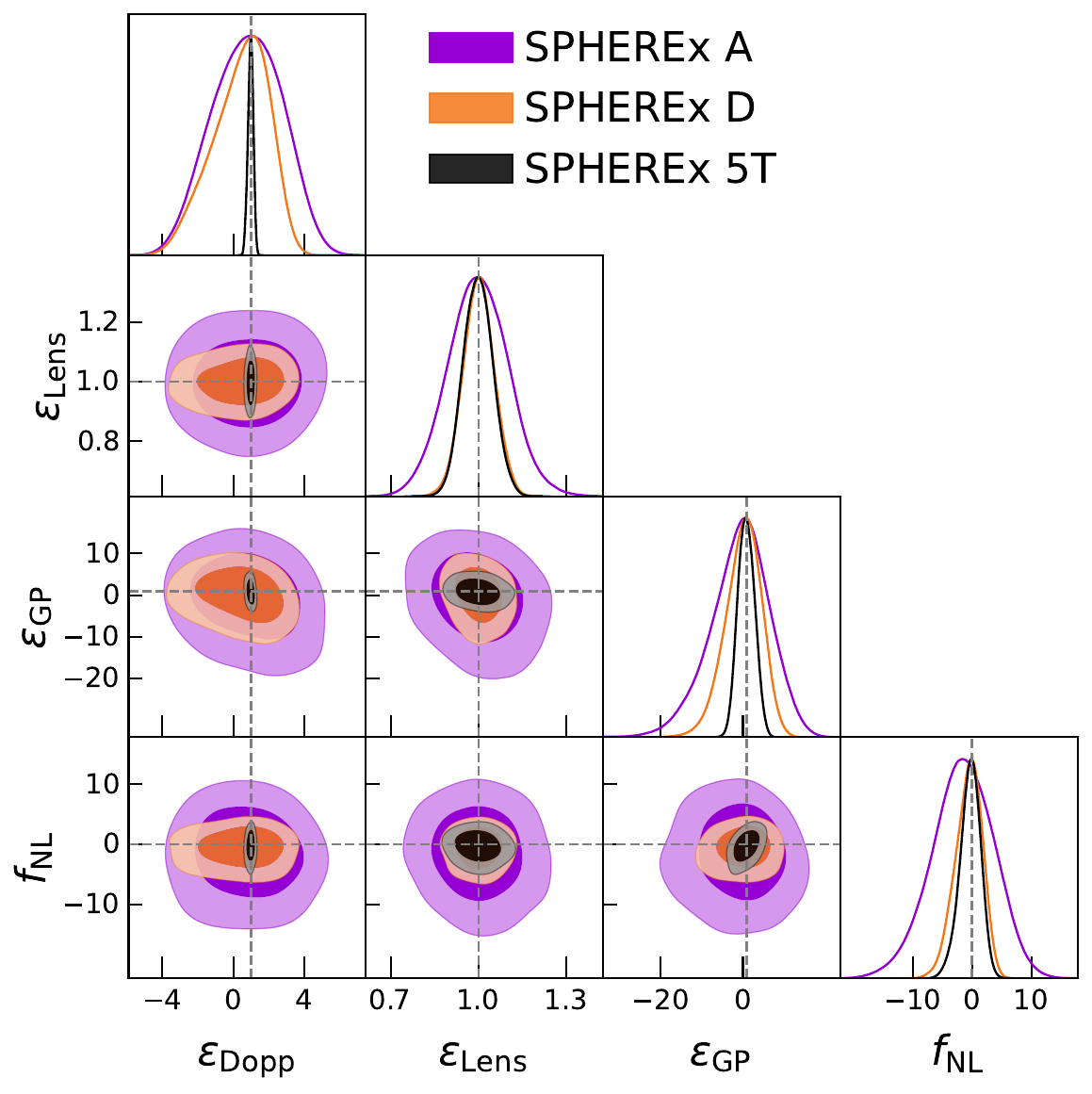}}
\caption{Joint posteriors of
$\{\epsilon_{\rm Dopp},\,\epsilon_{\rm Lens},\,\epsilon_{\rm GP},\,f_{\rm NL}\}$
for SPHEREx A and D single-tracer cases, as well as the combined five-tracer case (5T) for SPHEREx. Same setup as \cref{fig:amp-EQA}.}
\label{fig:amp-AD}
\end{figure}

\begin{figure}
\centering
\includegraphics[width=0.9\hsize]{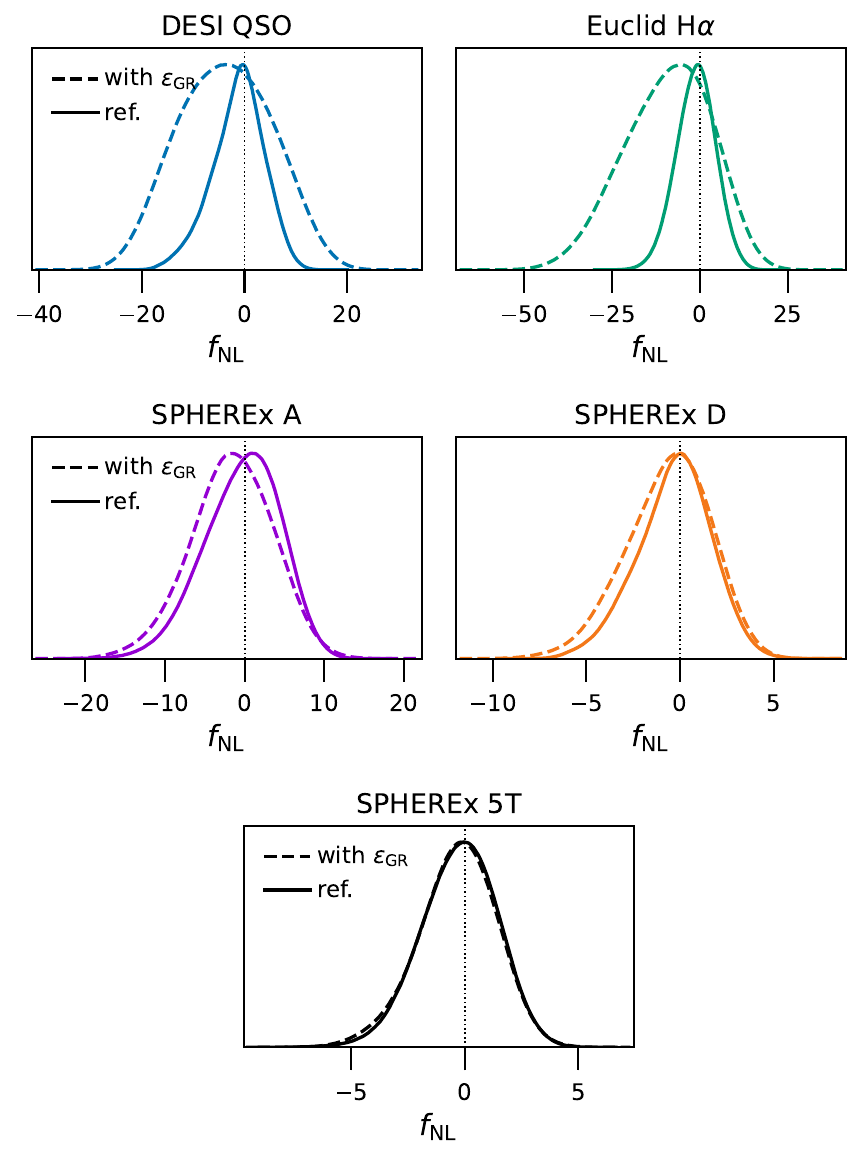}

\vspace{0.8em}

\begin{tabular}{lcc}
\hline
 & \multicolumn{2}{c}{$\sigma(f_{\rm NL})$} \\
\cline{2-3}
Tracer & Reference & With $\epsilon_{\rm GR}$ \\
\hline
DESI QSO & $5.1$ & $9.3$  \\
Euclid H$\alpha$ & $5.6$ & $12.6$  \\
SPHEREx A & $4.8$ & $5.2$ \\
SPHEREx D & $2.0$ & $2.3$ \\
SPHEREx 5T & $1.6$ & $1.7$ \\
\hline
\end{tabular}

\caption{One-dimensional marginalized constraints on $\fNL$ for single-tracer cases with the GR amplitude parameters fixed or marginalized. For each tracer, the solid line indicates the reference case with all GR amplitudes fixed, while the dashed line shows the $\fNL$ constraint where $\epsilon_{\rm GR}\equiv\{\epsilon_{\rm Dopp},\,\epsilon_{\rm Lens},\,\epsilon_{\rm GP}\}$ are marginalized over. The table gives the corresponding values of $\sigma(f_{\rm NL})$. The $\fNL$ constraints for DESI QSO and Euclid H$\alpha$ show significant degradation due to the degeneracy between the PNG and GP terms.}
\label{fig:amp-fnl-1D}
\end{figure}

\subsubsection{Degeneracy between $\fNL$ and GR terms}

Examining \cref{fig:amp-EQA,fig:amp-AD}, we find that the lensing and Doppler amplitudes are not noticeably degenerate with $\fNL$ for any tracer, as expected, since these terms have different angular and radial dependence in SFB space compared to the local PNG term \cite{24GR_SFB}. There is also little degeneracy among the three relativistic terms themselves, reflecting their different angular and Fourier patterns. The most noticeable degeneracy is between the gravitational-potential amplitude $\epsilon_{\rm GP}$ and $\fNL$, especially for DESI QSO and Euclid H$\alpha$, as shown in \cref{fig:amp-EQA}.

This degeneracy is expected based on \cref{eq:GR,eq:bias-PNG}. Both the local PNG and GP terms are proportional to the Bardeen potentials, up to tracer-dependent biases. Since the potentials scale as $k^{-2}$ on large scales, the two effects produce similar low-$k$ enhancements in the power spectrum, with similar angular--radial patterns in SFB space. As a result, marginalizing over $\epsilon_{\rm GP}$ can substantially degrade the $f_{\rm NL}$ constraint for DESI QSO and Euclid H$\alpha$, by around a factor of two (\cref{fig:amp-fnl-1D}).

However, this degeneracy between the PNG and GP terms is not exact. Even for these two tracers, $\fNL$ remains constrained. For the two SPHEREx samples shown in \cref{fig:amp-AD,fig:amp-fnl-1D}, there is almost no degeneracy between the two terms, with only around a $10\%$ degradation in $\sigma(f_{\rm NL})$. Therefore, the PNG and GP constraints do not significantly degrade each other for the SPHEREx samples.

This degeneracy is not exact because the PNG term depends on the potential at the matter-dominated epoch, whereas the GP contribution depends on the potential at the source redshift for the non-integrated terms, and on the integrated potentials between the source and the observer for the ISW and Shapiro terms. The two terms therefore carry different tracer-dependent prefactors. Even though the large-scale $k$-dependence is similar, the radial/redshift dependence is not identical, and can in principle be used to distinguish the two effects.

The redshift dependence of the GP and PNG terms depends on the fiducial magnification and evolution biases, so the degree of degeneracy between the two terms will be tracer-dependent as well. This plausibly explains both the different degeneracy directions between $\epsilon_{\rm GP}$ and $\fNL$ for DESI QSO and Euclid H$\alpha$ in \cref{fig:amp-EQA}, and the much weaker degeneracy for the SPHEREx samples in \cref{fig:amp-AD}.

\begin{table}
  \centering
\begin{tabular}{lccc}
    \hline\hline
    Tracer            & $\sigma(\epsilon_{\rm Dopp})$ & $\sigma(\epsilon_{\rm Lens})$ & $\sigma(\epsilon_{\rm GP})$ \\
    \hline
    DESI LRG          &  8.6  &  0.33   &  26   \\
    DESI QSO          &  8.9  &  0.81   &  51   \\
    \hline
    Euclid H$\alpha$  &  57   &  0.092  &  16   \\
    \hline
    SPHEREx A         &  1.9  &  0.10   &  7.3  \\
    SPHEREx B         &  1.4  &  0.075  &  5.5  \\
    SPHEREx C         &  1.3  &  0.066  &  4.8  \\
    SPHEREx D         &  1.6  &  0.053  &  4.5  \\
    SPHEREx E         &  1.0  &  0.064  &  5.0  \\
    SPHEREx 5T        &  0.15 &  0.050  &  2.0  \\
    \hline\hline
\end{tabular}
\caption{Summary of the forecasted $1\sigma$ constraints on the Doppler, lensing, and gravitational-potential amplitudes for DESI, Euclid, and SPHEREx ($z\in[0,2]$) tracers, with linear biases and $\fNL$ marginalized. The fiducial values of these amplitudes are set to unity. The results for DESI ELG are given in \cref{fig:amp-DESI-multi}.}
  \label{tab:gr_amplitudes}
\end{table}

\begin{table}
  \centering
\begin{tabular}{lccc}
\hline\hline
$s_{\rm fid}$ & $\sigma(\epsilon_{\rm Dopp})$ & $\sigma(\epsilon_{\rm Lens})$ & $\sigma(\epsilon_{\rm GP})$ \\
\hline
0.1 & 1.9 & 0.14   & 7.4 \\
0.4 & 9.2 & $-$    & 72  \\
0.7 & 2.6 & 0.13   & 9.3 \\
1.0 & 1.3 & 0.066  & 4.8 \\
\hline\hline
\end{tabular}
\caption{The forecasted $1\sigma$ constraints on the Doppler, lensing, and GP amplitudes for SPHEREx Sample C, with the fiducial magnification bias varied from $s=0.1$ to  $1.0$. The constraints strongly depend on the chosen magnification bias.}
  \label{tab:gr_amp_spherex_c}
\end{table}

\begin{figure}
\centering
\includegraphics[width=0.93\hsize]{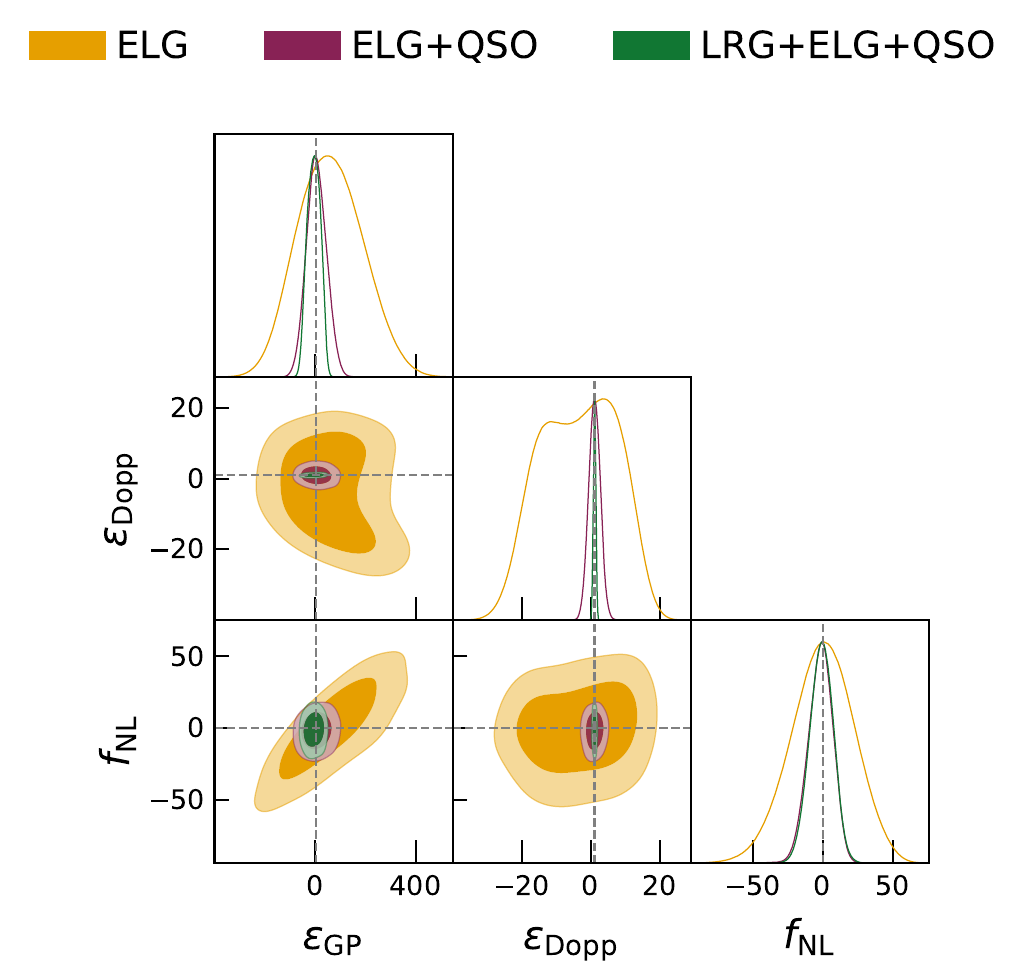}

\vspace{0.8em}

\begin{tabular}{lccc}
\hline
Tracer(s)  & Redshift & $\sigma(\epsilon_{\rm Dopp})$ & $\sigma(\epsilon_{\rm GP})$ \\
\hline
ELG  &  $[0.8,1.6]$ & $11$ & $130$ \\
ELG+QSO & $[0.8,1.6]$ & $1.7$ & $39$ \\
LRG+ELG+QSO & $[0.8,1.1]$ & $0.36$ & $25$ \\
\hline
\end{tabular}

\caption{Measurability of the Doppler and gravitational-potential terms for the DESI multi-tracer configurations. We restrict the analysis to the redshift range where the tracers overlap. The lensing amplitude is not shown because, for the fiducial DESI ELG magnification bias, the lensing prefactor $5s-2$ vanishes, making the lensing term unmeasurable in these configurations. Nevertheless, the lensing amplitude is marginalized over in all cases. The multi-tracer analyses significantly improve the constraints on both the Doppler and gravitational-potential terms compared to the single-tracer constraints from DESI ELG shown here and from DESI LRG and QSO in \cref{tab:gr_amplitudes}.}
\label{fig:amp-DESI-multi}
\end{figure}

\subsubsection{Detectability of GR terms}

We now explore whether the GR terms themselves can be detected. The results are illustrated in \cref{fig:amp-EQA,fig:amp-AD} for selected tracers and summarized for every sample in \cref{tab:gr_amplitudes}. The lensing term is much easier to measure for tracers whose fiducial magnification bias gives a non-negligible lensing prefactor. For Euclid H$\alpha$, we find $\sigma(\epsilon_{\rm Lens})=0.092$, corresponding to a signal-to-noise ratio (SNR) of around 11. Across all five SPHEREx samples the lensing term is measured at SNRs of roughly 10 to 20, the highest being SPHEREx D. The DESI samples give weaker constraints — $\sigma(\epsilon_{\rm Lens})=0.33$ for LRG and $0.81$ for QSO — reflecting the lower redshift coverage of LRG and the smaller effective lensing prefactor for QSO. The measurability of lensing strongly depends on the fiducial magnification bias of the sample, as shown in \cref{tab:gr_amp_spherex_c} with SPHEREx C as an example. 

In contrast, the Doppler and gravitational-potential amplitudes remain hard to isolate in single-tracer analyses. They are not measured at $1\sigma$ for any single tracer (\cref{tab:gr_amplitudes}), except for SPHEREx E barely achieving $\sigma(\epsilon_{\rm Dopp})=1.0$. The tightest single-tracer GP constraint comes from SPHEREx D, with $\sigma(\epsilon_{\rm GP})=4.5$. For SPHEREx tracers, where the degeneracy between PNG and GP terms is weak, $\sigma(\epsilon_{\rm GP})$ is roughly twice the corresponding \(\sigma(\fNL)\) values in \cref{tab:fnl_baseline}. This reflects the approximate factor of two between $b_{\rm e}$ (setting the amplitude of the GP term) and $b_\phi$ (setting the PNG term) in their equivalence relation of \cref{eq:bphi-be-equiv}. Thus, while lensing is expected to be detected at high significance for tracers whose magnification bias is sufficiently different from $0.4$, the Doppler and GP terms remain difficult to measure with a single tracer.

The Doppler and GP constraints can improve significantly in multi-tracer analyses. We illustrate this with the DESI multi-tracer combinations in \cref{fig:amp-DESI-multi}. Adding QSO to ELG over the same redshift range improves the Doppler constraint by a factor of $6.5$ and the GP constraint by a factor of $3.3$\footnote{This gain is not simply due to the QSO sample dominating the combination: despite being restricted to the overlapping range $z\in[0.8,1.6]$, the ELG+QSO constraint is significantly stronger than the Doppler constraint from the single-tracer QSO in \cref{tab:gr_amplitudes}, which uses the much larger full QSO range $z\in[0.8,3.5]$. This shows that the improvement is driven by genuine multi-tracer information.}. The full LRG+ELG+QSO combination further improves the Doppler constraint by a factor of $4.6$ relative to ELG+QSO, yielding a $2.8\sigma$ measurement of the Doppler amplitude. The GP constraint also improves in the three-tracer case, although it remains far from detection. These improvements are especially notable because the three-tracer combination uses the smaller overlapping redshift range $z\in [0.8,1.1]$, therefore improving constraints despite the smaller volume compared to the single-tracer ELG case.

Similar improvements appear in the SPHEREx multi-tracer configurations, as shown in \cref{fig:amp-AD} and \cref{tab:gr_amplitudes}. Going from the single-tracer SPHEREx E sample to the five-tracer combination sharpens the Doppler constraint by more than a factor of $6$, yielding a detection of the Doppler amplitude at more than $6\sigma$. The GP constraint also improves by more than a factor of $2$, from $\sigma(\epsilon_{\rm GP})=4.5$ in SPHEREx D to $\sigma(\epsilon_{\rm GP})=2.0$ in the multi-tracer case.

These improvements in the Doppler and GP constraints reflect multi-tracer variance cancellation, similar to the case for local PNG \cite{09Seljak_multi}: tracers with different linear biases, magnification biases, and evolution biases respond differently to the same underlying cosmological field. For a single tracer, the Doppler and GP terms are small relativistic corrections on top of the much larger density and RSD contributions, and are therefore strongly sample-variance limited on large scales. Multi-tracing compares these different tracer responses mode by mode, partially canceling the shared Newtonian fluctuation while retaining the differential Doppler and GP contributions. Therefore, additional gain is obtained from the contrast between the different response coefficients of tracers. 

Our results in \cref{fig:amp-DESI-multi} should be viewed as a baseline using the available DESI tracer samples as they are. Further gains may be possible with targeted sample optimization or sub-sample splits \cite{14PBonvinAsymmetryCorrelation,23Bonvin_Dipole,24Montano_Doppler_split} designed to maximize differences in linear bias, magnification bias, and evolution bias, thereby enhancing the relative Doppler and GP responses between tracers to further improve their measurability.

\begin{table*}
\centering
\begin{tabular*}{\textwidth}{@{\extracolsep{\fill}}lccccccc}
\hline
Parameter 
& $b_{1,0}^{X}$ 
& $b_{1,1}^{X}$ 
& $b_{1,2}^{X}$ 
& $s_{,0}^{X}$ 
& $b_{\phi,0}^{X}$ 
& $p^{X}$ 
& $\fNL$ \\
\hline
Prior 
& $\mathcal{U}(0,3)$ 
& $\mathcal{U}(-2,2)$ 
& $\mathcal{U}(-2,2)$ 
& $\mathcal{U}(-10,10)$ 
& $\mathcal{U}(-50,50)$ 
& $\mathcal{U}(-50,50)$ 
& $\mathcal{U}(-100,100)$ \\
\hline
\end{tabular*}
\caption{Uniform priors used in nested sampling runs. Here $X$ labels each tracer. Each parameter is assigned an independent prior. The $b_{\phi,0}^{X}$ and $p^X$ priors correspond to two alternative parametrizations of redshift evolution for the LPNG bias, as discussed in Sec.~\ref{sec:bphi-implication}, and are never used simultaneously. Except for $\{b_{\phi,0}^{X}, p^X, \fNL\}$, all posteriors are well within the prior boundaries. For PNG-related parameters, the posteriors are also prior-independent in the well-constrained cases, while finite prior ranges are needed when the product degeneracy between the LPNG bias and $\fNL$ would otherwise allow the sampler to explore arbitrarily large individual parameter values, such as the left panel of \cref{fig:p-bphi-QSO}.}
\label{tab:nautilus-priors}
\end{table*}

\begin{figure}
\centerline{\includegraphics[width=0.75\hsize]{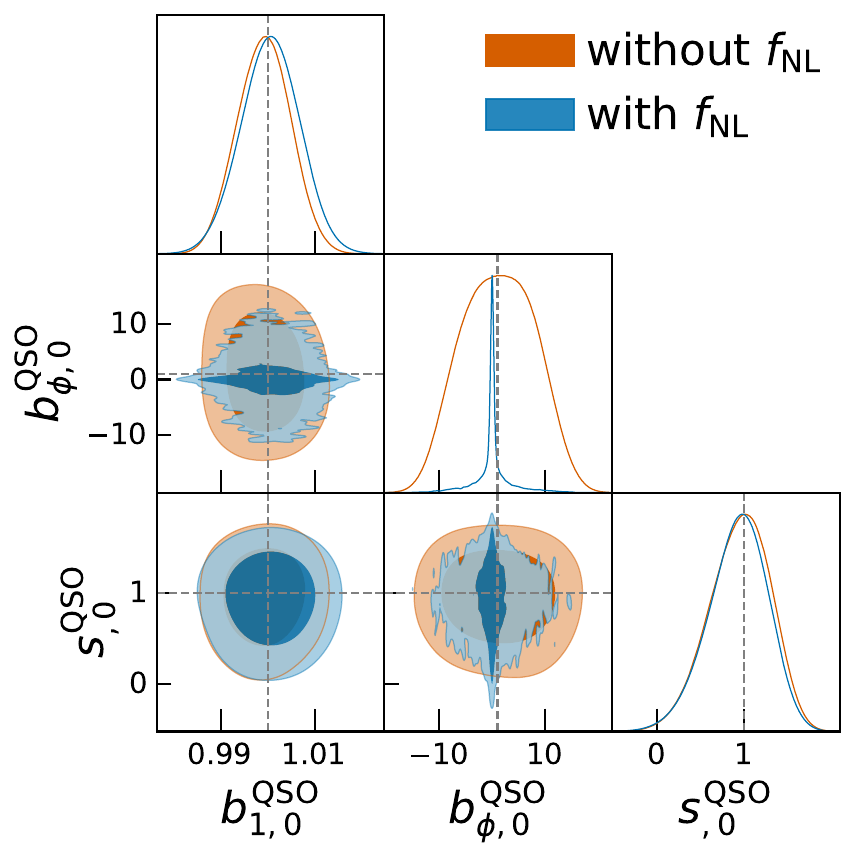}}
\caption{Constraints on \(s\) and \(b_\phi\) for the DESI QSO sample, shown for the cases with \(\fNL\) fixed and marginalized. The magnification bias is relatively well constrained and largely unaffected by PNG marginalization, while the local PNG bias constraint is strongly affected by whether \(\fNL\) is varied. Although the two cases have comparable spread in \(b_{\phi,0}\) in the two-dimensional marginalized contours, their one-dimensional marginalized constraints differ significantly, indicating strong projection effects in the \(b_{\phi,0}\) posterior.}
\label{fig:sbphi-QSO}
\end{figure}

\begin{table}
  \centering
\begin{tabular}{lcc}
\hline\hline
Tracer & $\sigma(s_{,0})$ & $\sigma(b_{\phi,0})$ \\
\hline
DESI LRG         & 0.19  & 12   \\
DESI ELG         & 0.31  & 14   \\
DESI QSO         & 0.35  & 6.8  \\
Euclid H$\alpha$ & 0.052 & 9.9  \\
SPHEREx A        & 0.060 & 7.7  \\
SPHEREx E        & 0.040 & 5.1  \\
\hline\hline
\end{tabular}
\caption{Measurability of the magnification bias $s$ and LPNG bias $b_\phi$ for DESI, Euclid, and SPHEREx samples, with $\fNL=0$ fixed and the linear-bias parameters marginalized. We show forecasted $1\sigma$ constraints on the zeroth-order Chebyshev coefficients, which control the overall amplitude of the baseline bias functions and are normalized to one for their fiducial values. Here we fix $\fNL$ to quantify the constraining power on $b_\phi$ relativistic effects, without the impacts of projection effects due to degeneracy between $b_\phi$ and $\fNL$.}
  \label{tab:sbphi}
\end{table}

\begin{figure*}
\centering
\makebox[\textwidth][c]{%
\begin{minipage}[t]{0.35\textwidth}
\includegraphics[width=\linewidth]{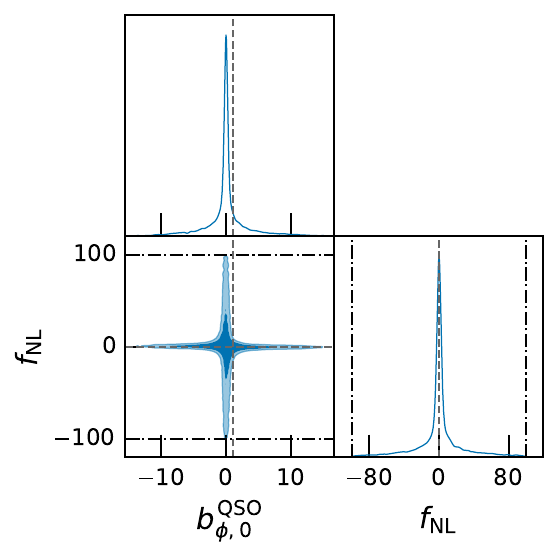}
\end{minipage}\hspace{0.15\textwidth}%
\begin{minipage}[t]{0.35\textwidth}
\includegraphics[width=\linewidth]{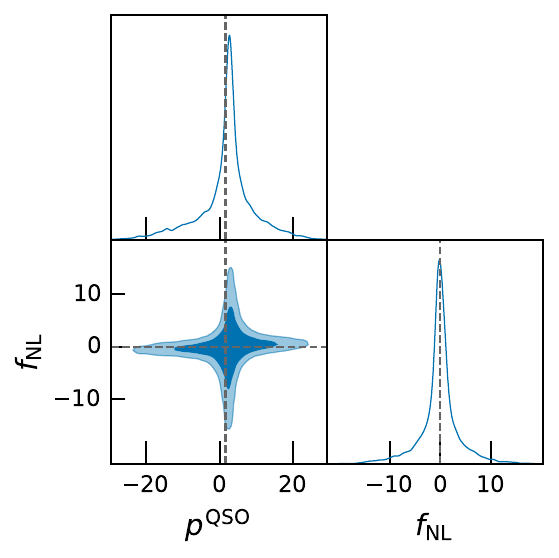}
\end{minipage}%
}
\caption{Constraints of $b_\phi$ and $\fNL$ under two different parametrizations of $b_\phi$ for DESI QSO, with linear and magnification bias parameters marginalized. The left plot uses $b_{\phi,0}$ as an amplitude parameter modulating the fiducial $b_\phi$ function (\cref{eq:bphi-cheb}), while the right plot varies $p$ in the modified universality relation (\cref{eq:bphi_universal}). The black dot-dashed lines indicate the uniform prior used for $\fNL$ in sampling, indicating the constraints on $\fNL$ being  bounded by its prior in the $b_{\phi,0}$ parametrization. In comparison, one can obtain meaningful bounds on $\fNL$ under the $p$ parametrization. Under both cases, the 1D marginalized $\fNL$ posterior suffers from projection effects, where the constraints are highly sensitive to the prior volume of $b_\phi$. }
\label{fig:p-bphi-QSO}
\end{figure*}

\subsection{Implication for $b_{\phi}$ }\label{sec:bphi-implication}

We now consider the joint constraints on $\{s,b_{\phi},\fNL\}$, with the evolution bias $b_{\rm e}$ set by $b_{\phi}$ through their equivalence relation. As discussed in Sec.~\ref{sec:bias-param}, we vary the zeroth-order Chebyshev coefficients for $s$ and $b_\phi$, which act as amplitude parameters multiplying their fiducial bias functions. For the LPNG bias, we also consider a second parametrization where the slope parameter $p$ in the universality relation of \cref{eq:bphi_universal} is varied in inference. 

The magnification bias is generally well constrained, as shown in \cref{tab:sbphi}. It is not strongly degenerate with PNG, and marginalizing over $\fNL$ has little impact on its constraint (\cref{fig:sbphi-QSO}). Most of the constraining power on $s$ comes from the lensing term, which can be measured at high SNR. Therefore, relativistic clustering provides an independent way to measure $s$, complementary to estimates based on perturbing galaxy fluxes and re-running the target selection. This provides a useful cross-check between the two approaches and suggests that the magnification-bias measurement can be robust. Since $s$ is not strongly degenerate with either $\fNL$ or $b_\phi$, we leave it aside in the rest of this section and focus on the two PNG-related parameters.

In linear-order PNG power-spectrum analyses, $b_\phi$ and $\fNL$ are completely degenerate in the absence of relativistic effects. One can only constrain the product $b_\phi \fNL$, so an informative prior on $b_\phi$ is required to obtain any constraint on $\fNL$. Refs.~\cite{20Barreira_bphi_impact_constraint,22Barreira_bphi_BOSS} have studied the impact of uncertainties in the slope parameter $p$ on $f_{\rm NL}$ constraints under Newtonian modeling\footnote{One caveat is that Ref.~\cite{20Barreira_bphi_impact_constraint} assumes the same value of $p$ for different samples in their multi-tracer analyses. This produces different $b_\phi$ values through different $b_1$ values, but introduces only one additional degree of freedom. In this work, we allow $p$ to vary independently for each sample.}.

With relativistic effects included, the observed power spectrum also becomes sensitive to the evolution bias $b_{\rm e}$. If $b_{\rm e}$ is set by $b_\phi$ through their equivalence relation, the relativistic terms provide an additional dependence on $b_\phi$ alone beyond just $b_\phi \fNL$. This can potentially break the $b_\phi$--$\fNL$ degeneracy and, in principle, yield a constraint on $\fNL$ without imposing an external prior on $b_\phi$. Using relativistic terms to infer $b_{\rm e}$, and hence $b_\phi$, requires assuming the $\Lambda$CDM model under GR, as discussed in Sec.~\ref{sec:measurability-GR}. 

We therefore sample directly in $\{b_\phi,\fNL\}$ and do not use the product parametrization $b_\phi\fNL$. This differs from Newtonian clustering analyses, where $b_\phi\fNL$ is a natural parameter for inference. Once relativistic terms are included and $b_{\rm e}$ is tied to $b_\phi$, $b_\phi$ will enter the data vector independently. In a single-tracer analysis one could formally sample $\{b_\phi\fNL,b_\phi\}$, but sampling $\{b_\phi\fNL,\fNL\}$ is ill-defined near $\fNL=0$, where $b_\phi=(b_\phi\fNL)/\fNL$ diverges. In multi-tracer analyses, the product parametrization is even less stable because each tracer has its own $b_\phi$ while $\fNL$ is shared, requiring divisions by $\fNL$ to recover the individual $b_\phi$ values. For this reason, we are left with the sampling choice $\{b_\phi,\fNL\}$.

\begin{figure}
\centering
\includegraphics[width=0.85\hsize]{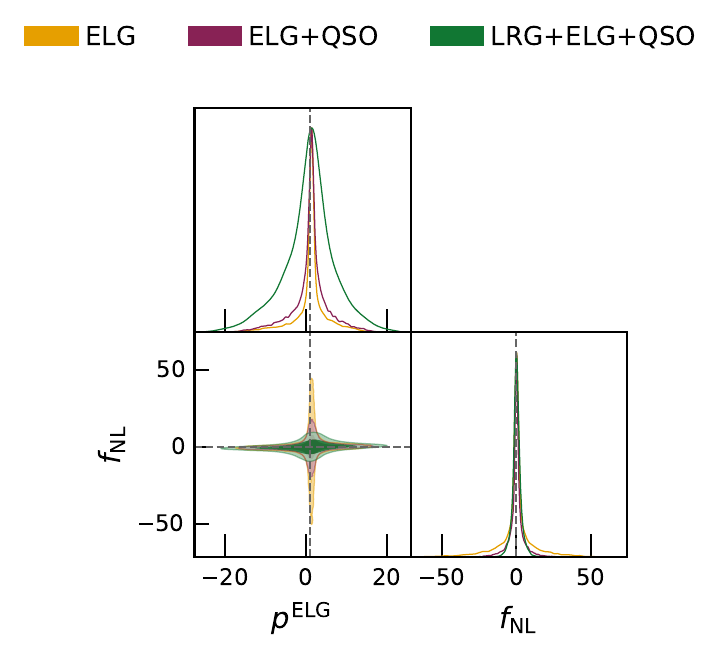}
\caption{Joint constraints on \(\fNL\) and the ELG tracer's \(p\)  under the DESI multi-tracer configurations. Multi-tracer information substantially reduces, but does not eliminate, the degeneracy between \(b_\phi\) and \(\fNL\) in the \(p\) parametrization. As a result, the 1D marginalized constraints remain dominated by projection effects, and should therefore not be interpreted as reliable physical constraints on the individual parameters. The two-dimensional posteriors are needed to assess the actual information gained from multi-tracer analyses.
}
\label{fig:pfnl-DESI-multi}
\end{figure}

\begin{figure*}
\centering
\makebox[\textwidth][c]{%
\begin{minipage}[t]{0.35\textwidth}
\includegraphics[width=\linewidth]{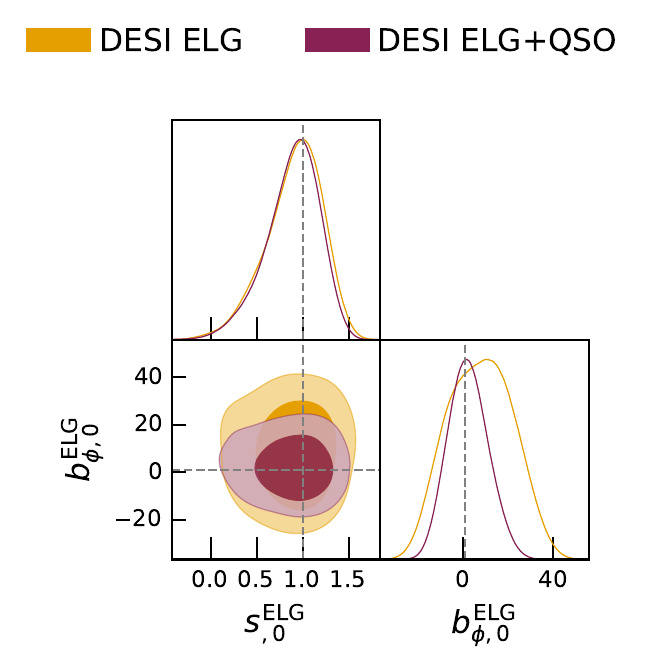}
\end{minipage}\hspace{0.1\textwidth}%
\begin{minipage}[t]{0.34\textwidth}
\includegraphics[width=\linewidth]{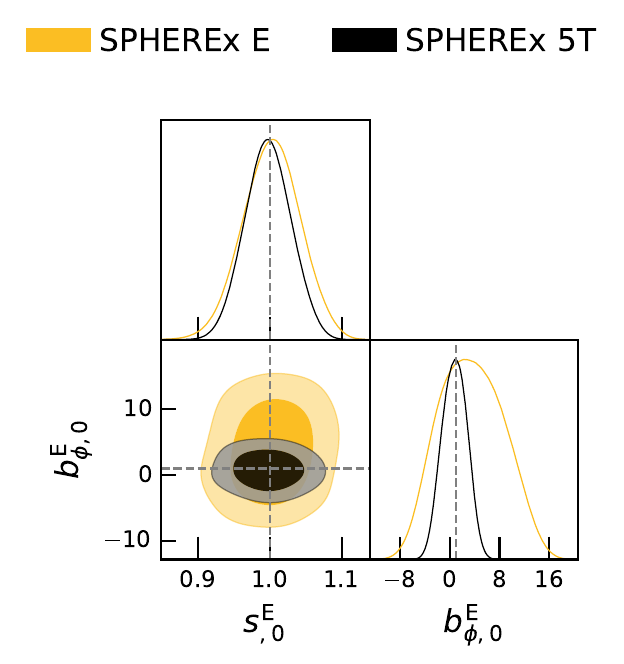}
\end{minipage}%
}
\caption{Constraints on $s_{,0}$ and $b_{\phi,0}$, the amplitude parameters for the magnification bias and LPNG bias functions, for the DESI ELG+QSO multi-tracer sample and the combined five-tracer SPHEREx sample. We use the same setup as in \cref{tab:sbphi}, with $\fNL$ fixed to zero. Multi-tracer analyses noticeably improve the constraining power on $b_\phi$: we find $\sigma(b_{\phi,0})=8.9$ for ELG in ELG+QSO and $\sigma(b_{\phi,0})=2.0$ for SPHEREx E in SPHEREx 5T, each improving over the corresponding single-tracer constraint by about a factor of two.} 
\label{fig:s-bphi-multi}
\end{figure*}

We show the joint constraints for local PNG under the two parametrizations for $b_\phi(z)$ in \cref{fig:p-bphi-QSO}, using DESI QSO as an example. In general, the degeneracy breaking provided by relativistic effects is clearly present but weak: \(b_\phi\) is constrained only at the level of \(\mathcal{O}(10)\), leaving a strong star-shaped degeneracy between \(b_\phi\) and \(\fNL\). The resulting joint constraints depend sensitively on how \(b_\phi(z)\) is parametrized. In the \(b_{\phi,0}\) parametrization of \cref{eq:bphi-cheb}, \(\fNL\) remains effectively unconstrained because values near \(b_{\phi,0}=0\) are allowed. When \(b_{\phi,0}=0\), the local PNG bias vanishes across the redshift range, so that \(b_\phi(z)\fNL=0\) independently of \(\fNL\). The likelihood is then insensitive to \(\fNL\), and the posterior of $\fNL$ is only limited by the prior, as seen in the left panel of \cref{fig:p-bphi-QSO}. This behavior persists for all single-tracer and multi-tracer cases considered in this work, since none of these configurations can rule out \(b_{\phi,0}=0\), as demonstrated in \cref{tab:sbphi} and \cref{fig:s-bphi-multi}.

By contrast, in the \(p\) parametrization of \cref{eq:bphi_universal}, \(b_\phi(z)\) is tied to the redshift evolution of \(b_1(z)\), and varying \(p\) changes the response without generically allowing \(b_\phi(z)\) to vanish over the full redshift range. The signal will therefore always retain some sensitivity to \(\fNL\). Despite the strong degeneracy, \(\fNL\) remains bounded as shown in the right panel of \cref{fig:p-bphi-QSO}. 

For tracers considered in this work, relativistic effects do not constrain \(b_\phi\) sufficiently well to make the inferred \(\fNL\) constraint independent of the assumed form of \(b_\phi(z)\). The parametrization acts effectively as a prior on the allowed tracer response. Although tracers with vanishing \(b_\phi(z)\) over an extended redshift range are likely unphysical, especially for all tracers in a multi-tracer analyses, excluding such cases effectively requires prior knowledge on galaxy formation and evolution, obtained either from theory or data. 

Even in the \(p\) parametrization, where \(\fNL\) remains bounded, the 1D marginalized posterior of \(\fNL\) is dominated by the projection effect, also known as the prior-volume effect \cite{25Guachalla_prior}. This arises from the star-shaped degeneracy between \(p\) and \(\fNL\), which causes the marginalized posterior to depend strongly on the volume of poorly constrained parameter space. We therefore do not report the 1D marginalized \(\fNL\) constraint even in the $p$ case. 

The 2D marginalized contours nevertheless provide useful information about the joint constraining power on \(p\) and \(\fNL\). Examining the DESI multi-tracer cases shown in \cref{fig:pfnl-DESI-multi}, we see that multi-tracer analyses substantially reduce the degeneracy between \(p\) and \(\fNL\), leading to tighter bounds on \(\fNL\) in the 2D contours, although the degeneracy is not completely removed.

One can alternatively quantify the multi-tracer gain on $b_\phi$ by fixing $\fNL=0$, so that the degeneracy between $\fNL$ and $b_\phi$ is eliminated and the one-dimensional marginalized constraint is not affected by projection effects. In this setup, we only quantify how well $b_\phi$ (or equivalently $b_{\rm e}$) can be measured from the relativistic contributions. From \cref{fig:s-bphi-multi}, we see that multi-tracer analyses significantly improve the constraining power on $b_{\phi,0}$ for individual tracers, with roughly a factor of two improvement over the corresponding single-tracer constraints for cases plotted. Even in the best case, however, the constraint only reaches $\sigma(b_{\phi,0})=2.0$ for SPHEREx E in the SPHEREx 5T sample, still a factor of a few away from ruling out $b_{\phi,0}=0$ and thereby preventing unbounded $\fNL$ constraints. Nevertheless, these results show that multi-tracer analyses may still make it possible to constrain $b_\phi$, and hence obtain bounded $\fNL$ constraints from power spectrum alone without external priors, for more optimized samples or future surveys.

\section{Conclusion and discussion}\label{sec:conclusion}

In this work, we present forecasts for local PNG and linear-order relativistic effects in the SFB power spectrum for DESI, Euclid, and SPHEREx surveys. Assuming a Gaussian distribution for redshift errors, we model the redshift smearing effects exactly in the SFB basis and use Chebyshev decompositions to accelerate the SFB power-spectrum calculations. This makes Bayesian inference with the full set of linear relativistic effects practical, enabling both MCMC and nested-sampling analyses.

We find that neglecting linear relativistic effects can produce non-negligible biases in local PNG inference when surveys reach \(\sigma(f_{\rm NL})\sim 5\). For Euclid H\(\alpha\) and SPHEREx, the induced shifts in \(f_{\rm NL}\) can reach the \(1\)--\(3\sigma\) level. The size of the shift depend strongly on the fiducial magnification bias, which sets the lensing contribution to the observed number count fluctuations. The impact of relativistic corrections therefore depends on the survey configuration and sample definitions, requiring a tracer-by-tracer assessment.

From the joint constraints on PNG and relativistic amplitudes, we confirm that the lensing and Doppler terms are not strongly degenerate with local PNG in the SFB basis: lensing differs mainly in its angular dependence compared to local PNG, while Doppler differs mainly in its Fourier dependence. However, neglecting these two terms can still bias \(f_{\rm NL}\). The gravitational potential term is more strongly degenerate with the local PNG, but the degeneracy is not exact because the two contributions have different redshift dependence. The resulting degradation of \(\sigma(f_{\rm NL})\) is strongly tracer-dependent due to the different redshift evolutions set by the tracer's biases, ranging from a few percent to nearly a factor of two for samples considered.

We next quantify the detectability of the individual relativistic terms. The detectability is also strongly impacted by the fiducial magnification bias value of a sample. The lensing amplitude is measurable at high significance for tracers with a large lensing prefactor, reaching \({\rm SNR}\gtrsim 10\) for Euclid H\(\alpha\) and SPHEREx. The Doppler and gravitational-potential amplitudes are much harder to isolate with a single tracer, but multi-tracer analyses significantly improve the Doppler constraint. For the DESI LRG+ELG+QSO combination, even when restricted to the overlapping redshift range $z\in[0.8,1.1]$, the Doppler term can be measured at roughly \(3\sigma\). For the SPHEREx five-tracer combination, the significance can reach around \(6\sigma\). The gravitational-potential amplitude remains undetected in the configurations considered here, although its constraint also improves substantially in the multi-tracer cases. These results imply that targeted sub-sample splits or optimized tracer selections could further improve the measurability of these terms.

Assuming general relativity, we then consider the complementary problem of constraining the tracer bias parameters. Relativistic effects tightly constrain the magnification bias \(s\), and provide some information on \(b_\phi\) through its connection to the evolution bias \(b_{\rm e}\). This breaks the exact \(b_\phi f_{\rm NL}\) product degeneracy present in Newtonian linear power-spectrum analyses, allowing, in principle, the power spectrum to constrain \(f_{\rm NL}\) without imposing an external prior on \(b_\phi\). In practice, however, the \(b_\phi\) constraints from relativistic effects are found to be generally weak for current surveys' configurations. The marginalized \(f_{\rm NL}\) posterior remains highly sensitive to projection effects, and to the assumed parametrization for the redshift evolution of \(b_\phi\), which effectively acts as a prior. 

Even if \(b_\phi\) is not measured with high significance through relativistic clustering, this joint PNG--GR inference provides a way to propagate its uncertainty consistently into the \(f_{\rm NL}\) constraint. External information on \(b_\phi\) through galaxy-formation simulations, or measurements of tracer evolution from data, remains crucial for any competitive $\fNL$ constraint. Nevertheless, measuring $b_\phi$ through relativistic clustering remains an interesting avenue to pursue, especially for more optimized multi-tracer samples and future galaxy surveys.

Through the simulated inference, we have demonstrated that our ultra-large scale modeling of galaxy clustering is now ready for data application. More generally, the theoretical modeling and inference can be extended to cross-correlations across surveys and tracers, including tracers that cover different redshift ranges, as well as correlations with other fields such as CMB lensing, cosmic shear, and intensity mapping surveys. Such cross-correlations will be important for mitigating sample and survey-dependent systematics and for further tightening constraints on both local PNG and relativistic contributions. In general, optimized multi-tracer analyses will be essential to fully exploit ultra-large-scale clustering as a probe of both local PNG and general relativity on horizon scales.

\section*{Acknowledgements} 
\copyright 2026 California Institute of Technology. Government sponsorship acknowledged. We thank Uros Seljak, James Sullivan, Neal Dalal, Boryana Hadzhiyska, Hai-yang Wang, Yun-ting Cheng, and the SPHEREx cosmology team for useful discussions. We acknowledge support from the SPHEREx project under a contract from the NASA/GODDARD Space Flight Center to the California Institute of Technology. Part of this work was done at Jet Propulsion Laboratory, California Institute of Technology, under a contract with the National Aeronautics and Space Administration (80NM0018D0004). RYW further acknowledges support through the Canada Graduate Research Scholarship-- Doctoral program (CGRS D) from the Natural Sciences and Engineering Research Council of Canada (NSERC).

\appendix
\section{Redshift Error Modeling}\label{sec:redshift-error}
We consider a linear integral operator acting on the density field
\begin{align}
\delta^{\mathcal K}(\br)
=\int d^3\bx\,\mathcal K(\br,\bx)\delta(\bx)\label{eq:convolution}\,,
\end{align}
where $\mathcal{K}(\br,\bx)$ is a generic kernel, $\br$ denotes the observed coordinate, and $\bx$ the underlying true coordinate. We allow $\br$ and $\bx$ to be defined on different radial domains, e.g. $\br\in[r_{\rm min},r_{\rm max}]$ and $\bx\in[x_{\rm min},x_{\rm max}]$, since objects selected in the observed domain can originate from a broader true domain due to redshift errors.

We distinguish the SFB basis functions on the two different radial domains as
\begin{subequations}
\begin{align}
    e^{\mu}(\br)&\equiv g_{n_{\mu}\ell_{\mu}}(r)Y_{\ell_{\mu}m_{\mu}}(\hat{\br})\\
\widetilde{e}^{\mu}(\bx)&\equiv \widetilde{g}_{n_{\mu}\ell_{\mu}}(x)Y_{\ell_{\mu}m_{\mu}}(\hat{\bx})\,,
\end{align}
\end{subequations}
where the tilde indicates functions and indices on the underlying $\bx$-domain. We can then write the bivariate SFB transform and its associated expansion for the kernel function as
\begin{subequations}
\begin{align}
\mathcal K_{\rho;\widetilde{\mu}}
&=\int d^3\br\, e_{\rho}(\br)\int d^3\bx\,\widetilde e_{\mu}(\bx)\mathcal K(\br,\bx),\\
\mathcal K(\br,\bx)
&=\sum_{\rho,\mu}\mathcal K_{\rho;\widetilde{\mu}}e^{\rho}(\br)\widetilde e^{\mu}(\bx).
\end{align}
\end{subequations}

Using orthogonality of the SFB basis, the SFB modes of the transformed field satisfy\footnote{The effect of a linear integral
operator on the SFB basis was considered in Ref.~\cite{26Wen_PSM}, where the same radial domain is used for both $\br$ and $\bx$. We now generalize to two different radial domains, each with its own set of radial basis functions.}
\begin{align}
(\delta^{\mathcal K})_{\rho}
=\sum_{\mu}\mathcal K_{\rho;}^{\phantom{*}\widetilde{\mu}}\delta_{\widetilde{\mu}}\,.
\end{align}
The SFB PS of the transformed field is therefore
\begin{align}
\langle(\delta^{\mathcal K})_{\rho}(\delta^{\mathcal K})^{\lambda}\rangle
=\sum_{\mu,\nu}
\mathcal K_{\rho;}^{\phantom{*}\widetilde{\mu}}
\mathcal K^{\lambda;}_{\phantom{*}\widetilde{\nu}}
\langle\delta_{\widetilde{\mu}}\delta^{\widetilde{\nu}}\rangle\,.
\end{align}

\begin{figure}[tbp]
\centerline{\includegraphics[width=0.77\hsize]{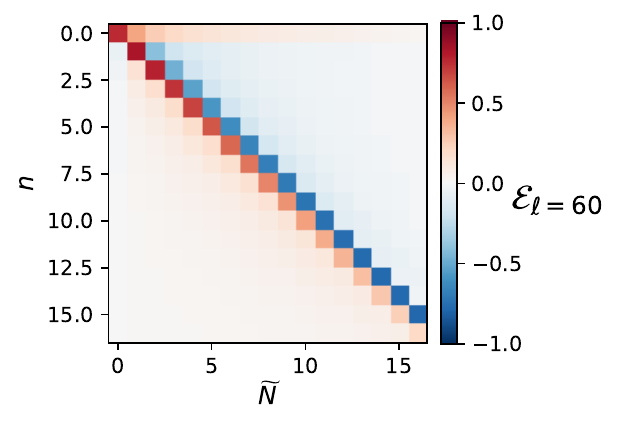}}
\caption{Illustration of the radial mixing matrix induced by redshift errors in the SFB basis. We consider an observed redshift range $z\in[0.2,0.5]$ with a constant radial uncertainty of $\sigma_r=10\,{\rm Mpc}/h$. The index $n$ labels radial modes of the observed range, while $\tilde{N}$ denotes radial modes of the underlying (true) range. The mixing is shown up to $k_{\rm max}=0.1\,h/{\rm Mpc}$ for $\ell=60$. The matrix is predominantly diagonal, and the suppression due to redshift uncertainties becomes more pronounced for smaller-scale (larger-$n$) radial modes. In the absence of redshift errors and radial selection effects, and for matched observed and true radial domains, the mixing matrix would reduce to the identity.}
\label{fig:fog_radial_mixing}
\end{figure}

We now specialize to redshift uncertainties, where $\br$ corresponds to the observed galaxy positions (including redshift error) and $\bx$ to the true positions in redshift space. The bivariate kernel for redshift error may be written as
\begin{align}
E(\br,\bx)
=E(\br-\bx|\bx)=\frac{1}{x^2}p(r-x|\bx)\delta^{\rm D}(\hat{\br}-\hat{\bx})\label{eq:redshift-error-def}\,,
\end{align}
where $p(r-x|\bx)$ is the probability density function (pdf) for observing $r$ given the true position $\bx$. The bivariate SFB transform of above kernel is
\begin{align}
E_{\rho;}^{\phantom{*}\widetilde{\mu}}
&=\int d^3\bx\,\frac{\widetilde e^{\mu}(\bx)}{x^2}
Y_{\ell_{\rho}m_{\rho}}(\hat{\bx})\nonumber\\
&\qquad
\int dr\,r^2 g_{n_{\rho}\ell_{\rho}}(r)p(r-x|\bx)\,.
\label{eq:general-redshift}
\end{align}

If a galaxy's redshift error depends only on its true radial distance instead of its angular position on the sky, that is $p(r-x|\bx)=p(r-x|x)$, the kernel simplifies to
\begin{align}
E_{\rho;}^{\phantom{*}\widetilde{\mu}}
=\delta^{\rm K}_{\ell_{\rho}\ell_{\mu}}
\delta^{\rm K}_{m_{\rho}m_{\mu}}
\mathcal E_{\ell_{\rho}n_{\rho}\widetilde n_{\mu}}\,,
\label{eq:radial-dep-redshift}
\end{align}
where the redshift-error radial mixing matrix is
\begin{align}
\mathcal E_{\ell_{\rho}n_{\rho}\widetilde n_{\mu}}
&=\int dx\,\widetilde g_{\ell_{\rho}n_{\mu}}(x)\nonumber\\
&\qquad\int dr\,r^2 g_{\ell_{\rho}n_{\rho}}(r)p(r-x|x)\label{eq:redshift-error-mixing-SFB}\,.
\end{align}
The above expression is the discrete analogue of the result in Ref.~\cite{95Heavens_SFB} for the continuous SFB basis, which depends only on the Bessel function of the first kind. We give an example of the mixing matrix for redshift error in \cref{fig:fog_radial_mixing}. 

The theoretical SFB PS under redshift error independent of angular positions is therefore
\begin{align}
C^{\rm E}_{\ell n_1 n_2}
=\sum_{\widetilde N_1,\widetilde N_2}
\mathcal E_{\ell n_1\widetilde N_1}
\mathcal E_{\ell n_2\widetilde N_2}
C_{\ell\widetilde N_1\widetilde N_2}\label{eq:redshift-error-mixing-SFB-PS}\,.
\end{align}
The above redshift errors only induce radial-mode mixing at each angular multipole, which demonstrates the separation of angular and radial modes offered by the SFB basis.

We emphasize that the above formalism also applies to RSD, where we can take $\br$ as the redshift-space position and $\bx$ as the real-space position. Earlier work \cite{95Heavens_SFB,95Fisher_SFB} modeled the effects of linear RSD on the SFB PS through radial-mode mixing, as expressed in \cref{eq:redshift-error-mixing-SFB-PS}. In contrast, we model linear RSD through the angular kernels in \cref{eq:dSFB_compute,eq:D_R}, following more recent SFB work \cite{21Gebhardt_SuperFab,14Nicola_SFB} and consistent with the approach adopted for angular power spectrum \cite{13DiDio_classgal,19CCL,20Fang_FFTlog}.

Besides redshift error, \cref{eq:redshift-error-mixing-SFB,eq:redshift-error-mixing-SFB-PS} are useful for modeling non-linear RSD effects such as Fingers-of-God (FoG) \cite{72Jackson_FOG}, where random virial motions within galaxy clusters induce additional velocity dispersion along the line of sight. Such line-of-sight velocity dispersion can also be modeled through the pdf $p(r-x|x)$. Mathematically, redshift error and FoG act on the underlying field in the same way through radial convolution, and therefore cannot be distinguished through clustering statistics alone. We have ignored FoG effects for our forecasts in the main text.

Our above expressions apply to any form of redshift-error or FoG pdf. Common analytical choices include Gaussian and Lorentzian\footnote{\Cref{eq:L-pdf} is technically a Laplace distribution. The Fourier transform of \cref{eq:L-pdf} yields the Lorentzian damping function in Fourier space, as shown in \cref{eq:L-k-pdf}. We use ``Lorentzian'' for consistency with the FoG literature.} forms \cite{20Gebhardt_non-linear}
\begin{subequations}
\begin{align}
p^{\rm Gau.}(r-x|x)&=\frac{1}{\sqrt{2\pi}\sigma_{r\label{eq:FoG-pdf}}(x)}
\exp\left[-\frac{(r-x)^2}{2\sigma_{r}(x)^2}\right]\,,\\
p^{\rm Lor.}(r-x|x)&=\frac{1}{\sqrt{2}\sigma_{r}(x)}
\exp\left[-\frac{\sqrt{2}|r-x|}{\sigma_{r}(x)}\right]\label{eq:L-pdf}\,,
\end{align}
\end{subequations}
where $\sigma_r(x)$ denotes the uncertainty in comoving distance at the true distance $x$. It is related to the redshift uncertainty $\sigma_z(z)$ by
\begin{align}
\sigma_r(x(z))=\frac{c}{H(z)}\sigma_z(z)\,,
\end{align}
where $H(z)$ is the Hubble parameter and $c$ is the speed of light.

For a sample defined over the observed comoving distance range $[r_{\rm min},r_{\rm max}]$, redshift uncertainties cause the true radial coordinates to scatter outside this interval. When the redshift error or FoG effect is small, this leakage is negligible, and one may adopt the same radial domain for both the true and observed coordinates. However, for large photometric-like redshift errors, the support of the true radial coordinate must be extended:
\begin{align}
    [x_{\rm min},x_{\rm max}]&=[r_{\rm min}-\Delta,r_{\rm max}+\Delta]\nonumber\\
    &\supset[r_{\rm min},r_{\rm max}]\label{eq:x_support}\,,
\end{align}
where $\Delta$ is chosen to encompass the support of the redshift-error pdf $p(r-x|r)$, ensuring that almost all galaxies selected in the observed range $[r_{\rm min},r_{\rm max}]$ are included in the true range.

Besides the redshift error pdf, the radial mixing matrices in \cref{eq:redshift-error-mixing-SFB} can also incorporate radial selection functions
\begin{align}
\mathcal E_{\ell_{\rho}n_{\rho}\widetilde n_{\mu}}
&=\int dx\, \widetilde g_{\ell_{\rho}n_{\mu}}(x)A(x)\nonumber\\
&\qquad\int dr\,r^2 g_{\ell_{\rho}n_{\rho}}(r)p(r-x|x)B(r)\label{eq:redshift-error-mixing-SFB-radial-select}\,,
\end{align}
where the radial selection can be applied either in true-redshift space via $A(x)$ or in observed-redshift space through $B(r)$. If one simply counts galaxies as a function of observed redshift, the resulting selection is $B(r)$. In comparison, methods such as clustering redshifts \cite{08Newman_clusteringz,13Menard_clusteringz} allow one to measure the actual radial distribution $A(x)$ over the true redshift. The above equation naturally accommodates both cases. In this work, we use the true redshift distribution $A(x)$ from \cite{SPHEREx_data} for SPHEREx samples. Similarly, the redshift-error pdf can also be taken as $p(r-x|r)$, where the redshift error $\sigma(r)$ is measured as a function of the observed distance $r$ instead of the true distance.

\begin{figure}[tbp]
\centerline{\includegraphics[width=0.97\hsize]{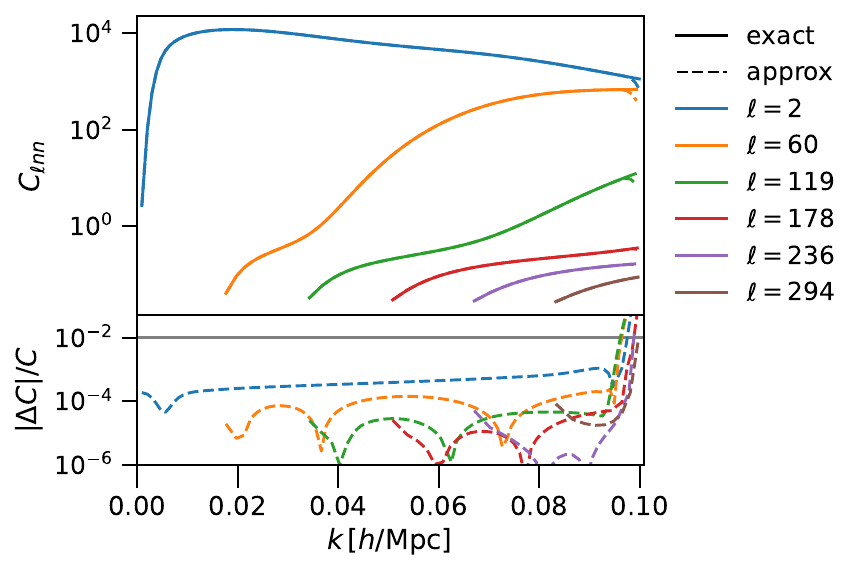}}
\caption{Validation of the convolution approach for incorporating radial selection functions into the SFB power spectrum, compared to its direct inclusion in the SFB kernel. In the legend, “approx” denotes the convolution approach, while “exact” denotes the direct-integration calculation. We adopt the specifications for SPHEREx Sample A over $z=0$–$2$, with a strongly non-uniform radial selection function (shown in \cref{fig:SPHEREx_Spec}) that decreases exponentially over orders of magnitude as a stress test. The upper panel shows the diagonal SFB PS $C_{\ell nn}$ ($n_1=n_2=n$) for representative angular multipoles computed with both methods, while the lower panel displays the relative errors. The two approaches agree to high precision across most Fourier scales, well below the percent level indicated by the grey band. Deviations in the convolution method appear only near the imposed cutoff scale $k_{\rm max}=0.1\,h/{\rm Mpc}$, chosen for the underlying theoretical SFB power spectrum prior to convolution, because the mode mixing induced by radial selection functions receives contributions from modes beyond the cutoff.}
\label{fig:radial_convolve_accuracy}
\end{figure}

In the limit of no redshift errors $p(r-x|x)=\delta^{\rm D}(r-x)$, the radial mixing matrix will only contain radial selection effects
\begin{align}  
\mathcal{E}^{\rm R}_{\ell_{\rho}n_{\rho}\widetilde{n}_{\mu}}&=\int_{r_{\rm min}}^{r_{\rm max}} dr\,r^2 g_{\ell_{\mu}n_{\rho}}(r)\widetilde{g}_{\ell_{\mu}n_{\mu}}(r)R(r)\,,
\label{eq:radial-radial-mixing-SFB}
\end{align}
with $A(x)=B(r)=R(r)$. Here the integration is restricted to the narrower (observed) redshift range. \Cref{eq:radial-radial-mixing-SFB,eq:redshift-error-mixing-SFB-PS} therefore offer an alternative way to include radial selections in modeling SFB PS compared to \cref{eq:Wnlq_kernel}. One may either include the radial selection directly into the evaluations of SFB kernels or convolve the theoretical SFB PS using the radial mixing matrices. 

The two approaches are equivalent, as demonstrated numerically in \cref{fig:radial_convolve_accuracy}. For the convolution method, the SFB power spectrum remains accurate over most of the $k$-range and only shows deviations near the maximum wavenumber $k_{\rm max}^{\rm conv}$ adopted in the underlying theoretical SFB calculation. To maintain accuracy up to the desired scales, one should therefore choose a slightly larger $k_{\rm max}^{\rm conv}$ in the theoretical input. The convolution approach is particularly convenient for the Chebyshev-based method in the multi-tracer case, as discussed in the next section.

\begin{figure}[tbp]
\centerline{\includegraphics[width=0.97\hsize]{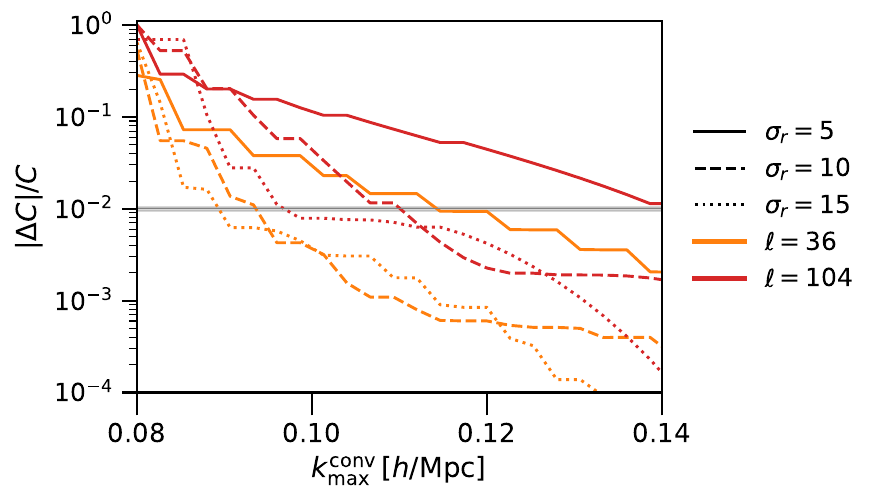}}
\caption{Convergence of the non-perturbative redshift-error convolution with respect to the maximum Fourier scale $k_{\rm max}^{\rm conv}$ included in the underlying theoretical SFB power spectrum. For each value of $k_{\rm max}^{\rm conv}$, we truncate the theory $C_{\ell n_1 n_2}$ to modes lower than that scale before applying the radial mixing matrices, and compare the resulting observed (convolved) spectrum at a fixed diagonal mode $k \approx 0.08\,h/\mathrm{Mpc}$ to the reference obtained using the full $k_{\rm max}^{\rm conv} = 0.16\,h/\mathrm{Mpc}$. Results are shown for two representative multipoles ($\ell=2$ and $\ell=104$) and three values of the radial displacement $\sigma_r = 5,10,15\,\mathrm{Mpc}/h$, with the grey band marking the $1\%$ level. Smaller $\sigma_r$ requires a higher $k_{\rm max}^{\rm conv}$ for convergence, reflecting the narrower radial mixing kernel that couples more higher $k$ modes into each observed mode.}
\label{fig:fog_kmax_convergence}
\end{figure}

For our forecast results, we use the radial convolution method to incorporate radial selection functions in the multi-tracer DESI cases, and to include both radial selections and redshift errors for all SPHEREx samples. To ensure accuracy, we compute the underlying theoretical SFB PS over a slightly extended $k$ range, with a buffer of $\Delta k = 0.02\,h/{\rm Mpc}$~\footnote{This buffer choice is justified by the accuracy of radial convolution demonstrated in \cref{fig:radial_convolve_accuracy,fig:fog_kmax_convergence}.}. For the Fisher forecasts, we therefore use $k_{\rm max}^{\rm conv}=0.10\,h/{\rm Mpc}$ for the underlying theoretical SFB PS, yielding convolved spectra accurate up to $k=0.08\,h/{\rm Mpc}$. For the simulated inference, we use $k_{\rm max}^{\rm conv}=0.08\,h/{\rm Mpc}$, yielding convolved spectra accurate up to $k=0.06\,h/{\rm Mpc}$.

\begin{figure*}[tbp]
\centerline{\includegraphics[width=0.93\hsize]{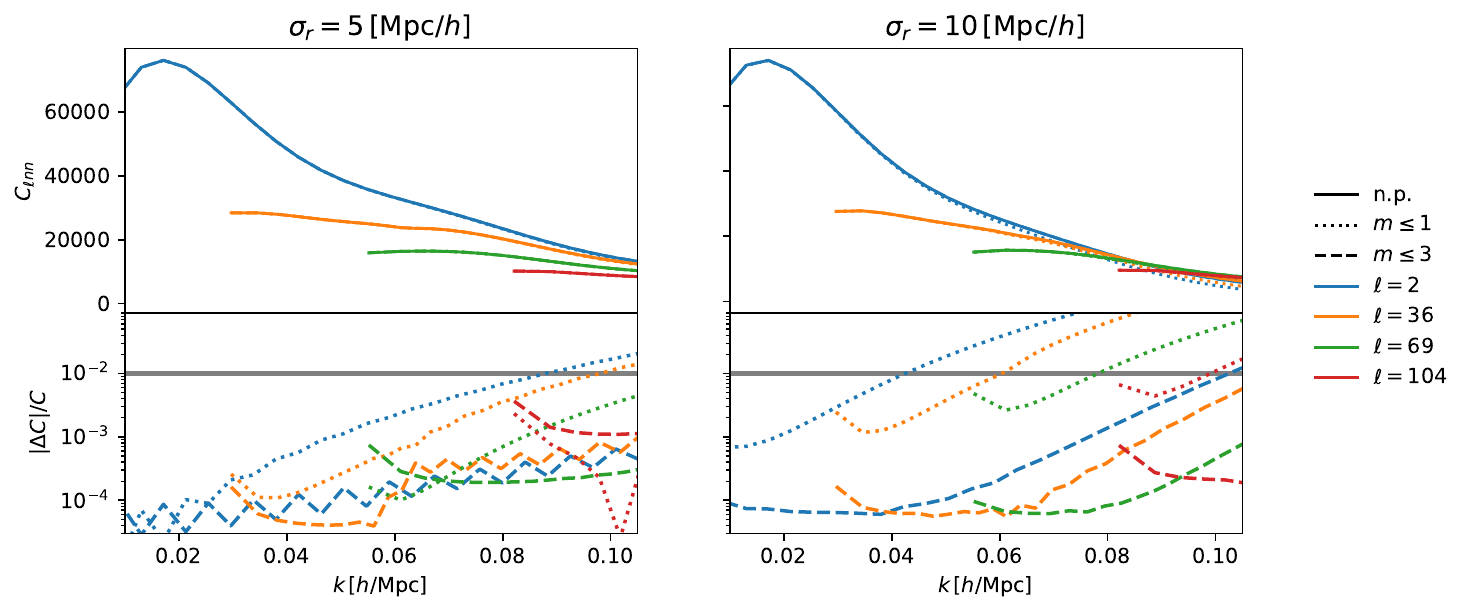}}
\caption{Comparison of perturbative and non-perturbative treatments of FoG/redshift-error effects in the SFB power spectrum for an observed redshift range $z\in[0.2,0.5]$ with $b_1=1.5$. The upper panels show the diagonal SFB PS for several angular multipoles, while the lower panels display the relative error of the perturbative approximations with respect to the non-perturbative results. The non-perturbative case is abbreviated as n.p. in the legend. The perturbative order shown in the legend is $m$, corresponding to the Taylor expansion order $n=2m$ of redshift error pdfs in \cref{eq:general-p-expansion}, since odd orders vanish by symmetry. For the non-perturbative approach, we compute and convolve the theoretical SFB power spectrum up to $k_{\rm max}=0.16\,h/{\rm Mpc}$. For a small FoG dispersion, $\sigma_r = 5\,{\rm Mpc}/h$, the leading-order approximation remains accurate at the sub-percent level for $k \leq 0.08\,h/{\rm Mpc}$. In contrast, for a larger error of $\sigma_r = 10\,{\rm Mpc}/h$, the leading-order treatment becomes significantly less accurate and requires higher-order perturbative corrections.}
\label{fig:fog_5_10}
\end{figure*}

\subsection{Perturbative approximation}

We compare the exact radial convolution treatment of redshift error/FoG with the perturbative approximation introduced in Refs.~\cite{21Gebhardt_SuperFab,23Gebhard_SFB_eBOSS}. We review the derivation of the perturbative method and then discuss the complementarity of the two approaches.

We first demonstrate how the redshift-error operator defined in \cref{eq:convolution,eq:redshift-error-def} in configuration space is represented in Cartesian Fourier space. The observed density contrast in \cref{eq:GR} can be written as the inverse Fourier transform \cite{23WAGR}
\begin{align}
\delta(\bx)
&=\int_{\bq} e^{i\bq\cdot\bx}
\mathcal{F}_{\rm DRSD}(q,x,\hat{\bq}\cdot\hat{\bx})
\delta_{m0}(\bq)\label{eq:iF-DRSD}\,,
\end{align}
where $\int_{\bq}\equiv(2\pi)^{-3}\int d^3\bq$, and the following Fourier kernel encodes the linear density and RSD effects
\begin{align}
\mathcal{F}_{\rm DRSD}(q,x,\hat{\bq}\cdot\hat{\bx})=
\left[b_1(x)+f(x)(\hat{\bq}\cdot\hat{\bx})^2\right]D(x)\nonumber\,.
\end{align}
We ignore the linear relativistic effects here\footnote{Strictly speaking, the pdf-based FoG prescription should not be applied to linear relativistic terms, since the FoG modeling is neither relativistic nor linear.}, since GR contributions decay with increasing radial mode numbers \cite{24GR_SFB}, while the damping effects from FoG become increasingly important at smaller radial scales (as shown in \cref{fig:fog_radial_mixing}). Consequently, the regimes where these two effects are non-negligible are well-separated.

Substituting \cref{eq:iF-DRSD} into \cref{eq:convolution}, and formally extending the integration domain\footnote{This extension is proper since the radial integration domain $[x_{\rm min},x_{\rm max}]$ in \cref{eq:x_support} is chosen to contain the full support of the redshift-error kernel, so the integrand will become negligible outside this region.} from the spherical shell of $[\bx_{\rm min},\bx_{\rm max}]$ to $\mathbb{R}^3$, we obtain
\begin{align}
&\delta^{\rm E}(\br)=\int_{\bx,\bq}e^{i\bq\cdot\br}E(\br-\bx|\br)\mathcal{F}(q,x,\hat{\bq}\cdot\hat{\bx})e^{i\bq\cdot(\bx-\br)}\delta_{m0}(\bq)\nonumber \\
&\approx\int_{\bq}e^{i\bq\cdot\br}\mathcal{F}(q,r,\hat{\bq}\cdot\hat{\br})\delta_{m0}(\bq)\int_{\bx-\br}E(\bx-\br|\br)e^{i\bq\cdot(\bx-\br)}\nonumber\\
&=\int_{\bq}e^{i\bq\cdot\br}\mathcal{F}(q,r,\hat{\bq}\cdot\hat{\br})E(-\bq|\br)\delta_{m0}(\bq)\label{eq:redshift-error-Fourier-general}\,.
\end{align}
The approximation occurs when we replace $\mathcal{F}(q,x,\hat{\bq}\cdot\hat{\bx})\approx \mathcal{F}(q,r,\hat{\bq}\cdot\hat{\bx})$, neglecting the differences between evaluating radial functions (such as the linear bias and growth factor) at the true coordinate $x$ and at the observed coordinate $r$.  This replacement is justified provided the FoG/redshift-error kernel is sufficiently narrow, such that the variation of these cosmological functions across the kernel width is negligible\footnote{These radial cosmological functions are fundamentally functions of the galaxies' true redshift. 
For redshift errors, they are therefore defined at the true coordinate $x$ (prior to redshift uncertainty), whereas for FoG effects they are defined at the observed coordinate $r$, since the galaxies' redshift already includes the FoG contribution.}. This condition holds for typical FoG dispersions of a few ${\rm Mpc}/h$ and for small redshift uncertainties of $\mathcal{O}(10^{-3})$.

With the redshift error pdf in the form of \cref{eq:redshift-error-def}, its Fourier transform over $\bDx\equiv \bx-\br$, the difference between the two coordinates, is
\begin{align}
E(\bk|\br)&\equiv\int_{\bDx}e^{-i\bk\cdot\bDx}E(\bDx|\br) \nonumber\\
&=\int_{\bx}e^{-i\bk\cdot\bDx}\frac{1}{x^2}p(x-r|\br)\delta_D(\hat{\br}-\hat{\bx})\nonumber\\
&=\int_{x}e^{-i(\bk\cdot\hat{\br})(x-r)}p(x-r|\br)=p(k\mu_{r}|\br)\label{eq:redshift-error-1DFourier} \,,
\end{align}
where $\mu_r\equiv\hat{\bk}\cdot\hat{\br}$, and $p(k\mu_r|\br)$ denotes the corresponding 1D Fourier transform (over $x-r$) of the radial pdf. For the Gaussian and Lorentzian FoG models in \cref{eq:FoG-pdf}, this 1D Fourier transform becomes \cite{20Gebhardt_non-linear}
\begin{subequations}
\begin{align}
p^{\rm Gau.}(k\mu_r|r)&=
\exp\left[-\frac{1}{2}\sigma_r(r)^2k^2\mu_r^2\right]\label{eq:G-k-pdf}\,,\\
p^{\rm Lor.}(k\mu_r|x)&=\frac{1}{1+\frac{1}{2}\sigma_r(r)^2k^2\mu_r^2}
\label{eq:L-k-pdf}\,.
\end{align}
\end{subequations}

Therefore, \cref{eq:redshift-error-Fourier-general} becomes
\begin{align}
  \delta^{\rm E}(\br)&=\int_{\bq}e^{i\bq\cdot\br}\mathcal{F}(q,r,\hat{\bq}\cdot\hat{\br})p(q\mu_r|r)\delta_{m0}(\bq) \label{eq:redshift-error-Fourier-1D}\,,
\end{align}
which expresses the FoG/redshift-error effects in Cartesian Fourier space. The above expression coincides with the starting point of Sec.~II B in Ref.~\cite{21Gebhardt_SuperFab}. 
Following their derivation and carrying out the expansion in the SFB basis, the SFB kernels in \cref{eq:Wnlq_kernel} are modified by FoG/redshift errors as
\begin{align}
\mathcal{W}_{n\ell}^{E}(q)&\equiv\sqrt{\frac{2}{\pi}}q\int_{x_{\rm min}}^{x_{\rm max}}dx\,x^2g_{n\ell}(x)R(x)\nonumber\\
&\quad p(-iq\partial_{qx}|x)\Delta^{\rm DRSD}_{\ell}(x,q)\label{eq:Wnlq_kernel_fog}\,,
\end{align}
where the factor $\mu$ is replaced by derivatives with respect to $qx$. In general, operators that are functions of derivatives are difficult to evaluate exactly. However, their action on functions can be treated perturbatively by expanding the Fourier-space pdf around $k\mu=0$:
\begin{align}
p(-iq\partial_{qx}|x)
=\sum_{n=0}^{\infty}
\frac{(-iq)^n}{n!}
\left.
\frac{\partial^n p(y|x)}{\partial y^n}
\right|_{y=0}\partial^n_{qx}\,.
\label{eq:general-p-expansion}
\end{align}

In particular, for symmetric pdfs such as Gaussian and Lorentzian with only even powers contributing, it is convenient to define the perturbative order in terms of $n=2m$. 
They yield the same leading-order expression
\begin{align}
&p^{\rm Gau./Lor.}(-iq\partial_{qx})j^{(d)}_\ell(qx)\approx
j^{(d)}_\ell(qx)\nonumber\\
&\qquad+\frac12\sigma_{r}(x)^2q^2\, j^{(d+2)}_\ell(qx)\label{eq:FoG-Bessel-perturb}\,.
\end{align}
At leading order, the Gaussian and Lorentzian prescriptions are therefore indistinguishable. This resulting correction term matches the 1-loop counterterm that absorbs FoG effects in the effective field theory of large-scale structure \cite{20Ivanov_BOSS}.

We quantify the accuracy of the perturbative approximations for FoG/redshift-error modeling of SFB PS in \cref{fig:fog_5_10}. For a small velocity dispersion of $5\,{\rm Mpc}/h$, the leading-order approximation remains accurate at the sub-percent level for $k \leq 0.08\,h/{\rm Mpc}$. In contrast, for a larger dispersion of $\sigma_r = 10\,{\rm Mpc}/h$, the leading-order treatment becomes significantly less accurate, and higher-order perturbative corrections are required, as shown in \cref{fig:fog_5_10}. The perturbative expansion becomes less reliable for photometric-like redshift errors, where going to higher-order terms becomes cumbersome. In such cases, non-perturbative approaches are preferable due to their accuracy. We therefore adopt the non-pertubative approach for redshift error modeling of all SPHEREx samples.

For typical galaxy FoG dispersions of a few ${\rm Mpc}/h$, however, the leading-order approximation is sufficiently accurate. We use it for modeling the redshift error of the Euclid H$\alpha$ sample. The leading-order approximation is computationally fast and can be straightforwardly incorporated into the Chebyshev method as discussed in Appendix~\ref{sec:angular-kernel}. This makes it well-suited for inference, where FoG parameters can be varied and marginalized over. By contrast, the non-perturbative radial mixing matrices in \cref{eq:redshift-error-mixing-SFB} are generally too expensive to recompute at each likelihood evaluation, which prevents their practical use when FoG/redshift-error parameters are allowed to vary.

For small velocity dispersions, the radial convolution method requires evaluating the underlying theoretical SFB power spectrum up to significantly higher $k_{\rm max}^{\rm conv}$ in order to achieve a desired accuracy at a given $k$, as shown in \cref{fig:fog_kmax_convergence}. This substantially increases the computational cost and, in practice, makes the method both slower and less efficient than the perturbative approach due to the need for going to higher $k_{\rm max}^{\rm conv}$. The situation improves for larger velocity dispersions, where the required $k_{\rm max}^{\rm conv}$ is considerably reduced.  The two methods are therefore complementary, each being optimal in different regimes of dispersion.

For FoG dispersions and small redshift errors, the perturbative expansion is accurate and can be directly incorporated into likelihood analyses. For large redshift errors, where the convolution method is required, a hybrid strategy allows one to retain both modeling accuracy and parameter marginalization. Importantly, for Gaussian redshift-error PDFs, successive convolutions remain Gaussian, with variances that add linearly. We can therefore decompose the convolution with a large redshift error into two components: (i) a dominant redshift-error contribution treated non-perturbatively via radial convolution; and
(ii) a residual FoG/redshift-error component that is sufficiently small to be handled perturbatively. In practice, the bulk redshift error of a given sample can typically be measured and calibrated with good precision and incorporated through the non-perturbative convolution. The remaining uncertainty in the redshift error is then small enough to be modeled perturbatively and marginalized over during inference.

\section{Chebyshev Decomposition}\label{sec:Chebyshev-method}

\subsection{Motivation}

In the SFB PS computation of \cref{eq:dSFB_compute,eq:Wnlq_kernel}, the main computational bottleneck lies in filling the SFB kernel $\mathcal{W}_{n\ell}(q)$, which involves radial integrals over spherical Bessel functions. Ref.~\cite{23Gebhard_SFB_eBOSS} proposed the Iso-qr integration method for constructing the SFB kernels under Newtonian RSD, which was later extended to include GR effects in Ref.~\cite{24GR_SFB}. The method exploits the fact that spherical Bessel functions depend only on $qr$, the product of comoving distance and Fourier mode. By discretizing $q$ and $r$ such that these iso-qr curves coincide with grid points, the spherical Bessel functions $j_{\ell}(qr)$ can be efficiently precomputed and cached, enabling extensive reuse across $(q,r)$ pairs and substantially reducing the cost of kernel evaluations.

With this approach, the evaluation of the SFB PS takes seconds on a single CPU under Newtonian RSD, enabling inference through MCMC using the SFB PS \cite{23Gebhard_SFB_eBOSS}. However, when the full set of GR effects is included, the runtime increases to tens of seconds per evaluation, which becomes prohibitive for inference and motivates the need for further computational speed-up.

The key idea is therefore to precompute the expensive radial integrations by separating the dependence on cosmological and astrophysical (bias) parameters from the radial functions. This is achieved by expanding the relevant radial functions in a suitable basis, such that the parameter dependence enters only through a small number of expansion coefficients. The radial integrals over the basis functions can then be computed once and stored, allowing subsequent evaluations of the SFB kernel to be performed through fast linear combinations.

This idea underlies the FFTLog-based approaches used for angular power spectrum (TSH) \cite{20Fang_FFTlog}, where radial functions are decomposed into power-law components that can be integrated analytically. However, FFTLog is best suited to cases where the integrand varies smoothly over a wide radial range. In the present context, the SFB power spectrum for spectroscopic surveys involves discrete redshift bins with sharp radial boundaries, which can reduce the robustness of FFTLog-based methods. 

We therefore adopt Chebyshev polynomials as the decomposition basis, which provide a numerically stable and efficient representation of smooth functions on a finite interval. Chebyshev polynomials have been used in Ref.~\cite{24Chiarenza} for the computation of angular power spectra, where the Fourier matter power spectrum is expanded in a Chebyshev basis and the corresponding Fourier integrals are precomputed. In comparison, our approach decomposes the radial dependence and precomputes the associated radial integrals.

Our use of Chebyshev polynomials is also closely related to their application in cosmographic studies, where they have been employed to reconstruct redshift-dependent functions such as the expansion history or dark energy equation of state \cite{18Capozziello_chebyshev_cosmographic,19Capozziello_cosmography,12Shafieloo_crossing,24DESI_DE_reconstruct}. Chebyshev polynomials provide stable and rapidly convergent reconstructions for smooth functions, and the Chebyshev coefficients used in our calculations can be naturally incorporated into inference for reconstructing such redshift-dependent cosmological functions. Indeed, we already employed Chebyshev polynomials to parametrize the redshift-dependent bias functions for our inference (Sec.~\ref{sec:bias-param}).

\subsection{Chebyshev polynomials} \label{sec:Chebyshev-back}
The Chebyshev polynomials are defined as
\begin{equation}
    T_n(\cos\theta) = \cos(n\theta),
\end{equation}
or, equivalently, by the recursion relation
\begin{equation}
\begin{aligned}
    T_0(t) &= 1, \\
    T_1(t) &= t, \\
    T_{n+1}(t) &= 2t\, T_n(t) - T_{n-1}(t).
\end{aligned}
\end{equation}
They form a complete and orthonormal basis over the interval $[-1,1]$, and are widely used in approximation theory due to their favorable convergence properties and near-optimal approximation performance \cite{19Trefethen,01Boyd}. 

In general, a function that is (Lipschitz) continuous on $t\in[-1,1]$ admits a unique Chebyshev series expansion
\begin{equation}\label{eqn:a1}
    u(t) = \sum_{n=0}^{\infty} a_n T_n(t),
\end{equation}
with coefficients given by
\begin{equation}\label{eqn:cheb1}
    a_n \equiv \frac{2}{\pi} \int_{-1}^1\mathrm{d}t\,\frac{u(t)T_n(t)}{\sqrt{1-t^2}}\,.
\end{equation}
In practice, however, computing these coefficients directly through numerical quadrature is both computationally expensive and numerically less robust due to the $1/\sqrt{1-t^2}$ integrand weight. Instead, it is more efficient to construct a finite-degree Chebyshev interpolant using values of the function sampled at Chebyshev points, which yields a near-minimax polynomial approximation and allows the expansion coefficients to be obtained stably through a discrete cosine transform.

We approximate the function using interpolation at the Chebyshev points\footnote{Here we use the Chebyshev--Lobatto points (including the endpoints of the interval) instead of the Chebyshev--Gauss points (excluding the endpoints) \cite{16Xu_chebyshev1st} for interpolation, since Chebyshev--Lobatto points align more naturally with the discrete Fourier transform. We follow the second-kind interpolation scheme described in Ref.~\cite{02Mason_Chebyshev}.}
\begin{equation}\label{eqn:chebpoints}
    t_k = \cos{\left(\frac{k}{N-1}\pi\right)},\;k=0,...,N-1.
\end{equation}
The interpolating polynomial constructed from 
$N$ Chebyshev points can be written as \cite{02Mason_Chebyshev}
\begin{equation}\label{eqn:cheb_pk}
    \tilde{u}(t)=\sum_{n=0}^{N-1} u_nT_n(t)=\sum_{n=0}^{N-1} w_nc_nT_n(t),
\end{equation}
where the weights satisfy $w_n=1$ for $1\leq n\leq N-2$ and $w_0=w_{N-1}=1/2$ for the end points. The Chebyshev coefficients are given by
\begin{align}
c_n = \frac{2}{N-1}\sum_{k=0}^{N-1} w_k u_k \cos(\frac{\pi n k}{N-1})\,,
\label{eq:cheby-interp-transform}
\end{align}
where $u_k\equiv u(t_k)$ is the function evaluated at Chebyshev points. 

\Cref{eq:cheby-interp-transform} corresponds to a discrete cosine transform of type-I (DCT-I) applied to the sampled values $[u_0,u_1,\ldots,u_{N-2},u_{N-1}]$ \cite{00Trefethen}. The DCT-I transform can be efficiently implemented using a fast Fourier transform by symmetrically extending the data as $[u_0,u_1,\ldots,u_{N-2},u_{N-1},u_{N-2},\ldots,u_1]$ \cite{84WangFFT}, obtaining the $c_n$ coefficients. For the relatively small expansion orders required in our SFB power spectrum calculations, the difference between FFT-based and direct evaluation of \cref{eq:cheby-interp-transform} is negligible.

For functions defined on a finite radial interval $x\in [x_{\rm min},x_{\rm max}]$, we can rescale the function coordinates to the standard Chebyshev domain $[-1,1]$ via
\begin{equation}
t(x)=\frac{2x - (x_{\max} + x_{\min})}{x_{\max} - x_{\min}} \label{eq:coordinate-mapping}\,,
\end{equation}
which allows application of the above Chebyshev interpolation scheme.

\begin{figure}[tbp]
\centerline{\includegraphics[width=0.92\hsize]{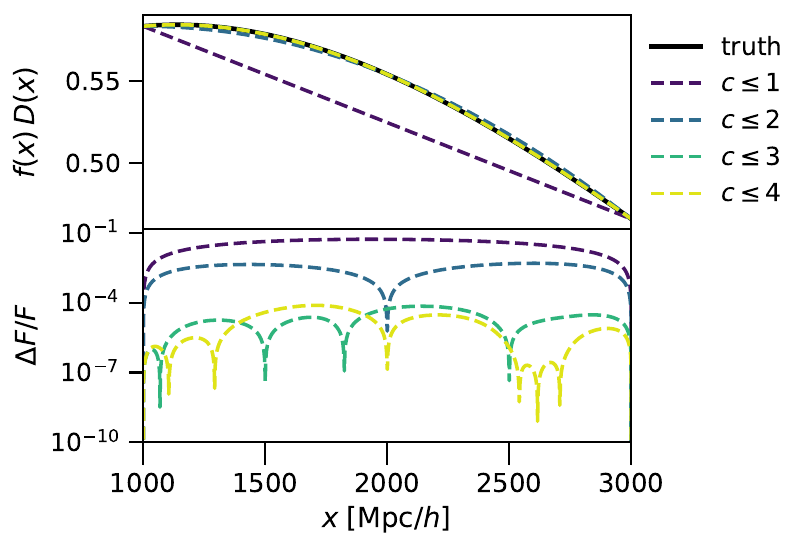}}
\caption{Example of Chebyshev interpolation of a radial function in angular kernels. Here we choose $f(x)D(x)$ for the RSD term in \cref{eq:E_R}. The upper panel shows the exact function together with Chebyshev interpolants of increasing expansion order (labelled by $c$). The lower panel shows the corresponding relative error, illustrating the rapid convergence of the approximation as the expansion order is increased.}
\label{fig:cheby_fd}
\end{figure}

\subsection{Decomposing angular kernels}\label{sec:angular-kernel}
We summarize the angular kernels $\Delta_{\ell}(x, q)$ corresponding to all linear-order GR effects in 
\cref{eq:GR} (excluding observer terms). Following Refs.~\cite{11Bonvin_GR,13DiDio_classgal,24GR_SFB}, these kernels are
\begin{widetext}
\begin{subequations}\label{eq:angular_kernel}
\begin{align}
&\Delta_{\ell}(x, q)^{\rm PNG}= \frac{f_{\rm NL}b_{\phi}(x)}{\alpha(q,x)}D_{\rm m}(x,q)j_{\ell}(qx)\label{eq:D_PNG}\\
&\Delta_{\ell}(x, q)^{\rm DRSD}=b_1(x)D_{\rm m}(x, q)j_{\ell}(q x)+\frac{q}{\mathcal{H}(x)}v(x, q) j_{\ell}''(q x)\label{eq:D_R}\\
&\Delta_{\ell}(x, q)^{\rm Doppler}=\mathcal{A}_1(x) v(x, q)j_{\ell}'(qx) \label{eq:D_D} \\
&\Delta_{\ell}(x, q)^{\rm NIP}=\Bigg[\left(\mathcal{A}_1(x)+1\right)\Psi(x, q)-(2-5s(x)) \Phi(x, q)+\frac{1}{\mathcal{H}(x)}\dot{\Phi}(x, q)+\left(b_{\rm e}(x)-3\right)\mathcal{H}(x)V(x, q)\Bigg]j_{\ell}(q x)\label{eq:D_G}\\
&\Delta_{\ell}(x, q)^{\rm Lensing}=\ell(\ell+1)\frac{2-5 s(x)}{2}\int_{0}^{x}dr\,\frac{x-r}{x r}(\Phi(r, q)+\Psi(r, q))j_{\ell}(q r) \label{eq:D_L}\\
&\Delta_{\ell}(x, q)^{\rm Shapiro}=\frac{2-5 s(x)}{x}\int_{0}^{x}dr\, (\Phi(r, q)+\Psi(r, q)) j_{\ell}(q r)\label{eq:D_S}\\
&\Delta_{\ell}(x, q)^{\rm ISW}=\mathcal{A}_1(x)\int_{0}^{x}dr\, \left(\dot{\Phi}(r, q)+\dot{\Psi}(r, q)\right)j_{\ell}(q r)\,.\label{eq:D_I}
\end{align}
\end{subequations}
\end{widetext}
Here $T(x(z),q)$ represents the transfer function for the quantity $T$ (e.g. $D_{\rm m}$ for matter density or $V$ for the velocity potential), defined relative to the present-time matter perturbation $D_{{\rm m},0}(q)$. In late-time $\Lambda$CDM, where anisotropic stress from radiation and neutrinos can be neglected, the linear transfer functions are separable into a time-dependent part (parameterized by comoving distance $x(z)$) and a $q$-dependent part. This separability simplifies our Chebyshev scheme, since cosmological and bias parameters enter only through redshift-dependent radial functions. Therefore, we only need to decompose radial functions that are independent of $q$ under this late-time $\Lambda$CDM approximation.

We first consider the non-integrated terms. Under the late-time $\Lambda$CDM cosmology, all their angular kernels in \cref{eq:angular_kernel} can be factorized as the following
\begin{align}
    \Delta^{\rm E}_{\ell}(x, q)=K^{\rm E}(q)F^{\rm E}(x)\omega^{\rm E}(x)j_{\ell}^{(i^{\rm E})}(q x)\label{eq:NI-factor}\,,
\end{align}
where $K^{\rm E}(q)$ encodes the Fourier dependence, $\omega^{\rm E}(x)$ stores the radial functions that are independent of cosmological or bias parameters and will be pre-computed, and $i^{\rm E}$ is the order of Bessel derivatives. Only $F^{\rm E}(x)$ depends on the parameters, which need to be decomposed with Chebyshev polynomials

\begin{subequations}\label{eq:F-NI}
\begin{align}
    &F^{\rm D}(x)=b_1(x)D(x)\label{eq:E_D}\\
    &F^{\rm RSD}(x)=-f(x)D(x)\label{eq:E_R}\\
    &F^{\rm PNG}(x)= f_{\rm NL}b_{\phi}(x)\frac{3H_0^2\Omega_{\rm m0}}{2\tilde{D}(x)}D(x)\label{eq:E_PNG}\\
    &F^{\rm Dopp}(x)=\mathcal{A}_1(x)v(x)x\label{eq:E_Dopp}\\
    &F^{\rm NIP}(x)=\Bigg[\left(\mathcal{A}_1(x)+1\right)\Psi(x)-(2-5s(x)) \Phi(x)\nonumber\\
    &\qquad+\frac{1}{\mathcal{H}(x)}\dot{\Phi}(x)+\left(b_{\rm e}(x)-3\right)\mathcal{H}(x)v(x)\Bigg]x\,,\label{eq:E_NIP}
\end{align}
\end{subequations}
where $D(x)=D_{\rm m}(x,q)$ is the linear growth factor normalized at the current time, and
\begin{subequations}\label{eq:approx-transfer}
 \begin{align}
&v(x)=-f(x)\mH(x)D(x)\label{eq:approx-v}\\
&\Phi(x)=\Psi(x)=-\frac{3}{2}\mH^2(x)\Omega_{\rm m}(x)D(x)\label{eq:approx-Phi}\\
&\dot{\Phi}(x)=\dot{\Psi}(x)=-\frac{3}{2}\mH^3(x)\Omega_{\rm m}(x)(f(x)-1)D(x)\label{eq:approx-dot-Phi}
\end{align}
\end{subequations}
which give the radial parts of the transfer functions for velocity and potentials under the late time $\Lambda$CDM
\cite{03Dodelson,22CatorinaGR-P,23WAGR}. The Fourier part and the Bessel derivative order of the kernels are 
\begin{subequations}
\begin{alignat}{3}
K^{\rm D}(q)&= 1          &\qquad i^{\rm D}   &= 0 \\
K^{\rm RSD}(q) &= 1         &\qquad i^{\rm RSD} &= 2 \\
K^{\rm PNG}(q) &= q^{-2}T^{-1}(q) &\qquad i^{\rm PNG} &= 0 \\
K^{\rm Dopp}(q) &= q^{-1} \qquad & i^{\rm Dopp} &= 1 \\
K^{\rm NIP}(q)  &= q^{-2} \qquad & i^{\rm NIP}     &= 0\,.
\end{alignat}
\end{subequations}

The parameter-independent parts of the radial functions are
\begin{align}
&\omega^{\rm  D}(x)=\omega^{\rm  RSD}(x)=\omega^{\rm  PNG}(x)=1\\
&\omega^{\rm  Dopp}(x)=\omega^{\rm  NIP}(x)=\frac{1}{x}
\end{align}
For Doppler and NIP terms, we extract an additional $1/x$ from the radial functions because the Doppler/ISW amplitude parameter $\mathcal{A}_1(x)$ (\cref{eq:A1}) contains $1/x$. When the survey radial range is $[0,x_{\rm max}]$, decomposing $1/x$ with Chebyshev polynomials becomes unstable, so we keep $1/x$ in the precomputation part (which is regulated by the $x^2$ weight in the radial integral). As a result, \cref{eq:E_Dopp,eq:E_NIP} contain an extra factor of $x$.

Using the Chebyshev interpolation scheme described in \cref{sec:Chebyshev-back}, we decompose the parameter-dependent radial functions as
\begin{align}
   F^{\rm E}(x)=\sum_{c}F^{\rm E}_c T_c(t(x))\label{eq:radial-decompose-Cheby}\,,
\end{align}
with the coordinate remapping of \cref{eq:coordinate-mapping}.

Therefore, the SFB kernels of \cref{eq:Wnlq_kernel} for the non-integrated terms are linear combinations of precomputed SFB templates weighted by the Chebyshev coefficients
\begin{align}
    \mathcal{W}^{\rm NI}_{n\ell}(q)=\sum_{c}\sum_{\rm E}F^{\rm E}_cK^{\rm E}(q)\mathcal{W}_{n\ell}^{c,i^{\rm E}}(q)\label{eq:kernel-NI-sum}\,,
\end{align}
where the SFB kernel template is
\begin{align}
    W_{n\ell}^{c,i^{\rm E}}(q)&=\sqrt{\frac{2}{\pi}}q\int_{x_{\rm min}}^{x_{\rm max}}dx\,x^2g_{n\ell}(x)R(x)\nonumber\\
    &\quad T_{c}(t(x))\omega^{\rm E}(x)j_{\ell}^{(i^{\rm E})}(q x)\label{eq:kernel-NI-precompute}\,.
\end{align}

The leading-order perturbative expansion for FoG/redshift error in \cref{eq:FoG-Bessel-perturb,eq:Wnlq_kernel_fog} can be easily included in the above Chebyshev decomposition approach by introducing the following additional radial functions that require Chebyshev decompositions
\begin{subequations}
   \begin{align}
&F^{\rm D, FoG}(x) = \frac{1}{2}b_1(x)D(x)\sigma_{r}(x)^2\\
&F^{\rm RSD, FoG}(x) = -\frac{1}{2}f(x)D(x)\sigma_{r}(x)^2
\end{align} 
\end{subequations}
with Fourier part being $K(q)=q^2$ and Bessel derivative orders being $i^{\rm D,FoG}=2$ and $i^{\rm RSD, FoG}=4$ respectively. 

For the integrated terms, we can organize in the following form
\begin{align}
&\Delta^{\rm E}_{\ell}(x, q)=K^{\rm I}(q)F^{\rm E}(x)I_{\ell}^{\rm E}[Y^{\rm E}](x)\,,
\end{align}
where we introduced the following integral operator acting on an arbitrary scalar function $Y(r)$
\begin{subequations}
\begin{align}
    I_\ell^{\rm Shap}[Y](x) &= I_\ell^{\rm ISW}[Y]= \frac{1}{x}\int_{0}^{x}dr\,Y(r)j_{\ell}(q r)\\
    I_\ell^{\rm Lens}[Y](x) &= \frac{\ell(\ell+1)}{2}\Bigg(\int_{0}^{x}dr\,\frac{1}{r}Y(r)j_{\ell}(q r)\nonumber\\
    &\quad-I_\ell^{\rm Shap}[Y](x)\Bigg)\,,
\end{align}
\end{subequations}
and the integrands for each of these effects are simply
\begin{subequations}
\begin{align}
    &Y^{\rm Lens}(r)=Y^{\rm Shap}(r)=\Phi(r)+\Psi(r)\label{eq:Y_S}\\
    &Y^{\rm ISW}(r)= \dot\Phi(r)+\dot\Psi(r)\,.\label{eq:Y_ISW}
\end{align}
\end{subequations}
All integrated terms share the same Fourier dependence $K^{\rm I}(q)=q^{-2}$, and the radial dependence functions are
\begin{subequations}\label{eq:F-I}
\begin{align}
    &F^{\rm Lens}(x)=F^{\rm Shap}(x)=2-5 s(x)\label{eq:E_S}\\
    &F^{\rm ISW}(x)= \mathcal{A}_1(x)x\,.\label{eq:E_ISW}
\end{align}
\end{subequations}
Similar to the Doppler and NIP terms with dependence on $\mathcal{A}_1(x)$, we also precompute the $1/x$ factor for the ISW term to avoid decomposing the inverse function.

For the integrated terms, besides decomposing the outer radial functions $F^{\rm E}(x)$ as done in \cref{eq:radial-decompose-Cheby}, we also need a separate decomposition of the inner integrand, which is defined on $[0,x_{\rm max}]$ rather than the tracer range $[x_{\rm min},x_{\rm max}]$
\begin{align}
    Y^{\rm E}(r)= \sum_{d}Y_{d}^{\rm E}T_{d}(t_0(r))\label{eq:radial-decompose-Cheby2}\,,
\end{align}
where $t_0$ denotes the mapping in \cref{eq:coordinate-mapping} but for the interval $[0,x_{\rm max}]$.

Then the SFB kernel for integrated term is
\begin{align}
    \mathcal{W}^{\rm I}_{n\ell}(q)=K^{\rm I}(q)\sum_{c,d}\sum_{\rm E}F^{\rm E}_cY_d^{\rm E}\mathcal{W}_{n\ell}^{cd,{\rm E}}(q)\label{eq:kernel-I-sum}\,,
\end{align}
and the template kernels for integrated term will be 
\begin{align}
    W_{n\ell}^{cd,{\rm E}}(q)&=\sqrt{\frac{2}{\pi}}q\int_{x_{\rm min}}^{x_{\rm max}}dx\,x^2g_{n\ell}(x)R(x)\nonumber\\
    &\quad T_{c}(t(x))I^{\rm E}_\ell[T_d(t_0)](x)\label{eq:kernel-I-precompute}\,.
\end{align}

\begin{figure}[tbp]
\centerline{\includegraphics[width=0.95\hsize]{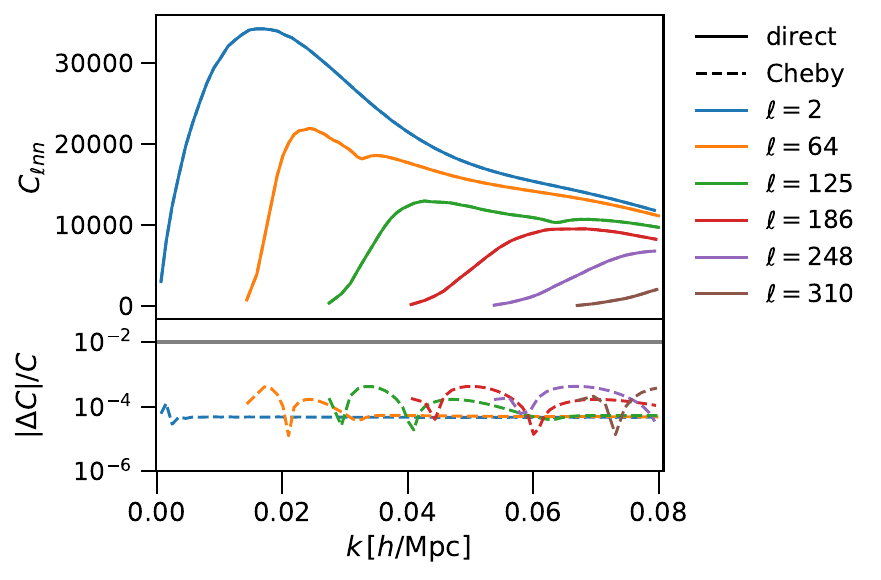}}
\caption{Accuracy of the Chebyshev method for the linear density and RSD terms in the diagonal SFB power spectrum $C_{\ell nn}$ under the specifications of the DESI QSO sample. We compare the direct computation against the Chebyshev decomposition method for several angular multipoles $\ell$. The Chebyshev method achieves great numerical precisions with relative errors at $O(10^{-4})$.}
\label{fig:cheby_drsd}
\end{figure}

\begin{figure}[tbp]
\centerline{\includegraphics[width=0.95\hsize]{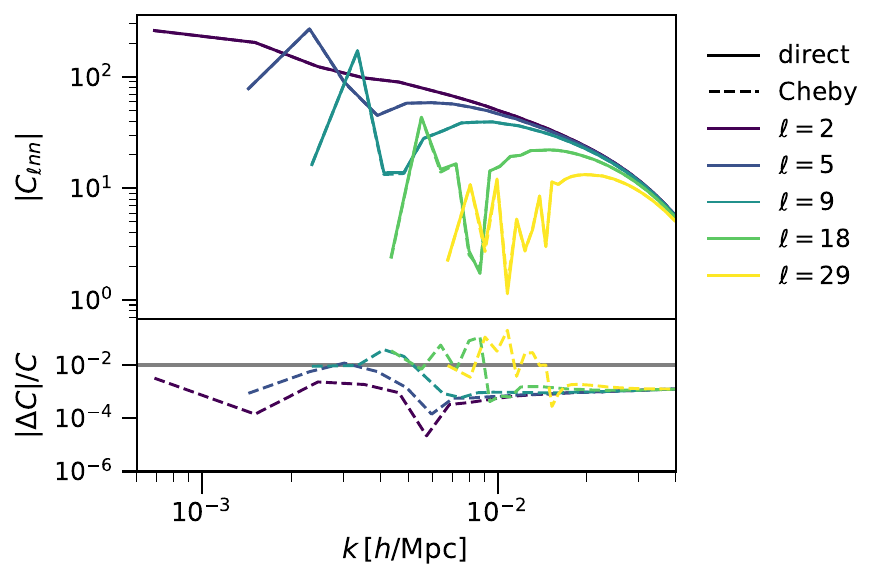}}
\caption{Accuracy of the Chebyshev method for the linear relativistic effects (excluding density and RSD contributions) in the diagonal SFB PS $C_{\ell nn}$ for the DESI QSO sample. The Chebyshev method reproduces the GR terms at the percent level compared to their direct evaluation, which is adequate for parameter inference given that these contributions are subdominant compared to the density and RSD terms. The non-smooth features in the signals are not numerical artifacts, but arise mainly from the lensing-density cross term. At small $n$ (large radial modes), the oscillations are caused by the combined effects of sign changes, radial-mode mixing induced by the QSO radial selection function, and the parity dependence of the lensing effect on the radial-mode index $n$ (i.e. odd versus even $n$; see Sec.~V D of Ref.~\cite{24GR_SFB}).}
\label{fig:cheby_GR}
\end{figure}

\subsection{Implementaion}

For the Chebyshev decompositions in \cref{eq:radial-decompose-Cheby,eq:radial-decompose-Cheby2}, only a few coefficients are needed because the functions in \cref{eq:F-NI,eq:F-I} are smooth (see \cref{fig:cheby_fd} as an example). We also require the bias parametrization in Sec.~\ref{sec:bias-param} to have fewer degrees of freedom than the Chebyshev expansion orders used here. For the non-integrated terms, we take $c=0$-$4$ to obtain high accuracy for the density and RSD terms. Since the integrated (GR) terms require less relative accuracy, we use $c,d=0$-$3$. As a result, we use 25 templates for the non-integrated terms (including leading-order FoG) and 32 templates for the integrated terms. 

To precompute the SFB kernel templates, we still use the Iso-qr integration scheme described in Refs.~\cite{23Gebhard_SFB_eBOSS,24GR_SFB}. Computing the full template set can take tens of minutes on a single CPU. These templates can then be stored; during inference one simply uses \cref{eq:kernel-NI-sum,eq:kernel-I-sum} to reconstruct the kernels and then performs the Fourier integration in \cref{eq:SFB-discrete-PS}. For a single tracer, the runtime is sub-second, and the main computational bottleneck becomes the linear summations in \cref{eq:kernel-NI-sum,eq:kernel-I-sum}. Since the template kernels carry indices for both SFB modes $(n,\ell)$ and Fourier modes $q$, they are large, and these summations become memory-bandwidth-limited.

Here we do not decompose the radial selection function $R(x)$, since they can be significantly less smooth than the cosmological and bias functions and would require significantly higher Chebyshev orders. For a single tracer with a known selection function, we directly incorporate $R(x)$ into the precomputed templates. For multi-tracer analyses with the same redshift range (the case considered in this work), we instead use the radial mixing matrices in \cref{eq:radial-radial-mixing-SFB} to convolve the theoretical SFB power spectrum, so that we still require only a single set of templates for all tracers, significantly reducing the memory cost for saving the templates.

We present the accuracy of our Chebyshev-based method for computing the SFB power spectrum in \cref{fig:cheby_drsd} for the Newtonian density and RSD contributions, and in \cref{fig:cheby_GR} for the GR contributions. The Chebyshev results are benchmarked against direct evaluations of the SFB kernels using the Iso-qr integration method. We find excellent agreement for the Newtonian terms and achieve percent-level accuracy for the GR terms. 

Our method has assumed the separability between Fourier and radial dependence as factorized in \cref{eq:NI-factor}, and we use the late-time $\Lambda$CDM approximations in \cref{eq:approx-transfer} for the linear transfer functions. For more general cosmologies, these transfer functions must be computed from Boltzmann codes such as \texttt{CAMB} and \texttt{CLASS} and will generally be not separable. They also become non-separable in the non-linear regime (e.g. for the non-linear matter power spectrum). In such cases, one has to perform Chebyshev decompositions of the radial functions for each $q$, schematically
\begin{align}
    D(x,q)=\sum_{a}D_{a}(q)T_a(x)\,,
\end{align}
and the rest of our methods still hold. We leave such more general implementations to future works when considering cosmologies beyond $\Lambda$CDM.

\bibliography{refs}

\end{document}